\documentclass[a4paper,12pt]{article}

\usepackage{hyperref}
\usepackage{mathtools}
\usepackage{physics}
\usepackage{dsfont}
\usepackage{bbm}
\usepackage{amsmath}
\usepackage{mathrsfs}
\usepackage{amsthm}
\usepackage{xcolor}
\usepackage{mathrsfs} 
\usepackage{amssymb}
 \usepackage{color}
\usepackage[margin=1.6cm]{geometry}
\usepackage{amsfonts}
\usepackage{amsthm}
\usepackage{amscd}
\usepackage[latin2]{inputenc}
\usepackage{t1enc}
\usepackage[mathscr]{eucal}
\usepackage{indentfirst}
\usepackage{graphicx}
\usepackage{graphics}
\usepackage{pict2e}
\usepackage{epic}
\usepackage[affil-it]{authblk}
\usepackage{authblk}

\numberwithin{equation}{section}

\providecommand{\keywords}[1]
{
  \small	
  \textbf{\textit{Keywords---}} #1
}

\newcommand\cH{{\mathcal H}}
\newcommand\bR{{\mathbb R}}

\newtheorem{theorem}{Theorem}[section]
\newtheorem{corollary}{Corollary}[theorem]
\newtheorem{proposition}{Proposition}[theorem]
\newtheorem{lemma}[theorem]{Lemma}
\newtheorem{remark}[theorem]{Remark}

\newcommand{\Cov}{\mathrm{Cov}}
\newcommand{\diag}{\mathrm{diag}}

\newcommand\ttr{{\mathtt r}}
\newcommand\ttp{{\mathtt p}}
 \newcommand{\Span}{\mathrm{Span}}

\begin{document}

    \title{Hydrodynamic limit for a disordered quantum harmonic chain }
    \author{Amirali Hannani \thanks{hannani@ceremade.dauphine.fr \\ https://orcid.org/0000-0002-9516-1223}}
	\affil{CEREMADE, UMR-CNRS 7534, Universit\'{e} Paris Dauphine-PSL Research
University, 75775 Paris cedex 16, France }        
    \date{\today}
     
    \maketitle    
    \abstract{We study the hydrodynamic limit, in the hyperbolic space-time scaling, for a one-dimensional unpinned chain of quantum harmonic oscillators with random masses. To the best of our knowledge, this is among the first examples where one can prove the hydrodynamic limit for a quantum system rigorously. In fact, we prove that after hyperbolic rescaling of time and space, the distribution of the elongation, momentum, and energy averaged under the proper locally Gibbs state converges to the solution of the Euler equation. Moreover, our result indicates that the temperature profile is frozen in any space-time scale; in particular, the thermal diffusion coefficient vanishes. There are two main phenomena in this chain that enable us to deduce this result. First is the Anderson localization, which decouples the mechanical and thermal energy, providing the closure of the equation for energy. The second phenomenon is similar to some sort of decay of correlation phenomena, which let us circumvent the difficulties arising from the fact that our Gibbs state is not a product state due to the quantum nature of the system.}
   

     \section{Introduction} \label{intro}
     Obtaining the macroscopic evolution of conserved quantities and their corresponding currents for a "physical" system from its microscopic dynamics, also known as hydrodynamic limit, is a matter of interest both in Physics and Mathematics communities. In physics literature, heuristic arguments as well as general assumptions such as ergodicity, linear response, and local equilibrium lead to the formal derivation of such equations (see \cite{herbert} and references therein). In recent years the former idea has been adapted to integrable systems with infinite conserved quantities via the theory of generalized hydrodynamic and introducing the generalized Gibbs ensemble (GGE) \cite{PRL},\cite{PRX},\cite{Spohn2020}. \\
      Rigorous mathematical treatment of this problem is usually much harder due to difficulties in proving the ergodic properties of the system, and it has been subjected to much research in recent decades.
 (see \cite{kipnis}, \cite{anna} and references therein). 
      One of the most interesting cases of this program is when the underlying microscopic dynamics is given by the fundamental law of nature, i.e., either Newtonian/Hamiltonian dynamics or Schr\"{o}dinger/Heisenberg dynamics for classical and quantum systems, respectively. In classical systems, having a mathematical proof is still quite challenging; in fact, one of the main strategies in the case of classical systems is to obtain ergodicity by adding certain stochastic noise to kill all the conserved quantities except the desired ones (see \cite{stefano93},\cite{stefano14}, \cite{stefano16} as examples of this strategy). One of the main mathematical tools for controlling the macroscopic evolution in these examples is the relative entropy method \cite{kipnis}. It is worth mentioning that in certain limits when we have energy conservation, this method does not work, and methods based on Wigner distributions can be exploited (see e.g., \cite{stefano18}). \\
      
      The situation is different when the underlying dynamics is given by quantum mechanics. Although there are several works where one derives the effective macroscopic evolution equation for a many-body quantum system such as proving the Non-linear Schr\"{o}dinger equation  from Quantum many-body dynamics, usually assuming low-density regime, (see e.g., \cite{erdos}, \cite{erdos2}, and \cite{niels} and references therein), these body of works are different from "Hydrodynamic Limits". In fact, to the best of our knowledge, there are no examples where one rigorously proves a hydrodynamic limit for a deterministic quantum system, except in \cite{BN03}, where they adapted the relative entropy method to quantum systems. However, their work is based on an ergodicity assumption (assumption III), which is quite challenging to prove for physical systems. We should also mention the recent progress in studying the Euler space-time scale of Hamiltonian quantum systems in \cite{doyon}, where the author shows the \textbf{linearised Euler equation} (the Euler equation for correlation functions, in contrast to our case, where here we show
the full Euler equation), in its most general form, involving all conserved quantities admitted by the model for a general class of one-dimensional spin chains with short-range interactions.
Note that contrary to our situation, the problem of closing the equation, i.e., showing that there are finite (in our case three) conserved quantities in the macroscopic equation, remains open.
      \\ 
      
      One of the main purposes and novelties of this work is proving the hydrodynamic limit for a simple interacting quantum system (one dimensional unpinned chain of quantum harmonic oscillator with random masses). Formally, this system can be described by the following Hamiltonian :
       
  \begin{equation}  \label{Ham00}
	H_n=\frac{1}{2}\sum_{x=1}^n \Big(\frac{p_x^2}{m_x} + (q_{x+1}-q_{x})^2\Big),
 \end{equation}
     where $q_x$ is the position (multiplication) operator acting on particle $x$, $p_x$ is the corresponding momentum operator with $[q_x,p_x]=i$, where we take $\hbar=1$. Moreover, $m_x$ denotes the mass of particle $x$, where they will be taken as i.i.d random variables. Finally, we have free boundary conditions $q_0=q_1$ and $q_n=q_{n+1}$.  \\
	    
	    The time evolution is given by the Heisenberg dynamics generated by $H_n$, i.e.  
\begin{equation} \label{equationofmotionq0}
	\dot{p}_x= i[H_n,p_x]=(\Delta q)_x, \: \: \: \: \: \dot{q}_x=i[H_n,q_x]=\frac{p_x}{m_x}, 
\end{equation}   
   where $\Delta$ is the discrete gradient defined in\eqref{productofgrad}.\\
  
   This dynamic has $n$ conserved quantities; however, we are only interested in the following three main ones: Total energy: $H_n:=\sum_{x=1}^n e_x$, total momentum: $\sum_{x=1}^n p_x$, and total elongation $\sum_{x=1}^{n-1} r_x$, where $r_x=q_{x+1}-q_x$ and $e_x=\frac12(\frac{p_x^2}{m_x}+r_x^2)$. Notice that total energy $H_n$ and total momentum $\sum_{x=1}^n p_x$ are truly conserved. In contrast, total elongation $\sum_{x=1}^{n-1} r_x$ is locally conserved, and global conservation of elongation breaks at the boundary due to boundary conditions $q_0=q_1$, and $q_n=q_{n+1}$.   We discuss the other conserved quantities in Remark \ref{OCQ}. \\
	
	First, we let the chain be in a locally Gibbs state corresponding to the aforementioned conserved quantities. This state is out of thermal equilibrium with a smooth temperature profile $\beta \in C^0([0,1])$, and out of mechanical equilibrium with a smooth profile of momentum and elongation $\bar{p},\bar{r} \in C^1([0,1])$, with $\bar{r}(0)=\bar{r}(1)=0$. Furthermore, we assume the chain's macroscopic center of the mass is fixed i.e., 
	$\int_0^1 \bar{p}(y)dy=0$ (We discuss this assumption in Appendix \ref{appendix0} and Remark \ref{remarkdef}). The density operator of this state is denoted by $\rho^n_{\beta,\bar{p},\bar{r}}$, and is defined in \eqref{initalstate}. Then, we let the chain evolve in time, according to the Heisenberg equation of motion and denote  $r_x(t),p_x(t),e_x(t)$ to be the solution of this  equation, i.e. the solution to the following local conservation laws corresponding to our conserved quantities:  
\begin{equation}
\begin{split} 
& \dot{r}_x(t) = \frac{p_{x+1}(t)}{m_{x+1}} -\frac{p_x(t)}{m_x}, \quad
		\dot{p}_x(t) = r_x(t) -r_{x-1}(t), \\
	&\dot{e}_x(t)= \frac{p_{x+1}(t)r_x(t)+r_x(t)p_{x+1}(t)}{2m_{x+1}} - 
\frac{p_x(t)r_{x-1}(t)+r_{x-1}(t)p_{x}(t)}{2m_x},
\end{split}
\end{equation}	
	 	where in the second equation, we emphasized the fact that $p_{x+1}r_x \neq r_xp_{x+1}$.\\
  Correspondingly, denoting $\bar{m}= \mathbb{E}(m_x)$, the macroscopic profiles evolve according to the following conservation laws with proper boundary conditions: 
	\begin{equation} \label{macro}
	\begin{split}		
		&\partial_t \mathrm{r}(y,t)= \frac{1}{\bar{m}}\partial_y \mathrm{p}(y,t), \quad
		\partial_t \mathrm{p}(y,t)=\partial_y \mathrm{r}(y,t), 
			\quad	
		 \partial_t \mathrm{e}(y,t)=\frac{1}{\bar{m}}\partial_y (\mathrm{r}(y,t) \mathrm{p}(y,t)), \\
		 &\mathrm{r}(y,0)=\bar{r}(y), \qquad \mathrm{p}(y,0)=\bar{p}(y), \qquad \mathrm{e}(y,0)=\frac12\Big(\frac{\bar{p}(y)^2}{\bar{m}}+\bar{r}(y) \Big)+\mathrm{f}_{\beta}^{\mu}(y),\\
		 & \mathrm{r}(0,t)= \mathrm{r}(1,t)=0, 
		\end{split}		
	\end{equation}	 
			where $\mathrm{f}_{\beta}^{\mu}(y)$ is uniquely determined by the profile of temperature and the distribution of the masses, we discuss about this function in Section \ref{SLLNsection} and Appendix \ref{app1}, in particular its definition is given in \eqref{fdef} (See Remark \eqref{frmk} as well). \\
		 Notice that total momentum $\int_0^1 \mathrm{p}(y,t)dy$, is conserved by the evolution equation \eqref{macro}, thanks to the boundary condition $\mathrm{r}(0,t)= \mathrm{r}(1,t)=0$.
		   This fact further justifies our assumption $\int_0^1 \bar{p}(y)dy=0$.\\

		 The main result of this manuscript is that, after hyperbolic rescaling of time and space, the empirical density of elongation, momentum, and energy, i.e. $r,p,e$, averaged under the locally Gibbs state $\rho^n_{\bar{p},\bar{r},\beta}$, converge to the solution of the macroscopic equation \eqref{macro}. The precise statement of the result is given in Theorem \ref{maintheorem}.\\

    A similar result has been obtained for the classical counterpart of this system in \cite{BHO}. The first feature of the system which permits us to close the equation for the energy and prove the hydrodynamic limit is the localization phenomena expressed in Lemma \ref{localizationlemma} mathematically. In fact, models of disordered chains (both classical and quantum mechanical) have been studied extensively in the literature (see e.g., \cite{FA}, \cite{Theo}, \cite{Joel1}, \cite{Dhar}, \cite{Rubin1971} for classical and \cite{NSS} for quantum mechanical). Most of these models exhibit properties similar to the Anderson Insulator \cite{Anderson} i.e. certain eigenmodes of these chains are spatially localized. In our case, the random matrix appearing in the analysis of our system is $M^{-1}\Delta$, where $M$ is the diagonal matrix of the masses. We denote the set of eigenvectors of $M^{-1}\Delta$ by $\{ \psi^k \}_{k=0}^{n-1}$, where they are ordered according to their corresponding eigenvalues.  One can deduce from the conservation of momentum that the ground state of this matrix is fixed, also known as "symmetry protected mode" \cite{Halp}. Consequently, the localization length diverges as we approach the ground state, namely for $k \sim n^{\gamma}$, such that  $\gamma \in [0,\frac12)$ is chosen properly. In fact, one can observe that the localization length $\xi_k$ behaves asymptotically as $\xi_k \sim \omega_k^2 \sim (\frac{k}{n})^2$, as we take the limit $\frac{k}{n} \to 0$ properly (see e.g. \cite{FA}, \cite{Theo}), where $\omega_k$ denotes the eigenvalue of the clean chain with unit mass \footnote{By "clean" we refer to the chain with all the masses are equal.}, i.e. $\omega_k=|2\sin(\pi \frac{k}{n})|$. Moreover, high modes (when $k$ is not close to zero) are completely localized.  \\ 
    In the microscopic level, the modes with $k \ll \sqrt{n}$ remain extended, i.e., similar to the eigenmodes of the clean chain (Fourier modes). First, the macroscopic  evolution of $r,p$ follows this low modes, this is proven via a mass homogenization in Section \ref{masshom}. Moreover, in the microscopic and macroscopic level, we decompose the energy into the thermal and mechanical part. In the macroscopic level, by solving \eqref{macro}, the evolution of the energy is purely mechanical; in the microscopic level low modes transport the mechanical energy, this fact is proven in Lemma \ref{kinticlemma}. In fact, the convergence of the microscopic mechanical energy to the macroscopic one can be deduced from the results of Section \ref{WE}; in particular, the pointwise convergence of the momentum and elongation in Step 4 of Section \ref{WE}. In the macroscopic scale, the thermal part of the energy does not evolve in time, since the  macroscopic evolution of the energy is purely mechanical. In the microscopic level, the similar phenomenon can be proven thanks to the localization phenomena where we expressed in lemma \ref{localizationlemma}, \ref{localizationlemma2}, exploiting the estimates appeared in \cite{BHO}, \cite{FA}, \cite{Theo}. In fact, since the high modes, i.e. $k \gg \sqrt{n}$ are localized, the thermal part of the energy (Thermal fluctuations) does not evolve in time, this has been proven in Lemma   \ref{Thermallimitlemma}. Notice that the following Remark is a direct consequence of this lemma:

    \begin{remark}
    In Lemma \ref{Thermallimitlemma}, the proof can be adapted easily for other time scales, where, we rescale time by a factor  $n^{\alpha}$ with $\alpha \geq 1$. Therefore, the  temperature's profile does not evolve in time at any time scale $n^{\alpha}t$ for $\alpha \geq 1$, including diffusive timescale, which means the thermal diffusion coefficient vanishes.
	\end{remark}

     The main difference of our model with the classical counterpart \cite{BHO} stems from the fact that the Gibbs state $\rho_{\bar{p},\bar{r},\beta}^n$ \eqref{initalstate} is not a product state, since the energies of nearest neighbor particles do not commute with each other in the quantum case. This issue leads to some technical difficulties for obtaining certain bounds, which is treated by diagonalizing the pseudo-Hamiltonian $H_{\beta}^n$ \eqref{hamiltoniantemp}, appearing in the definition of $\rho_{\bar{p},\bar{r},\beta}^n$.\\
      However, the more fundamental issue arising here is that for a fixed realization of the masses, $\expval{\tilde{e}_x}_{\rho^n}$, i.e., the average of the thermal energy of the particle $x$, which is computed in \eqref{r2thermalaverage} and \eqref{p2thermalaverage}, depends on the whole configuration of the masses and the whole profile of the temperature ($\beta(\frac{1}{n}), \dots, \beta(\frac{n}{n})$). In contrast, in the classical case it was simply equal to $\beta_x^{-1}=\beta(\frac{x}{n})^{-1}$. This difference also reflects in the macroscopic equation \eqref{macro} in the function $\mathrm{f}_{\beta}^{\mu}$, which is defined in \eqref{fdef}. In contrast to the classical case, where it was equal to $\frac{1}{\beta(y)}$. Notice that the quantum nature of our system survives in the macroscopic limit only through this function. In fact, in the limit $\hbar \to 0$, this function converges to its classical counterpart  $\frac{1}{\beta(y)}$. \\
      Since for a fixed realization of the masses the microscopic average $\expval{\tilde{e}_{[ny]}}_{\rho^n}$ depends on the whole  configuration of the masses, one should think of it as a random variable. Moreover, a priori,  it is not clear that the desired limiting object $\lim_{n \to  \infty} \frac1n \sum_{x=1}^n f(\frac{x}{n})\expval{\tilde{e}_{[ny]}}_{\rho^n}$, will be deterministic. We devote Section \ref{SLLNsection} and Appendix \ref{app1} to this issue. In order to show that this limit is deterministic, we prove that at each point  $x$, the average energy's dependence on the mass of a particle $y$, far away from $x$, decays sufficiently fast. Then we use the Strong Law of Large Numbers for weakly dependent random variables. In order to prove this fact, we use arguments similar to the decay of correlation. Here, we use the fact that the function $\sqrt{z}\coth{\sqrt{z}}$ is analytic in a certain domain, then we expand the average expression, which can be represented in terms of this function, and we use the fact that the matrices $A_p^{\beta}, A_{r}^{\beta}$  appearing in the expansion are local, i.e. the mass of particle $x$ only appears on the entries close to the diagonal term $(A_{p}^{\beta})_{xx}$;  hence, the expectation of the first $\frac{|x-y|}{2}$ terms is factorized, and the rest is small.\\
      
\begin{remark} \label{frmk}
  Notice that when $n$ is finite, $\expval{\tilde{e}_{[ny]}}_{\rho}$ depends
  on the whole profile of temperature ($\beta(\frac{1}{n}), \dots, \beta(1)$).
  We prove that as $n \to \infty$, for any $y\in (0,1)$,
  $\mathrm{f}^{\mu}_{\beta}(y)$ depends only
  on the macroscopic temperature $\beta(y)$.
  In fact, in Proposition \ref{spro}, we prove that
  $\mathrm{f}^{\mu}_{\beta}(y)= \mathrm{f}^{\mu}(\beta(y))$ for all $y \in(0,1)$.
  Here, $\mathrm{f}^{\mu}(\beta_{eq})$ is the corresponding function in thermal equilibrium
  at inverse temperature $\beta_{eq}$, i.e., the case where $\beta(y)=\beta_{eq}$,
  for all $y \in [0,1]$.
  We define this function in \eqref{feqdef},
  and we observe that it is well defined (does not depend on $y$) in Corollary  \ref{feqwd}.
  Finally, notice that in the classical case $\mathrm{f}^{\mu}(\beta_{eq})=\frac{1}{\beta_{eq}}$,
  while  in the quantum case one can observe that this function depends on all the moments
  of the probability distribution of the masses $\mu$.
  Thanks to this observation, it is worth mentioning that in the classical case,
  the macroscopic equations only reflects the expectation of the masses $\mathbb{E}(m_x)$,
  whereas in the quantum case the whole distribution appears
  in the macroscopic picture through the  function $\mathrm{f}^{\mu}$.
\end{remark}
     
     \begin{remark}
       In the classical case, for a clean chain (all the masses are equal to $m$),
       in thermal equilibrium (constant profile of temperature)
       the microscopic dynamic converges to the solution of the Euler equation \eqref{macro},
       with $\mathrm{f}_{\beta}(y)= \frac{1}{\beta}$ (see \cite{BHO}
       Section 1.1).  This fact is still true in the quantum case, where
       one should modify the value of this constant function
       (See Remark \ref{rmkthermaleqavg2} and Remark \ref{rmkthermaleqavg3}):
     	 \begin{equation} \label{thermaleqcte}
     		\mathrm{f}_{\beta}= \beta^{-1}\int_0^1 \frac{\beta \hbar}{\sqrt{m}}\sin(\pi k) \coth(\frac{\beta \hbar}{\sqrt{m}} \sin(\pi k))dk.
     	\end{equation}
Notice that the later expression in the classical limit $\hbar \to 0$,  converges to the previous value of $\mathrm{f}_{\beta}=\beta^{-1}$. Moreover, the asymptotic of $\mathrm{f}_{\beta}$ for large masses or high temperatures  ($m\gg 1$ or $\beta^{-1}\gg 1$) is equal to $\beta^{-1}$ as well.  \\
     	On the other hand, out of thermal equilibrium with an inhomogenous temperature's profile,  one can observe that the randomness of the masses is essential. In fact, one can easily adapt the argument in Section 1.1 of \cite{BHO} to observe that for a clean chain, with a varying (inhomogenous) temperature's profile,  although we have the wave equation, the evolution of the 
        thermal energy is not autonomous. This is in contrast to the disordered case where
        thermal energy does not evolve in time.
        (See (1.22) in \cite{BHO} for more details in the classical case).       	 
     \end{remark}
 \begin{remark} \label{OCQ}
   Since other conserved quantities can be written as further gradients of $p$ and $r$ (See $I_n$ in \eqref{conserevedquantities} as an example), by using the same strategy as we used for the energy, we can decompose them into two parts: one involving mechanical contributions,  and the other involving thermal contributions, similar to \eqref{microdecomp}. The thermal terms will be constant in time, using the similar argument we used for the thermal energy, thanks to the localization. However, since the conserved quantities have been obtained by taking further gradient from $r$ and $p$, the 
   mechanical terms 
   vanish in hyperbolic scaling of time and space,
   and consequently, their macroscopic transport will be suppressed in this scaling. 
	 \end{remark}
	
 We conclude our introduction by mentioning that our result shed light on the transport properties of disordered unpinned chain in the hyperbolic space time scale; mechanical energy's transport is ballistic, while thermal energy 
 transport will be suppressed thanks to \textbf{Anderson Localization}.
 In addition, concerning the other scaling, we observe that the transport
 of the thermal energy will be suppressed at any larger time-scale.

     \section{Model Description And Results}

We set the following conventions: 
\begin{itemize}
 
\item  $\mathbb{I}_n:=\{ 1, \dots ,n\}$.
\item $\mathbb{I}_n^0 :=\{0,1,\dots,n\}$.
\item Denote the set of $n \times n$ real matrices by $\mathcal{M}_n(\mathbb{R})$, for $A \in M_n(\mathbb{R})$ define $A^{\dagger}$ to be the transpose of $A$. 
\item We denote the inner product in $\mathbb{R}^n$ by $\expval{,}_n$. Moreover, for $x \in \mathbb{I}_n$, $\ket{x} $ stands for the following member of the canonical basis of $\mathbb{R}^n$: $(0,\dots,0,1,0,\dots,0)$, where $1$ appears in the $x$th position. We usually denote vectors of $\mathbb{R}^n$ by Greek letters  $\ket{\psi}, \ket{\phi}, \ket{\varphi}, ...$.
\item Let $a$ be an operator (possibly unbounded), defined on a dense subset of the Hilbert space $\mathcal{H}_n=L^2(\mathbb{R}^{n-1})$, we denote the adjoint of $a$ by $a^*$.
\item  $\diag(\lambda_1,\dots , \lambda_n)$ denotes the diagonal matrix with values  $\lambda_i, i \in \mathbb{I}_n$ on the diagonal. 
\end{itemize}

		The finite volume system of size $n$ is defined  as follows: 
	  Let $\mathcal{H}_n$ be the Hilbert space $\mathcal{H}_n=L^2(\mathbb{R}^{n-1})=\bigotimes_{x=1}^{n-1} L^2(\mathbb{R})$, denote the elements of $\mathcal{H}_n$ by bold ket notation: $\pmb{\ket{\psi}}$, and $\pmb{\braket{\psi}{\phi}}$ stands for the usual inner product in $\mathcal{H}_n$.

	We denote the space variable by $\pmb{\xi} \in \mathbb{R}^{n-1}$. Let $\mathcal{S}(\mathbb{R}^{n-1})$ be the Schwartz space of functions from $\mathbb{R}^{n-1} $ to $\mathbb{C}$, which is dense in $\mathcal{H}_n$. For each $x \in \mathbb{I}_{n-1}$, define the elongation operator  $r_x$ on $\mathcal{S}(\mathbb{R}^{n-1})$ as follows:  $\forall \pmb{\xi} \in \mathbb{R}^{n-1}$, and $\pmb{\ket{\psi}} \in \mathcal{S}(\mathbb{R}^{n-1})$, we have 
	$$r_x \pmb{\ket{\psi(\xi_1,\dots,\xi_x,\dots,\xi_{n-1})}}=\pmb{\xi_x} \pmb{\ket{\psi(\xi_1,\dots,\xi_x,\dots,\xi_{n-1})}}.$$ 
	One could think of $r_x$ as the operator denoting the elongation of the spring between the particle $x$ and $(x+1)$. Moreover, for each $x \in \mathbb{I}_{n}$, define the momentum operator $p_x$ on $\mathcal{S}(\mathbb{R}^n)$ as: $p_x=-i(\frac{\partial}{\partial \pmb{\xi}_{x-1}}-\frac{\partial}{\partial \pmb{\xi}_x})$ i.e. for $\pmb{\ket{\psi}} \in \mathcal{S}(\mathbb{R}^{n-1})$, 
	$$ p_x \pmb{\ket{\psi(\xi_1,\dots,\xi_x,\dots,\xi_{n-1})}}=-i \Big(\frac{\partial}{\partial \pmb{\xi}_{x-1}}-\frac{\partial}{\partial \pmb{\xi}_x}\Big) \pmb{\ket{\psi(\xi_1,\dots,\xi_x,\dots,\xi_{n-1})}},$$
	where, we assume the free boundary condition
$r_0=r_n=0$. This means  $p_1=i\frac{\partial}{\partial_{\pmb{\xi}_1}}$, and $p_n=-i\frac{\partial}{\partial_{\pmb{\xi}_{n-1}}}$, or $\frac{\partial}{\partial_{\xi_0}}=\frac{\partial}{\partial_{\xi_n}}=0$.  This condition means that the center of mass momentum vanishes, i.e. $\hat{p}_o:=\sum_{x=1}^n p_x=0$. Notice that this boundary condition can be understood as $q_0=q_1$, and $q_n=q_{n+1}$, in the position picture (See Appendix \ref{appendix0}).

	 		Although this system is usually described in terms of the position and momentum operators as in \eqref{Ham00}, it is more convenient to work with elongation operators instead of position operators. For the reader's convenience, in Appendix \ref{appendix0} we illustrate the relation between these  two pictures.   \\
	 
  The canonical commutation relations (CCR) in this coordinates read: 
\begin{equation} \label{CCR1}
[r_x,r_y]=[p_x,p_y]=0, \quad [r_x,p_y]=i\big(\delta_{x,(y-1)}-\delta_{x,y}\big), \quad \forall x \in \mathbb{I}_{n-1}, y \in \mathbb{I}_n,
\end{equation}
where $[a,b]=ab-ba$.  

	In terms of these operators, Hamiltonian operator is defined on $\mathcal{S}(\mathbb{R}^{n-1})$ by
\begin{equation} \label{Ham0}
	H_n=\frac{1}{2}\sum_{x=1}^n \Big(\frac{p_x^2}{m_x} + r_x^2\Big),
\end{equation}
	where $\{ m_x \}_{x=1}^{\infty}$ are i.i.d positive random variables, defined on a probability space $(\Omega, \mathcal{F},\mathbb{P})$. We assume that the the law of these random variables have a smooth density $\mu(x)$, compactly supported on the set $[m_{min},m_{max}]$, where $m_{min}>0$. We denote  the expectation with respect to $\mathbb{P}$  by $\mathbb{E}$, and  $\mathbb{E}(m_x)=\bar{m}$. These assumptions inspired from \cite{BHO}, \cite{FA}, in order to facilitate the application of the results concerning \textit{Anderson Localization}.

	Furthermore, for any $x \in \mathbb{I}_n$, define the operator $e_x=\frac{1}{2}(\frac{p_x^2}{m_x}+r_x^2)$ as the energy of the particle $x$. It is well known that $p_x$, $r_x$, hence $e_x$ and $H_n$ are essentially self-adjoint \cite{simon}. Consequently we can consider their closure on $\mathcal{H}_n$, that we denote with the same symbols (see e.g., \cite{simon}, \cite{NSS}). The domain of $H_n$ will be denoted by $\mathcal{D}(H_n) \subset L^2(\mathbb{R}^{n-1})$.

	The time evolution of the chain is given by the so-called Heisenberg dynamics generated by the Hamiltonian $H_n$. Since $H_n$ is self-adjoint, by spectral theory  $e^{-iH_nt}$ is well defined for any $t \in \mathbb{R}$. Furthermore, using Stone's theorem, we define the one parameter group of authomorphism $\tau_t^n$ on $\mathcal{B}(\mathcal{H}_n)$ as follows: 
	\begin{equation} \label{heisenberg2}
		a(t):= \tau_t^n(a)=e^{itH_n}ae^{-itH_n},\quad  \forall a \in \mathcal{B}(\mathcal{H}_n),
	\end{equation}	  
where $\mathcal{B}(\mathcal{H}_n)$ denotes the set of bounded operators on $\mathcal{H}_n$. Notice that $a(t)$ is  the solution of Heisenberg equation:
\begin{equation} \label{heisenberg1}
	\dot{a} =i[H_n,a], \qquad \: a(0)=a_0,
\end{equation}
where $[a,b]=ab-ba$, and we use the notation $\dot{a}(t):=\partial_ta(t)$. This equation holds in the strong sense on the proper domain. \\
 Again by using Stone's theorem, one can extend the domain of this dynamic to certain unbounded operators. Here we can do this task for operators such as $r_x$, $q_x$, and $e_x=\frac{1}{2}(\frac{p_x^2}{m_x}+r_x^2)$ (see e.g. \cite{NSS},\cite{simon},\cite{RB1} \cite{RB2}). We  denote  $\tau_t^n(r_x)$, $\tau_t^n(p_x)$, and $\tau_t^n(e_x)$ by $r_x(t),p_x(t),e_x(t)$, respectively.  In particular, using the canonical commutation relations (CCR), these operators satisfy the following equations:

\begin{equation} \label{equationofmotionr0}
\begin{aligned}
&\dot{p}_x(t)=r_x(t)-r_{x-1}(t), \quad \forall x \in \mathbb{I}_n, \\ & 
\dot{r}_x(t)=\frac{p_{x+1}(t)}{m_{x+1}}-\frac{p_x(t)}{m_x}, \quad \forall x \in \mathbb{I}_{n-1},
\end{aligned}
\end{equation}
where one should recall the boundary condition $r_0=r_n=0,$ in the first equation. \\ 

Let $M = \diag(m_1,\dots,m_n))$ denotes matrix of masses, $\nabla_- \in \mathbb{R}^{n \times (n-1)}$ be the matrix of discrete gradient with fixed boundary condition, and  $\nabla_+ \in \mathbb{R}^{(n-1) \times n}$ be the discrete gradient with free boundary condition. These matrices have the following explicit form:
	
	\begin{equation} \label{nabla}
\begin{split}
&M(x,y)=m_x \delta_{x,y} \quad \quad 1 \leq x,y \leq n, \\
&\nabla_-(x,y)=
\begin{cases}
1 \quad x=y , 1 \leq x,y \leq n-1,\\
-1 \quad x=y+1 , 1 \leq y \leq n-1,\\
0 \quad otherwise, 1\leq x \leq n, 1\leq y \leq n-1,
\end{cases} \\ 
&\nabla_+=-(\nabla_-)^{\dagger},
\end{split}
	\end{equation} 
where $(.)^{\dagger}$ represents the transpose (in case of complex matrices complex conjugate) of a matrix. Formally, we have: for $f \in \mathbb{R}^{n-1}$, $(\nabla_-f)_x=f_x-f_{x-1}$, for $x \in \mathbb{I}_n$ with $f_n=0$, and  for $f\in \mathbb{R}^{n}$, $(\nabla_+ f)_x=f_{x+1}-f_{x}$, for every $x \in \mathbb{I}_{n-1}$. \\

	Taking advantage of these matrices, one can write the equation of motion \eqref{equationofmotionr0} in the following form:

\begin{equation} \label{equationofmotionr}
	\dot{p}= \nabla_-r, \quad \quad \dot{r}=\nabla_+ M^{-1}p,
\end{equation} 
where $p$ and $r$ denote the vector of momentum and elongation operators, i.e. $p=p(t)=(p_1(t),p_2(t),\dots,p_n(t))$, and $r(t)=(r_1(t),r_2(t),\dots,r_{n-1}(t))$. Notice that this equation can be solved explicitly, we do this task in the next section. In fact, given the explicit expression for $p(t)$ and $r(t)$ from \eqref{bogoliinv2}, one can check that \eqref{equationofmotionr} holds in strong sense on  $\mathcal{D}(H_n)$, and this solution can be extended to the proper domain \footnote{One can take  $\cap_{k=1}^{n-1} (\mathcal{D}(\hat{b}_k) \cap \mathcal{D}(\hat{b}_k^*)) $ as the proper domain, which is dense in $\mathcal{H}_n$.} by linearity. 
  \\
 Finally, define  $\Delta$, the discrete Laplacian matrix with free boundary condition, as
	\begin{equation} \label{productofgrad}
		\Delta=\nabla_- \nabla_+.
	\end{equation}  
	Hence, we have $\forall f \in \mathbb{R}^n, (\Delta f)_x=f_{x+1}-2f_x+f_{x-1}$, for $x \in \mathbb{I}_n$, with free boundary condition $f_{n+1}=f_n, f_0=f_1$.

 Given  $\bar{p}$, $\bar{r} \in C^1([0,1])$, such that $\bar{r}(0)=\bar{r}(1)=0$, 
 $\int_0^1 \bar{p}(y)dy=0$, and $\beta \in C^0([0,1])$, such that $\forall y \in [0,1], \: 0<\beta_{min} \leq \beta(y) \leq \beta_{max}$,
 correspondingly, we define the locally Gibbs state as the density matrix operator acting on $\mathcal{H}_n$ given by

	\begin{equation} \label{initalstate}
		\rho_{\bar{p},\bar{r},\beta}^n = \frac{1}{Z_n} \exp(-\frac{1}{2}\sum_{x=1}^n \Big[ \frac{\beta(\frac{x}{n})}{m_x} (p_x-\bar{p}(\frac{x}		{n})\frac{m_x}{\bar{m}})^2+\beta(\frac{x}{n})(r_x-\bar{r}(\frac{x}{n}))^2 \Big] ).
	\end{equation}
This means initially we let the chain to be in the locally Gibbs state such that the elongation and momentum be out of mechanical equilibrium, and their initial "average" coming from a smooth profile. Moreover, we have a smooth profile of temperature. We drop the subscript and superscripts of $\rho$ whenever it makes no confusion. Here $Z_n$ is a normalizing constant such that $\Tr(\rho^n_{\bar{p},\bar{r},\beta})=1$. Note that in this expression $\bar{p}(\frac{x}{n})$, $\bar{r}(\frac{x}{n})$ have been multiplied by the identity operator.  It is more convenient to define the density operator in terms of the following pseudo-Hamiltonian: 
\begin{equation} \label{hamiltoniantemp}
H_{\beta}^n=\frac{1}{2} \sum_{x=1}^n \Big(\frac{\beta_x}{m_x} \big(p_x-(\frac{m_x}{\bar{m}})\bar{p}_x\big)^2+\beta_x(r_x-\bar{r}_x)^2\Big),
\end{equation}
where we wrote the following terms in shorthanded manner: $\beta_x:=\beta(\frac{x}{n}), \bar{p}_x:=\bar{p}(\frac{x}{n}), \bar{r}_x:=\bar{r}(\frac{x}{n})$. Note that it would be more appropriate to write this operator as $H_{\beta(.)}^n$, since it actually depends on the function $\beta(.)$. However, we use the notation $H^n_{\beta}$, whenever it does not make any confusion. Then we have 

	\begin{equation}\label{initialstate2}
		\rho^n_{\bar{p},\bar{r},\beta}=\exp(-H_{\beta}^n).
	\end{equation}
First, observe that $H_{\beta}^n$ is essentially self-adjoint on 
$\mathcal{S}(\mathbb{R}^{n-1})$ (see e.g. \cite{simon}, \cite{NSS}, \cite{RB2}), and denote its closure with the same symbol. Furthermore,  one can check that $H_{\beta}^n$  has a discrete spectrum with non-negative eigenvalues. In fact, we can write $H_{\beta}^n$ in terms of the sum of free bosonic operators, and we can obtain the spectrum explicitly. We do this task in details in Section \ref{pb} . Hence, using spectral theory and properties of $\beta(.)$ in our assumption, one can observe that $\rho$ is well defined  and trace-class. 
\\
 	Therefore, for every operator $a$, if $a\rho$ is a trace class operator, we can define the "average of the observable $a$ in the state $\rho$", i.e. $\expval{a}_{\rho}$ as:
\begin{equation} \label{mixedaverage}
	\expval{a}_{\rho}=\Tr(\rho a).
\end{equation} 
In particular, one can observe that $\expval{p_x}_{\rho}$, $\expval{r_x}_{\rho}$, and $\expval{e_x}_{\rho}$ are well defined. 

Fix $T>0$,  and consider the following system of conservation laws: 

	\begin{equation} \label{pde1}	
		\begin{split}		
		&\partial_t \mathrm{r}(y,t)= \frac{1}{\bar{m}}\partial_y \mathrm{p}(y,t),  \\
		&\partial_t \mathrm{p}(y,t)=\partial_y \mathrm{r}(y,t),\\
		&\partial_t \mathrm{e}(y,t)=\frac{1}{\bar{m}}\partial_y (\mathrm{r}(y,t) \mathrm{p}(y,t)),
		\end{split}
	\end{equation}
where  $\mathrm{r}(y,t) \in C^1( [0,1] \times [0,T] )$, $\mathrm{p}(y,t) , \mathrm{e}(y,t) \in C^1([0,1] \times [0,T])$. We impose the following boundary condition 
	
	\begin{equation} \label{bc1}
	\forall t \in [0,T], \: \: \: \mathrm{r}(0,t)=\mathrm{r}(1,t)=0,
	\end{equation}
and we add the following initial datum:

	\begin{equation} \label{bc2}
	\mathrm{r}(y,0)=\bar{r}(y), \: \: \mathrm{p}(y,0)= \bar{p}(y), \: \: \mathrm{e}(y,0)=\frac{1}{2}\Big(\frac{\bar{p}(y)^2}{\bar{m}}+\bar{r}(y)^2\Big)		+\mathrm{f}_{\beta}^{\mu}(y).
	\end{equation}

Here $\bar{p}$ and $\bar{r}$ are the same functions that appeared in the definition of the Gibbs state \eqref{initalstate}, and
 $\mathrm{f}_{\beta}^{\mu}(y)= \mathrm{f}^{\mu}(\beta(y))$, where $\mathrm{f}^{\mu}(\beta_{eq})$ for $\beta_{eq} \in (0,\infty)$, is a function which can be determined uniquely by  the law of the distribution of the masses $\mu$. (See \eqref{fdef}, \eqref{thermaleqfunc} for the definition, Section \ref{SLLNsection} and Appendix \ref{app1}).\\
One should recall $\int_0^1 \bar{p}(y)dy=0$, and 
$\bar{r}(0)=\bar{r}(1)$. Observe that $\int_0^1 \mathrm{p}(y,t)dy$ is conserved by \eqref{pde1}, thanks to the boundary condition \eqref{bc1}, justifying the assumption 
$\int_0^1 \bar{p}(y)dy=0$ (See Remark \ref{initialstateremark}, \ref{remarkdef}).\\

Note that since the initial datum for $r$ and $p$ is regular, this equation has a unique classical solution in $r$ and $p$, (see e.g. \cite{evans}, Section 7).
 
Now we are prepared to present the main theorem of this manuscript, which states that the empirical distribution of the average of $(r,p,e)$ with respect to initial state $\rho$, after hyperbolic scaling of time and space converges to the solution of \eqref{pde1} with \eqref{bc1} and initial datum \eqref{bc2}. Precisely, we have: 
\begin{theorem} \label{maintheorem}
 Let $f \in C^0([0,1])$  be an arbitrary test function. Fix $T>0$ and let $t \in [0,T]$. Recall the definition of the initial state $\rho$ \eqref{initalstate}, and let $\bar{p}$, $\bar{r}$, and $\beta $ satisfy the assumptions stated in the definition of \eqref{initalstate}. Then, let $(r(nt),p(nt),e(nt))$  be the evolved operators in the Heisenberg picture with the dynamics generated by the Hamiltonian \eqref{Ham0}.  Moreover, let the $(\mathrm{r}(y,t),\mathrm{p}(y,t),\mathrm{e}(y,t))$ be the solution to \eqref{pde1} with boundary condition \eqref{bc1} and initial datum \eqref{bc2}. Then as $n \to \infty$, we have the following almost sure convergence with respect to distribution of the masses:
	
		\begin{align}
			\frac{1}{n} \sum_{x=1}^n f(\frac{x}{n}) \expval{r_x(nt)}_{\rho} \to \int_0^1 f(y)\mathrm{r}(y,t)dy, \label{rconv}\\
			\frac{1}{n} \sum_{x=1}^n f(\frac{x}{n}) \expval{p_x(nt)}_{\rho} \to \int_0^1 f(y)\mathrm{p}(y,t)dy, \label{pconv} \\
			\frac{1}{n} \sum_{x=1}^n f(\frac{x}{n}) \expval{e_x(nt)}_{\rho} \to \int_0^1 f(y)\mathrm{e}(y,t)dy.	\label{econv}	
		\end{align}
	 \end{theorem}    
The rest of this paper is devoted to the proof of Theorem \ref{maintheorem}. The sketch of the proof and organization of the paper is as follows: One can recall a similar theorem for the classical harmonic chain from (\cite{BHO}, Theorem 1), the main difference here lies in the fact that we cannot write our Gibbs state $\rho$ as a product state, since the energy of each site does not commute with its neighbors. In order to deal with this difficulty, we write $H_{\beta}^n$, which appeared in the definition of $\rho=\exp(-H_{\beta}^n)$, in terms of free bosonic operators (using Bogoliubov or quantum canonical transformation), and then compute the average of suitable operators in this new basis. These computations enable us to obtain appropriate bounds in order to prove \ref{maintheorem}. 

Moreover, we use the explicit form of the solution to the equation of motion \eqref{equationofmotionr} in our proof; hence, we use another Bogoliubov transformation in order to solve the equation of motion. We devote Section \ref{pb} to these transformations and corresponding bounds. 

Then we deal with the evolution of $(r,p)$ in Section \ref{WE}. This part is similar in spirit to the classical case, since all the operators can be written as the linear combination of bosonic operators. However, our proof will be different from the one in \cite{BHO}.

We devote the last two sections to prove \eqref{econv}. As we sketched before, to prove \eqref{econv} at $t=0$, we need SLLN (Strong Law of Large Numbers) for $\expval{e_x}_{\rho}$. In order to prove the SLLN, we need to show that the dependence of $\expval{e_x}$ on $m_z$ is exponentially decaying for $z$ being far away from $x$, we devote Section \ref{SLLNsection} to prove this fact. 

Finally, in Section \ref{EE}, we exploit the localization phenomenon in order to prove \eqref{econv} for any $t \in (0,T]$, as we explained in Section \ref{intro}.
\section{Preliminary Bounds} \label{pb}

\subsection{Hamiltonain Diagonalization} 
 Recall the Hamiltonian $H_n$ \eqref{Ham0}, and the operator $H_{\beta(.)}^n$ acting on $L^2(\mathbb{R}^{n-1})$. These operators have discrete spectrums, and the full set of their eigenvalues and eigenfunctions can be represented rather explicitly, thanks to the quantum canonical transformation also known as Bugoliubov Transformation, (see e.g. \cite{RB1},\cite{degro},\cite{NSS} for a through discussion). We recall here these transformations in details. The solution to the equation of motion is expressed in terms of this new coordinates. Furthermore, these transformations have the localization properties, enabling us to close the equation and prove the hydrodynamic limit.

 Let $\langle , \rangle_n$ denotes the inner product in $\mathbb{R}^n$, we  drop the subscript whenever it does not make any confusion. We express the canonical basis of $\mathbb{R}^n$ by $\ket{x}$, for $x \in \mathbb{I}_n$. 
 By abusing the notation, we use the same symbol for a linear combination of operators or for the product of two vector of commuting operators. Using this notation we have:
 \begin{equation} \label{Ham2}
 H_n=\frac{1}{2} \big( \langle p, M^{-1} p \rangle_n + \langle r, r \rangle_{n-1}\big).  
\end{equation}   
 Consider the following matrix: 
\begin{equation} \label{bogolimatrixp}
 A_p^0:=M^{-\frac{1}{2}}(-\Delta)M^{-\frac{1}{2}} \in \mathbb{R}^{n\times n},
\end{equation}  
  this matrix is symmetric, positive semidefinite and  almost surely, it has a non-degenerate spectrum (it is evident by using proposition II.1 of \cite{Kunz}). Let $0=\omega^2_0<\omega_1^2 <\dots <\omega_{n-1}^2$, be the set of eigenvalues of this matrix in the increasing order, and let $\{ \varphi^k \}_{k=0}^{n-1}$ be their corresponding eigenvectors, such that they form an orthonormal basis for $\mathbb{R}^n$, where we have $\langle \varphi^k, \varphi^j \rangle =\delta_{k,j}$. Observe that we have 
   $\varphi^0= (\sum_{x=1}^n m_x)^{-\frac{1}{2}}M^{\frac12} \ket{\mathbf{1}}$, where $\ket{\mathbf{1}}$  denotes the vector $(1,1,\dots,1)^{\dagger}$. 
   
   Define the operators $\hat{p}_k,  \forall k \in \mathbb{I}_{n-1}^0$ to be the following linear combination of $p_x$: 
   	\begin{equation} \label{bogolip}
	   \begin{split}
  	 		&\hat{p}_k= \langle \varphi^k, M^{-\frac12}p \rangle_n =\sum_{x=1}^{n} \frac{\varphi^k_x}{\sqrt{m_x}}p_x.
		\end{split}	
	\end{equation} 
	
	Taking into account the definition of $\varphi^0$, one can check that $\hat{p}_o:=\sum_{x=1}^np_x=0=  (\sum_{x=1}^n m_x)^{\frac12}\hat{p}_0$, thanks to the imposed boundary condition. \\
On the other hand let 
\begin{equation} \label{bogolimatrixr}
 A_r^0:=-\nabla_+ M^{-1} \nabla_-,
\end{equation} 
due to the positivity assumption on $M$, one can observe that $A_r^0$ is a $(n-1)\times (n-1)$ positive symmetric matrix. Moreover, if we let for $k \in \mathbb{I}_{n-1}, \: \:\phi^k:=\frac{1}{\omega_k} \nabla_+ M^{-\frac12} \varphi^k$, we have: 

	\begin{equation} \label{phi1}
		\begin{split}		
			A_r^0 \phi^k &= -\frac{1}{\omega_k}\nabla_+M^{-1}\nabla_-\big(\nabla_+ M^{-\frac12} \varphi^k\big)=\frac{1}{\omega_k}\nabla_+M^{-\frac12}\big(M^{-\frac{1}{2}}(-\Delta)M^{-\frac{1}{2}}\big) \varphi^k \\
			&=\frac{1}{\omega_k} \nabla_+M^{-\frac12}(\omega_k^2)\varphi^k=\omega_k^2 \phi^k, \\
\langle \phi^k , \phi^{k'} \rangle_{n-1}&= \frac{1}{\omega_k\omega_{k'}} \langle \nabla_+M^{-\frac12} \varphi^k , \nabla_+M^{-\frac12} \varphi^{k'} \rangle_{n-1} =\frac{1}{\omega_k \omega_{k'}} \langle \varphi^k,M^{-\frac{1}{2}}(-\Delta)M^{-\frac{1}{2}} \varphi^{k'} \rangle_n \\
&=\frac{\omega_{k'}}{\omega_k} \langle \varphi^{k} , \varphi^{k'} \rangle_n= \delta_{k,k'}.
 			\end{split}	
	\end{equation}    
Hence, $\{ \phi^k \}_{k=1}^{n-1}$ is the full set of eigenvectors for $A_r^0$ with the same set of eigenvalues as $A_p^0$: $\{ \omega_1^2,\dots \omega_{n-1}^2 \}$. Therefore, they form an orthonormal basis for $\mathbb{R}^{n-1}$. Now define $\hat{r}_k$, for every $k \in \mathbb{I}_{n-1}$ as 
	\begin{equation} \label{bogolir}
		\hat{r}_k:=\langle \phi^k, r \rangle_{n-1}.
	\end{equation}

	Let us denote the operator $s_x:=-i\partial/\partial_{\pmb{\xi}_x}$, so we have $p_x=-(s_x-s_{x-1})$, more precisely, thanks to the definition of $\nabla_-$ and the boundary condition $r_0=r_n=0,$ we have: 
	\begin{equation}
	p= - \nabla_- s. 
\end{equation}
The canonical commutation relation in the $s$ coordinates reads: 
\begin{equation}
	\forall x ,y \in \mathbb{I}_{n-1}, \quad [r_x,s_y]=i\delta_{x,y}. 
\end{equation}
Therefore, one can compute $[r_k,p_{k'}]$:
\begin{equation}
\begin{split}	
	[r_k,p_k']&=\Big[\expval{\phi^k,r}_{n-1},\expval{\varphi^{k'},M^{-\frac12}p}_n\Big]=\Big[\expval{\phi^k,r}_{n-1}, -\expval{\varphi^{k'},M^{-\frac12}\nabla_-s}_n\Big] \\
	&=\Big[\expval{\phi^k,r}_{n-1}, \expval{\omega_{k'}\nabla_+M^{-\frac12}\varphi^{k'},s}_{n-1}\Big]=\omega_{k'}\Big[\expval{\phi^k,r}_{n-1},\expval{\phi^{k'},s}_{n-1}\Big] \\
	&=\omega_{k'}\sum_{x,y=1}^{n-1} \phi^k_x \phi^{k'}_y [r_x,s_y]=\omega_{k'}\sum_{x,y=1}^{n-1} \phi^k_x \phi^{k'}_y i \delta_{x,y}=i\omega_{k'} \sum_{x=1}^{n-1} \phi^k_x \phi^{k'}_x= i \omega_k\delta_{k,k'}, 
\end{split}
\end{equation}
where we used the identities $(\nabla_-)^{\dagger}=-\nabla_+$ and $\nabla_+ M^{-\frac12} \varphi^k=\omega_k \phi^k$ as well as the fact that $\{ \phi^k \}_{k=1}^{(n-1)}$ is an orthonormal basis. Since $[r_x,r_y]=[p_x,p_y]=0$, we can sum up the commutation relation for our new coordinates as follows: 
\begin{equation} \label{ccrtimer}
\forall k,k' \in \mathbb{I}_{n-1}^0, \quad [\hat{r}_k,\hat{p}_{k'}]=i\omega_k\delta_{k,k'}, \qquad [\hat{r}_k,\hat{r}_{k'}]=[\hat{p}_k,\hat{p}_{k'}]=0.
\end{equation}

  Later, we benefit from the inverse of  \eqref{bogolip} and \eqref{bogolir}. Let $O$ and $\tilde{O}$ be orthogonal matrices of eigenvectors of $A_p^0$ and  $A_r^0$, respectively. Hence, $O^{\dagger}O=OO^{\dagger}=I_n$, $\tilde{O}^{\dagger}\tilde{O}=\tilde{O} \tilde{O}^{\dagger}=I_{n-1}$. In other words: $\sum_{k=0}^{n-1} \varphi^k_x \varphi^k_y=\sum_{k=1}^{n-1}\phi_x^k \phi_y^k=\delta_{x,y}$. Therefore, the inverse expressions read:
  	\begin{equation} \label{bogoliinv1}
  		\begin{split}
  		&p=M^{\frac12}O\hat{p}, \qquad p_x=\sqrt{m_x}\sum_{k=0}^{n-1} \varphi_x^k \hat{p}_k,	\\
  		&r=\tilde{O}\hat{r}, \quad \qquad \qquad r_x=\sum_{k=0}^{n-1} \phi_x^k \hat{r}_k.
		\end{split}
	\end{equation}   
	Here, $\hat{p},\hat{r}$ denote the vector of corresponding operators in the transformed coordinate. \\
The Hamiltonian $H_n$ can be written in terms of these  new coordinates: 
	\begin{equation} \label{boson1}
		\begin{split}
			H_n&= \frac{1}{2}( \langle p, M^{-1} p \rangle_n + \langle r , r\rangle_{n-1})=\frac12\big(\langle M^{\frac12}O\hat{p},M^{-\frac12}O\hat{p} \rangle_n + \langle \tilde{O}\hat{r},\tilde{O}\hat{r} \rangle_{n-1} \big) \\
		&= \frac12(\langle \hat{p},\hat{p} \rangle_n+
		\langle \hat{r}, \hat{r} \rangle_{n-1})= \frac12 \hat{p}_0^2 + \frac12\sum_{k=1}^{n-1} (\hat{p}_k^2+ \hat{r}_k^2)= \frac12 \sum_{k=1}^{n-1} (\hat{p}_k^2 + \hat{r}_k^2),		
		\end{split}
	\end{equation}
 where we used the identities $O^{\dagger}O=OO^{\dagger}=I_n, \tilde{O}^{\dagger}\tilde{O}=\tilde{O} \tilde{O}^{\dagger}=I_{n-1}$ and the fact that $\hat{p}_0=0$.  
We introduce the operators $\hat{b}_k$ with their adjoints  $\hat{b}_k^*$ as follows:
 
		\begin{align} 		
 		&\hat{b}_k= \frac{1}	{\sqrt{2\omega_k}} (\hat{r}_k+i\hat{p}_k),   \label{bosonoperator} \\
 		&\hat{b}_k^*=\frac{1}{\sqrt{2\omega_k}} (\hat{r}_k-i\hat{p}_k). \label{bosonadj}
	\end{align}
	    
Notice that \eqref{bosonadj} was deduced from the fact that $\hat{p}_k, \hat{r}_k$ are self-adjoint.  
  These operators are the bosonic creation and annihilation operators. Using \eqref{ccrtimer}, one can verify that they satisfy the annihilation-creation form of the canonical  commutation relation: 
\begin{equation} \label{ccrbosonic}
	[\hat{b}_k,\hat{b}_{k'}^*]=\delta_{k,k'} \quad [\hat{b}_k,\hat{b}_{k'}]=[\hat{b}_k^*,\hat{b}_{k'}^*]=0.
\end{equation}
	Furthermore, we can express the Hamiltonian as follows, using the identity $\omega_k \hat{b}^*_k \hat{b}_k=\frac12(\hat{r}_k^2+\hat{p}_k^2 -\frac{\omega_k}{2})$, thanks to  \eqref{ccrtimer}:
\begin{equation} \label{hambosontime}
	H_n=\sum_{k =1}^{n-1}\omega_k(\hat{b}_k^*\hat{b}_k+\frac12).
\end{equation}

 Having \eqref{hambosontime}, with the commutation relation \eqref{ccrbosonic}, full spectrum  of $H_n$ (which is discrete) and the corresponding eigenfunctions are quite  well known (see e.g. \cite{hall} Section 9 or \cite{NSS}).\\
  We describe this spectrum and its corresponding eigenfunctions as follows: recall the space variable $\pmb{\xi}\in \mathbb{R}^{n-1}$, similar to  \eqref{bogolir}, define the new space coordinate $\hat{\pmb{\xi}}_k:=\expval{\phi^k,\pmb{\xi}}_{n-1}$ for $k \in \mathbb{I}^0_{n-1}$.
	Let $\pmb{\ket{\Phi_0}} \in L^2(\mathbb{R}^{n-1})$ be the unique normalized solution of $$\forall k \in \mathbb{I}_{n-1}, \quad \hat{b}_k \pmb{\ket{\Phi_0}}=\frac{1}{\sqrt{2}}\Big(\frac{\hat{r}_k}{\sqrt{\omega_k}}+\sqrt{\omega_k}\partial/\partial_{\hat{\pmb{\xi}}_k}\Big)\pmb{\ket{\Phi_0(\hat{\pmb{\xi}}_1, \dots, \hat{\pmb{\xi}}_{n-1})}}=0.$$
	Then, $\pmb{\ket{\Phi_0}}$ will be the ground state of $H_n$ with the corresponding eigenvalue (energy) $E_0= \frac12\sum_{k=1}^{n-1} \omega_k=\frac12 \Tr(\omega^{\frac12})=\frac12\Tr((-\Delta M^{-1})^{\frac12})$, where $\omega=\diag(\omega_1^2,\dots,\omega_{n-1}^2)$. Let $\mathbb{N}$ be the set of nonnegative integers, for every $\bar{\theta} \in \mathbb{N}^{n-1}$, $\bar{\theta}=(\theta_1,\dots, \theta_{n-1})$ define $E_{\bar{\theta}}$ as:
	\begin{equation}
		E_{\bar{\theta}}=E_0+ \sum_{k=1}^{n-1} \omega_k \theta_k.
	\end{equation}	  
Then $\mathcal{E}=\{E_{\bar{\theta}} |\bar{\theta} \in \mathbb{N}^{n-1}\} $ is the set of eigenvalues of $H_n$ with the corresponding eigenfunctions $\pmb{\ket{\Phi_{\bar{\theta}}}}$, where they form an orthonormal basis for $\mathcal{H}_n$. One can write these eigenfunctions as follows:\footnote{One can describe $\pmb{\ket{\Phi_{\bar{\theta}}}}$ more precisely as $\pmb{\ket{\Phi_{\bar{\theta}}(\hat{\pmb{\xi}_1},\dots,\hat{\pmb{\xi}}_{n-1})}}=\prod_k f_{\theta_k}(\hat{\pmb{\xi}}_k)$, where $f_{\theta_k}(x)=\frac{1}{\sqrt{\theta_k!}} (\frac{\omega_k}{\pi})^{\frac14} e^{-\frac{\omega_kx^2}{2}}{H}_{\theta_k}(\sqrt{2\omega_k}x)$, and $H_j$ stands or the $j$th hermite polynomial. Note that the aforementioned set $\mathcal{E}$ is the full spectrum of $H_n$, since the collection of their corresponding eigenfunctions create a complete orthonormal basis for $L^2(\mathbb{R}^{n-1})$.}: 	
	\begin{equation} 
		\pmb{\ket{\Phi_{\bar{\theta}}}} = \prod_{k=1}^{n-1} \frac{(\hat{b}^*_k)^{\theta_k}}{\sqrt{\theta_k!}} \pmb{\ket{\Phi_0}}.
	\end{equation}

 Finally, we address the time evolution of the momentum and elongation operator. First, the time evolution of the bosonic operators $\hat{b}_k$ and $\hat{b}_k^*$, i.e. the action of the Heisenberg dynamic $\tau_t^n$ \eqref{heisenberg2} on these operators is given by: 
 \begin{equation} \label{bosonictimeevolution}
 	\tau_t^n(\hat{b}_k)=:\hat{b}_k(t)=e^{-i\omega_kt}\hat{b}_k(0), \quad \qquad 
		\tau_t^n(\hat{b}^*_k)=:\hat{b}_k^*(t)=e^{i \omega_kt}{b}_k^*(0),
 \end{equation}
 where the current form of the Hamiltonian \eqref{hambosontime}, and the commutation relations \eqref{ccrbosonic}, have been used, with $\hat{b}_k(0)=\hat{b}_k$ and $\hat{b}_k^*(0)=\hat{b}_k^*$. 
Note that $\tau_t^n(\hat{b}_k)=:\hat{b}_k(t)=e^{-i\omega_kt}\hat{b}_k(0)$ is the unique solution of the Heisenberg evolution equation $\dot{\hat{b}}_k=i[H_n,\hat{b}^k]=-i\omega_k\hat{b}_k$, where we used \eqref{ccrbosonic}. This equation holds in the strong sense on $\mathcal{D}(H_n)$ and the solution can be extended to $\mathcal{D}(\hat{b}_k)$ by linearity. \\
  Rewriting $\hat{p}_k(0)$, $\hat{r}_k(0),$ for $k \in \mathbb{I}_{n-1}$  in terms of creation and annihilation operators from \eqref{bosonoperator} and \eqref{bosonadj} gives: 
\begin{equation} \label{ladderoperator2}
\hat{p}_k(0)=i\sqrt{\frac{\omega_k}{2}} (\hat{b}_k^*(0)-\hat{b}_k(0)), \qquad \quad \hat{r}_k(t)=\sqrt{\frac{\omega_k}{2}}(\hat{b}_k(0)+\hat{b}_k^*(0)).
\end{equation}
 By applying \eqref{bosonictimeevolution}, we end up with: 

\begin{equation} \label{timeevolutionk}
	\begin{split}	
		&\hat{p}_k(t)=i\sqrt{\frac{\omega_k}{2}} (\hat{b}_k^*(0)e^{-i\omega_kt}-\hat{b}_k(0)e^{i\omega_kt}) =\cos(\omega_kt)\hat{p}_k(0)-\sin(\omega_kt) \hat{r}_k(0)
		\\& \quad \quad =\langle M^{-\frac12} \varphi^k , p(0) \rangle \cos(\omega_kt)- \langle \phi^k , r(0) \rangle \sin(\omega_kt), 
		\\ &\hat{r}_k(t)=\sqrt{\frac{\omega_k}{2}}(\hat{b}_k(0)e^{-i\omega_kt}+\hat{b}_k^*(0)e^{i\omega_kt})=\cos(\omega_kt)\hat{r}_k(0)+\sin(\omega_kt) \hat{p}_k(0) \\
		&\quad \quad = \langle \phi^k ,r(0) \rangle \cos(\omega_kt) + \langle M^{-\frac12} \varphi^k,p(0) \rangle \sin(\omega_kt),
	\end{split}
\end{equation}

Lastly, exploiting the relation \eqref{bogoliinv1} the time evolution of $p$ and $r$ i.e. $p(t):=\tau_t^n(p)$ and $r(t):=\tau_t^n(r)$ is given by:
\begin{equation} \label{bogoliinv2}
\begin{split}	
	&p(t)=\sum_{k=0}^{n-1}  M^{\frac12} \varphi^k \hat{p}_k (t) = \sum_{k=0}^{n-1} \big(\cos(\omega_kt)\hat{p}_k(0)-\sin(\omega_kt) \hat{r}_k(0)\big)M^{\frac12} \varphi^k \\
	& \quad \quad= \sum_{k=0}^{n-1} \big(\langle M^{-\frac12} \varphi^k , p(0) \rangle \cos(\omega_kt)- \langle \phi^k , r(0) \rangle\sin(\omega_kt)\big)M^{\frac12} \varphi^k, \\
&p_x(t)=\sum_{k=0}^{n-1}  \sqrt{m_x} \varphi^k_x \hat{p}_k (t) = \sum_{k=0}^{n-1} \big(\cos(\omega_kt)\hat{p}_k(0)-\sin(\omega_kt) \hat{r}_k(0)\big)\sqrt{m_x} \varphi^k_x,
	\\
	 &r(t)=\sum_{k=1}^{n-1} \phi^k r_k(t)=\sum_{k=1}^{n-1} \big(\cos(\omega_kt)\hat{r}_k(0)+\sin(\omega_kt) \hat{p}_k(0)\big)\phi^k
	 \\
&\quad \quad = \sum_{k=1}^{n-1}\big(\langle \phi^k ,r(0) \rangle \cos(\omega_kt) + \langle M^{-\frac12} \varphi^k,p(0) \rangle \sin(\omega_kt) \big) \phi^k, \\
&r_x(t)=\sum_{k=1}^{n-1} \phi^k_x r_k(t)=\sum_{k=1}^{n-1} \big(\cos(\omega_kt)\hat{r}_k(0)+\sin(\omega_kt) \hat{p}_k(0)\big)\phi^k_x.
\end{split}
\end{equation}

\begin{remark} \label{diagrmk}
	
It is worth mentioning that following \cite{NSS}, the process of rewriting a quadratic Hamiltonian in terms of free bosons can be done in a more general setting. In fact, this task is doable for any Hamiltonian of the form $H= \frac{1}{2}\expval{(q,p),A (q,p)}_{2n}$, such that $A \in \mathcal{M}_{2n}(\mathbb{R})$ is a positive symmetric matrix. This fact is a direct consequence of Williamson's theorem (see, \cite{degro} section 8.3). This theorem states that any positive symmetric matrix  $A \in \mathcal{M}_{2n}(\mathbb{R})$ can be diagonalized via a sympletic matrix $S$. Recall that $S$ is sympletic, if and only if we have $SJS^{\dagger} =J$, where 
\[J=
\begin{bmatrix}
    0 & I_n \\
    -I_n & 0
     \end{bmatrix}.
\]
	Using this theorem, one can define the new coordinates $(\hat{p},\hat{q})^{\dagger}=S^{-1}(p,q)^{\dagger}$. Here $v^{\dagger}$ denotes the transpose of the vector $v$. Thanks to the properties of $S$ (see Remark 2.4 of \cite{degro}) the canonical commutation relation for these new coordinates is evident i.e. $\forall x,y \in \mathbb{I}_n, \quad [\hat{p}_x,\hat{p}_y]=[\hat{q}_x,\hat{q}_y]=0$, and $[\hat{q}_x,\hat{p}_y]=i\delta_{x,y}$. Moreover, $S$ diagonalizes $A$, i.e. $SAS^{\dagger}=\lambda^2$,  where $\lambda^2= \diag(\lambda_1^2, \dots,\lambda_n^2, (\lambda'_1)^2,\dots, (\lambda'_n)^2)$. And the Hamiltonian reads: $H=\frac{1}{2}\sum_{x=1}^n(\lambda_x')^2( \hat{p}_x^2+ \gamma_x^2 \hat{q}_x^2),$ with $\gamma_x =\frac{\lambda_x}{\lambda_x'}$. Finally, the bosonic operators are given by $B_x=\frac{1}{\sqrt2}(\sqrt{\gamma_x} \hat{q}_x + \frac{i \hat{p}_x}{\sqrt{\gamma_x}} )$, where $[B_x,B_y]=[B_x^*,B_y^*]=0$, and $[B_x,B_y^*]=\delta_{x,y}$. Moreover, the Hamiltonian can be written as $H=\sum_{x=1}^n (\lambda_x')^2\gamma_x(B_x^* B_x+\frac12)$, which is sum of free bosons as we desired. \\ 

In a more physical setup, we have $H= \frac12(\expval{p,V_p p}_n + \expval{q, V_q q}_n)$, such that $V_p,V_q \in \mathcal{M}_n(\mathbb{R})$ are positive and symmetric. In this situation, we can express the desired transformation in the following explicit manner: In this case since $V_q^{\frac12}V_p V_q^{\frac12}$ is positive and symmetric, let $O$ be the orthogonal matrix such that $O^{\dagger} V_p^{\frac12}V_q V_p^{\frac12}O = \gamma^2=: \diag(\gamma_1^2,\dots,\gamma_n^2)$. Then $S$ (the sympletic transformation introduced above) has the following form: 
\[S=
\begin{bmatrix}
    V_p^{\frac12}O & 0 \\
    0 & V_p^{-\frac12}O
     \end{bmatrix}.
\]
Moreover, the Hamiltonian can be written as: $$ H= \frac{1}{2}\sum_{k=1}^n \hat{p}_k^2+\gamma_k^2\hat{q}_k^2= \sum_{k=1}^n\gamma_k(B_k^* B_k +\frac12), $$ where the bosonic operators are defined as before: $B_k:=\frac{1}{\sqrt2}(\sqrt{\gamma_k}\hat{q}_k+i\frac{\hat{p}_k}{\sqrt{\gamma_k}})$.  
In this remark, we followed the notation in \cite{Cramer}. Notice that in order to diagonalize $H_n$ and $H_n^{\beta}$, we adapted the same strategy to our setup, where we have the Hamiltonian in terms of $r$ coordinates. \footnote{This modification stems from the fact that in our case $V_x$ is not positive definite  and has a zero eigenvalue.}
\end{remark}

\subsection{Density operator diagonalisation} \label{Densityoperatordiag}
Recall the definition \eqref{initalstate} of locally Gibbs state $\rho= \exp(-H_{\beta}^n)$, where $H_{\beta}^n$ is defined in \eqref{hamiltoniantemp}. In this section, we recall the necessary transformation for rewriting $H_{\beta}^n$ in terms of free bosons, following the lines of Remark \ref{diagrmk}. This helps us to compute certain averages with respect to $\rho$.   \\
 We begin by defining the following operators: 
	\begin{equation} \label{thermalcoordinate00}
			\begin{split}
		&\tilde{\ttp}_x:=p_x-\frac{m_x}{\bar{m}}\bar{p}_x,  \qquad \forall x \in \mathbb{I}_n, \\
 &\tilde{\ttr}_x=r_x-\bar{r}_x,  \qquad \forall x \in \mathbb{I}_{n-1},
				\end{split}
	\end{equation}
Observe that $H_{\beta}^n$ can be written as:
	\begin{equation} \label{hamtemp2}
	H_{\beta}^n = \frac{1}{2}\big(\langle \tilde{\ttp},M_{\beta}^{-1} \tilde{\ttp} \rangle_n	+ 		\langle \tilde{\ttr}, \beta^o \tilde{\ttr} \rangle_{n-1}\big),
	\end{equation}
where $M_\beta=M\tilde{\beta}^{-1}$, with $\tilde{\beta}:= \diag(\beta(\frac{1}{n}),\dots,\beta(\frac{n}{n}))$ and $\beta^o:=\diag(\beta(\frac{1}{n}),\dots,\beta(\frac{n-1}{n}))$.
Define $A_{p}^{\beta}$ and  $A_r^{\beta}$, similar to $A_r^0$ and $A_p^0$ as: 
	\begin{equation} \label{thermalmatrices}
		A_p^{\beta}=M_{\beta}^{-\frac12}(-\nabla_-\beta^0 \nabla_+)M_{\beta}^{-\frac12}, \qquad A_r^{\beta}= (\beta^o)^{\frac12}(-\nabla_+ M_{\beta}^{-1} \nabla_-)(\beta^o)^{\frac12}.
	\end{equation}
Since $A_p^{\beta}$ is symmetric positive semidefinite with almost sure non-degenerate spectrum, let $\{ \psi^k \}_{k=0}^{n-1}$ be the orthonormal set of eigenvectors for $A_p^{\beta}$, such that they form a basis for $\mathbb{R}^n$. Then the corresponding set of increasing eigenvalues is given by $\{0=\gamma_0^2<\gamma_1^2<\dots<\gamma_{n-1}^2\}$. Denote the matrix of these eigenvectors by $O_{\beta}$. We have $O_{\beta}O_{\beta}^{\dagger}=O_{\beta}^{\dagger}O_{\beta}=I_n$, $\langle \psi^k, \psi ^{k'}\rangle=\sum_{x=1}^n \psi^k_x \psi^{k'}_x=\delta_{k,k'}$, and $\sum_{k=0}^{n-1}\psi^k_x\psi^{k'}_y=\delta_{x,y}$. Note that we have $\psi^0=(\sum_{x=1}^{n} \frac{m_x}{\beta_x})^{-\frac12}M_{\beta}^{\frac12}\ket{\textbf{1}}$. Moreover, $A_r^{\beta}$ is symmetric positive definite, and one can see if for $k \in \mathbb{I}_{n-1}$, 
$$ \tilde{\psi}^k:= \frac{1}{\gamma_k} (\beta^o)^{\frac12} \nabla_+M_{\beta}^{-\frac12} \psi^k,$$  
then $\{ \tilde{\psi}_k\}_{k=1}^{n-1}$  is the set of eigenvectors of $A_r^{\beta}$ with similar eigenvalues $\gamma_1^2<\dots<\gamma_{n-1}^2$. Hence, they form an orthonormal basis for $\mathbb{R}^{n-1}$. Denote the matrix of these eigenvectors by $\tilde{O}_{\beta}$, with $\tilde{O}_{\beta}\tilde{O}_{\beta}^{\dagger}=\tilde{O}_{\beta}^{\dagger}\tilde{O}_{\beta}=I_{n-1}$. This claim follows from the following computation:
$$A_r^{\beta}\tilde{\psi}_k=\frac{1}{\gamma_k}(\beta^o)^{\frac12}(-\nabla_+ M_{\beta}^{-1} \nabla_-)(\beta^o)^{\frac12}(\beta^o)^{\frac12} \nabla_+M_{\beta}^{-\frac12} \psi^k=\frac{1}{\gamma_k}(\beta^o)^{\frac12} \nabla_+ M_{\beta}^{-\frac12}A_p^{\beta} \psi^k =\gamma_k^2 \tilde{\psi}^k.$$ 
Define another set of coordinates $\tilde{\mathfrak{p}}_k,\tilde{\mathfrak{r}}_k$, similar to \ref{bogolip}, \ref{bogolir}, for $k \in \mathbb{I}^0_{n-1}$: 
	\begin{equation} \label{bogolitemp1}
		\begin{split}
			&\tilde{\mathfrak{p}}= O_{\beta}^{\dagger} M_{\beta}^{-\frac12}\tilde{\ttp}, \qquad \tilde{\mathfrak{p}}_k=\langle \psi^k,M_{\beta}^{-\frac12} \tilde{\ttp} \rangle_n = \sum_{x=1}^{n}  \sqrt{\frac{\beta_x}{m_x}}	\psi^k_x \tilde{\ttp}_x, \\
	&\mathfrak{\tilde{\ttr}}=\tilde{O}^{\dagger}_{\beta} (\beta^o)^{\frac12} \tilde{\ttr}, \qquad  \tilde{\mathfrak{r}}_k= \langle (\beta^o)^\frac12 \tilde{\psi}^k,\tilde{\ttr} \rangle_{n-1}.	
		\end{split}
	\end{equation}
Let us  define $\mathfrak{p}_k, \mathfrak{r}_k$ as  
\begin{equation} \label{notshifted}
	\mathfrak{p}_k =  \langle \psi^k,M_{\beta}^{-\frac12} p \rangle_n, \qquad 
	\mathfrak{r}_k= \langle (\beta^o)^\frac12 \tilde{\psi}^k,r \rangle_{n-1}.
\end{equation}
Notice that $\mathfrak{p}_k$ and $\mathfrak{r}_k$, differs with  $\tilde{\mathfrak{p}}$ and $\tilde{\mathfrak{r}}_k$ only by a constant, respectively. In particular, thanks to the expression of $\psi^0$ we have:  
\begin{equation} \label{pzero}
\tilde{\mathfrak{p}}_0 =  \left(\sum_{x=1}^n \frac{m_x}{\beta_x}\right)^
{-\frac12} \left(\sum_{x=1}^n p_x -\sum_{x=1}^n \bar{p}(\frac{x}{n})\frac{m_x}{\bar{m}} \right)=:- \lambda \Pi_0,
\end{equation}
where we take advantage of the fact $\sum_{x=1}^n p_x=0$, and we defined the (random) constants
\begin{equation} \label{lpdef}
\lambda:=\left(\sum_{x=1}^n \frac{m_x}{\beta_x}\right)^{-\frac12}, \qquad   
\Pi_0:=\sum_{x=1}^n \bar{p}(\frac{x}{n})\frac{m_x}{\bar{m}}.
\end{equation}
Similar to the previous section, we have the inverse transformations: 
	\begin{equation} \label{bogoliinvthermal}
		\begin{split}
			&\tilde{\ttp}=M_{\beta}^{\frac12} O_{\beta}\tilde{\mathfrak{p}}, \qquad \tilde{\ttp}_x= \sqrt{\frac{m_x}{\beta_x}}\sum_{k=0}^{n-1} \psi^k_x\tilde{\mathfrak{p}}_k,	\\
				&\tilde{\ttr}=(\beta^o)^{-\frac12} \tilde{O}_{\beta}\tilde{\mathfrak{r}}, \qquad \tilde{\ttr}_x= \frac{1}{\sqrt{\beta_x}}\sum_{k=1}^{n-1} \tilde{\psi}^k_x\tilde{\mathfrak{r}}_k.
		\end{split}
	\end{equation}	
	We  express $H_{\beta}^n$ in terms of these new coordinates, and obtain the desired diagonalization in terms of independent oscillators:
	\begin{equation} \label{hambosontemp}
		\begin{split}		
		H_{\beta}^n &= \frac12(\langle \tilde{\ttp},M_{\beta}^{-1} \tilde{\ttp} \rangle_n + \expval{\tilde{\ttr},\beta^o \tilde{\ttr}}_{n-1})=\frac12\Big(\langle M_{\beta}^{\frac12} O_{\beta}\tilde{\mathfrak{p}}, M_{\beta}^{-\frac12} O_{\beta}\tilde{\mathfrak{p}} \rangle+ \expval{(\beta^o)^{-\frac12} \tilde{O}_{\beta} \tilde{\mathfrak{r}},(\beta^o)^{\frac12}\tilde{O}_{\beta}\tilde{\mathfrak{r}}}_{n-1} \Big) \\
		  &=\frac{1}{2}(\expval{\tilde{\mathfrak{p}},\tilde{\mathfrak{p}}}_n+\expval{\tilde{\mathfrak{r}},\tilde{\mathfrak{r}}}_{n-1})= \frac{\tilde{\mathfrak{p}}_0^2}{2} +\frac12\sum_{k=1}^{n-1} (\tilde{\mathfrak{p}}_k^2+\tilde{\mathfrak{r}}_k^2)= 
	\frac{\lambda^2\Pi_0^2}{2}+\frac12 \sum_{k=1}^{n-1} (\tilde{\mathfrak{p}}_k^2+\tilde{\mathfrak{r}}_k^2),
		\end{split}
	\end{equation}
where we used the identities $O_{\beta}O_{\beta}^{\dagger}=I_n$ and $\tilde{O}_{\beta}\tilde{O}_{\beta}^{\dagger}=I_{n-1}$. Moreover, we take advantage of \eqref{pzero} to replace 
$\tilde{\mathfrak{p}}_0^2$ with a constant.
The fact that \eqref{hambosontemp} is sum of free uncoupled oscillators is a direct consequence of the following commutation relations:  
\begin{equation} \label{ccrtemp}
	\begin{split}	
	[\tilde{\mathfrak{p}}_k,\tilde{\mathfrak{p}}_{k'}]=[\tilde{\mathfrak{r}}_k,\tilde{\mathfrak{r}}_{k'}]=0, \quad   [\tilde{\mathfrak{r}}_k,\tilde{\mathfrak{p}}_{k'}] =i\gamma_k \delta_{k,k'}, \quad \forall k,k' \in \mathbb{I}_{n-1}.	 \end{split}
\end{equation}
The first relation in \eqref{ccrtemp} is evident from the definition. The second relation can be justified as follows: Recall the operator $s_x=-i\partial/\partial_{\pmb{\xi}_x}$, where we had $[r_x,s_y]=i\delta_{x,y}$. For any $v,w \in \mathbb{R}^{n-1}$ we have (here we drop the subscript $n-1$ in the $\expval{}_{n-1}$ ): 
\begin{equation} \label{ccrlema}
	[\expval{v,r},\expval{w,s}]= \sum_{x,y=1}^{n-1}v_xw_y [r_x,s_y]= \sum_{x,y=1}^{n-1}v_xw_yi\delta_{x,y}=i\sum_{x=1}^{n-1}v_x w_x = i \expval{v,w}_{n-1}.
\end{equation}
	Therefore, by using the relation $p= -\nabla_-s$, and thanks to\eqref{ccrlema}, the definition of $A_p^{\beta}$ and its eigenvectors $\psi^k$, we compute:
	\begin{equation}
		\begin{split}
			[\tilde{\mathfrak{r}}_k,\tilde{\mathfrak{p}}_{k'}]&=[\langle (\beta^o)^\frac12 \tilde{\psi}^k,r \rangle_{n-1},\langle \psi^{k'},M_{\beta}^{-\frac12} p \rangle_n]=-[\langle (\beta^o)^\frac12 \tilde{\psi}^k,r \rangle_{n-1},\langle\psi^{k'},M_{\beta}^{-\frac12} \nabla_-s \rangle_n] \\
			& =[\langle (\beta^o)^\frac12 \tilde{\psi}^k,r \rangle_{n-1},\langle\nabla_+M_{\beta}^{-\frac12} \psi^{k'},s\rangle_{n-1}]	=i\langle (\beta^o)^\frac12 \tilde{\psi}^k,\nabla_+M_{\beta}^{-\frac12} \psi^{k'} \rangle_{n-1} \\
			& = i\frac{1}{\gamma_k} \langle M_{\beta}^{-\frac12}(-\nabla_- \beta^o \nabla_+)M_{\beta}^{-\frac12} \psi^k,\psi^{k'} \rangle_{n}= \frac{i}{\gamma_k} \langle A_p^{\beta} \psi^k, \psi^{k'} \rangle_n= i\gamma_k \delta_{k,k'}.
		\end{split}
	\end{equation}
In the first equality we substitute $\tilde{\ttr}$ and $\tilde{\ttp}$ with $r$ and $p$, since they only differ in a constant. \\	
	
	 Now designate the bosonic operators $\tilde{\mathfrak{b}}_k, \tilde{\mathfrak{b}}_k^*$ for $k \in \mathbb{I}_{n-1}$ similar to \eqref{bosonoperator}. Their commutation relations, which can be deduced from \eqref{ccrtemp}, reads:
	  \begin{equation} \label{bosonoperatorthermal}
	  \begin{split}
	  	&\tilde{\mathfrak{b}}_k=\frac{1}{\sqrt{2 \gamma_k}}(\tilde{\mathfrak{r}}_k +i \tilde{\mathfrak{p}}_k), \qquad \tilde{\mathfrak{b}}_k^*=\frac{1}{\sqrt{2 \gamma_k}}(\tilde{\mathfrak{r}}_k -i \tilde{\mathfrak{p}}_k),\\
	  	&[\tilde{\mathfrak{b}}_k,\tilde{\mathfrak{b}}_{k'}^*]=\delta_{k,k'}, \quad [\tilde{\mathfrak{b}}_k,\tilde{\mathfrak{b}}_{k'}] =[\tilde{\mathfrak{b}}_k^*,\tilde{\mathfrak{b}}_{k'}^*]=0.
	  \end{split}
	  \end{equation}
 The expression of $H_{\beta}^n$ in terms of these operators is as follows: 
 \begin{equation} \label{hamiltoniantempbosonic}
 	H_{\beta}^n=\frac{\lambda^2 \Pi_0^2}{2} +\sum_{k=1}^{n-1} \gamma_k(\tilde{\mathfrak{b}}_k^*\tilde{\mathfrak{b}}_k +\frac12 ).
 \end{equation}
Let us denote the constant $\frac{\lambda^2 \Pi_0^2}{2} $ by $E_0$. 
Owing to \eqref{hamiltoniantempbosonic}, and the commutator relations in \eqref{bosonoperatorthermal}, we can treat $H_{\beta}^n$ as sum of independent oscillator. Therefore, we can deduce that this operator has a full discrete spectrum, which can be described along with their corresponding eigenfunctions explicitly; 
Let $\pmb{\ket{\Psi_0}} \in L^2(\mathbb{R}^{n-1})$ be the 
the ground state of $H_{\beta}^n$, 
with corresponding energy (eigenvalue) $\tilde{E}_0=E_0+\frac12 \sum_{k=1}^{n-1} \gamma_k$.  Then the set of eigenvalues and eigenfunctions can be labeled by $ \bar{\theta}:=(\theta_1,\dots,\theta_{n-1}) \in \mathbb{N}^{n-1}_0$:
\begin{equation} \label{energyterm}
	E_{\bar{\theta}}=\tilde{E}_0+\sum_{k=1}^{n-1}\theta_k \gamma_k, \qquad \pmb{\ket{\Psi_{\bar{\theta}}}}= \prod_{k=1}^{n-1} \frac{(\tilde{\mathfrak{b}}^*_k)^{\theta_k}}{\sqrt{\theta_k!}} \pmb{\ket{\Psi_0}}.
\end{equation}
 By abusing the notation, we  write $\pmb{\ket{\bar{\theta}}} \equiv\pmb{\ket{(\theta_1,\dots,\theta_{n-1})}}$ instead of $\pmb{\ket{\Psi_{\bar{\theta}}}}$. Here the eigenfunction  $\pmb{\ket{\bar{\theta}}}$ corresponds to the eigenvalue $E_{\bar{\theta}}$. Moreover, $\{ \pmb{\ket{\bar{\theta}}} | \bar{\theta} \in \mathbb{N}^{n-1} \} $ forms an orthonormal basis for $L^2(\mathbb{R}^{n-1})$. Notice that these functions can be obtained rather explicitly, similar to $\pmb{\ket{\Phi_{\bar{\theta}}}}$. Here, they are functions of $\tilde{\pmb{\xi}}_1,\dots \tilde{\pmb{\xi}}_{n-1}$, where $ \forall k \in \mathbb{I}_{n-1}, \tilde{\pmb{\xi}}_k$ can be defined similar to \eqref{bogolitemp1} as: $\tilde{\pmb{\xi}}_k:=\langle (\beta^o)^\frac12 \tilde{\psi}^k,\tilde{\pmb{\xi}} \rangle_{n-1}$, while in the previous case they are written as a function of $\hat{\pmb{\xi}}_1,\dots,\hat{\pmb{\xi}}_{n-1}$. \\
 Let us mention the fact that \eqref{hamiltoniantempbosonic} is sum of "shifted" harmonic oscillators, rather than normal harmonic oscillators, since all the bosonic operators are shifted by a constant. Therefore, obtaining the ground state  
 $ \pmb{\ket{\Psi_0}}$, and other eigenstates of this operator is slightly different from   $\pmb{\ket{\Phi_0}}$ and other eigenstates of $H_n$. Notice that, this shift \textit{does not} change the spectrum, and consequently it does not affect our computation. \\
  In fact, if one construct $\mathfrak{b}_k$, and $\mathfrak{b}_k^*$ from \eqref{notshifted}, 
  similar to \eqref{bosonoperatorthermal}, then $H_{\beta}^n$ is shifted version of $
  \tilde{H}_{\beta}^n =E_0+\sum_k \gamma_k (\mathfrak{b}_k^* \mathfrak{b}_k+\frac{1}{2})$. The ground state of the later is
  well-understood and can be obtained similar to $\pmb{\ket{\Phi_0}}$. Let us denote the
  ground state of $\tilde{H}_{\beta}^n$ by $\pmb{\ket{\tilde{\Psi}_0}}$, then applying
  the displacement operator we obtain $\pmb{\ket{\tilde{\Psi}_0}}=\mathcal{D}\pmb{\ket{\tilde{\Psi}_0}}$. Here the displacement operator $\mathcal{D}$ is a unitary operator which can be define as follow: let $b_k:=\tilde{\mathfrak{b}}_k-\mathfrak{b}_k $  be the displacement constant for $k$-th  oscillator. Then $\mathcal{D}_k=\exp(b_k \mathfrak{b}_k^*- b_k^*\mathfrak{b}_k)$ and 
  $\mathcal{D}=\mathcal{D}_1 \dots \mathcal{D}_{n-1}$.   

 \subsection{Ensemble average} \label{ensemble average} 
	Since we established the eigenfunctions and eigenvalues of $H_{\beta}^n$, using spectral theory one can write $H_{\beta}^n=\sum_{\bar{\theta} \in \mathbb{N}^{n-1}_0} E_{\bar{\theta}} \pmb{\dyad{\bar{\theta}}}$,  where $\pmb{\dyad{\bar{\theta}}}$ is the projection operator on the subspace generated by $\{ \pmb{\ket{\bar{\theta}}} \}$, for $\bar{\theta} \in \mathbb{N}^{n-1}_0$. Moreover, one can observe that $\exp(-H_{\beta}^n)$ is trace-class, thanks to the properties of $\gamma_k$s, and can be written as $\exp(-H_{\beta}^n)=\sum_{\bar{\theta} \in \mathbb{N}^{n-1}} e^{-E_{\bar{\theta}}} \pmb{\dyad{\bar{\theta}}}$, thanks to spectral theory. Recall the number operator $\tilde{\mathfrak{b}}_k^*\tilde{\mathfrak{b}}_{k}$, then we can compute the average of the bosonic operators and number operators in the locally Gibbs state $\rho$ (also known as thermal state). This computation is classical and one can see (Proposition 5.2.28 of \cite{RB2}).  
	\begin{lemma}  \label{avgboslem}
		Recall the density state $\rho$ \eqref{initalstate}, and the operators $\tilde{\mathfrak{b}}_k^* ,\tilde{\mathfrak{b}}_k$ defined in \eqref{bosonoperatorthermal}. Moreover, recall the definition of $\expval{}_{\rho}$ \eqref{mixedaverage}. Then, we have the followings  $\forall k \in \mathbb{I}_{n-1}$:
		\begin{equation} \label{bosonicthermalaverage}
			\expval{\tilde{\mathfrak{b}}_k^*}_{\rho}=\expval{\tilde{\mathfrak{b}}_k}_{\rho}=\expval{\tilde{\mathfrak{b}}_k\tilde{\mathfrak{b}}_{k'}}_{\rho}=\expval{\tilde{\mathfrak{b}}_k^*\tilde{\mathfrak{b}}_{k'}^*}_{\rho}=0, \quad  \expval{\tilde{\mathfrak{b}}_k^* \tilde{\mathfrak{b}}_{k'}}_{\rho}=  \frac{\delta_{k,k'}}{e^{\gamma_k}-1}, \quad   \expval{\tilde{\mathfrak{b}}_k \tilde{\mathfrak{b}}_{k'}^*}_{\rho}=  \frac{\delta_{k,k'}}{e^{\gamma_k}-1}+1.
		\end{equation}
	\end{lemma} 
	
	\begin{proof}  
		One can find the rather straightforward computation of this lemma in Appendix \ref{avgbos}. 
		\end{proof}

		In the rest of this section, we compute the average of certain observables namely momentum, elongation, and energy at each site at time zero, as an application of \eqref{bosonicthermalaverage}. Before proceeding, let us define for $x \in \mathbb{I}_n$: 
		\begin{equation} \label{ERROR!}
					\mathscr{E}_n^x:= \frac{m_x}{\beta_x} \lambda^2 \Pi_0 = 
					\frac{m_x}{\beta_x} \left(\sum_{x=1}^n \frac{m_x}{\beta_x} \right)^{-1}
					\left(\sum_{x=1}^n \bar{p}(\frac{x}{n}) \frac{m_x}{\bar{m}} \right),
		\end{equation}
		Note that $\frac{m_x}{\beta_x} \left(\sum_{x=1}^n \frac{m_x}{\beta_x} \right)^{-1}
		\leq \frac{C}{n}$, where $C= \frac{m_{max}\beta_{max}}{\beta_{min}m_{min}}$ is a
		constant independent of $n$. Therefore, thanks to the Strong Law of Large numbers and
		the assumption $\int_0^1 \bar{p}(y) dy =0$, we have for $y \in [0,1]$, 
		$\mathscr{E}_n^{[ny]} \to 0$ almost 
		surely:
		\begin{equation} \label{ERRORZero}
		 |\mathscr{E}_n^{[ny]} | \leq C \Big|\frac{1}{n} \sum_{x=1}^n \bar{p}(\frac{x}{n})\frac{m_x}
		 {\bar{m}}\Big| = C \frac{\Pi_0}{n} \to \int_0^1 \bar{p}(y) dy =0 ,
		\end{equation}
	almost surely as $n \to \infty$. 
	\begin{corollary} \label{meavg1}
	Recall the coordinates $\tilde{\ttp}_x(0)$ and $\tilde{\ttr}_x(0)$ from \eqref{thermalcoordinate00}, as a direct consequence of Lemma \ref{avgboslem} we have: 
\begin{equation} \label{momentumelongationaverage1}
\begin{split}	
	\forall x \in \mathbb{I}_n, \quad &\expval{\tilde{\ttp}_x(0)}_{\rho}=\expval{\tilde{\ttp}_x}_{\rho}=-\mathscr{E}_n^x, 
	\\ &\expval{\tilde{\ttr}_x(0)}_{\rho}=\expval{\tilde{\ttr}_x}_{\rho}=0.
\end{split}	
	\end{equation}
	Note that these expressions denote the average of momentum and elongation at time zero. This also implies: 
	\begin{equation} \label{momentumelongationaverage2}
 	\expval{p_x}_{\rho}=\frac{m_x}{\bar{m}} \bar{p}_x 
 	-\mathscr{E}_n^x, \quad \expval{r_x}_{\rho}=\bar{r}_x.
 	 \end{equation}
\end{corollary}
\begin{proof}
	By \eqref{bosonoperatorthermal}, and \eqref{bogoliinvthermal} $\forall x \in \mathbb{I}_n
	$, we can write $\tilde{\ttp}_x$ and $\tilde{\ttr}_x$ as linear combination of $
	\tilde{\mathfrak{b}}_k$s and $\tilde{\mathfrak{b}}_k^*$s; moreover, $\tilde{\mathfrak{p}}_0$
	appears in case of $\tilde{\ttp}_x$ as well.
	 Then, since $\expval{.}_{\rho}$  is a linear operator, we can deduce \eqref{momentumelongationaverage1}.\\
	To be more precise, from \eqref{bosonoperatorthermal} we have: 
	\begin{equation} \label{bogoliinvthermal1}
		\tilde{\mathfrak{p}}_k=i \sqrt{\frac{\gamma_k}{2}}(\tilde{\mathfrak{b}}_k^*-\tilde{\mathfrak{b}}_k), \qquad \tilde{\mathfrak{r}}_k= \sqrt{\frac{\gamma_k}{2}}(\tilde{\mathfrak{b}}_k^*+\tilde{\mathfrak{b}}_k).
		\end{equation}
Hence, if we  substitute  $\tilde{\mathfrak{p}}_k$ and $\tilde{\mathfrak{r}}_k$ with the later in \eqref{bogoliinvthermal},  use the definition $\psi^0_x =
\lambda \sqrt{\frac{m_x}{\beta_x}}$ as well as the expression \eqref{pzero}, we obtain 
	
	\begin{equation} \label{bosonicexpansionthermal}
		\begin{split}		
			&\tilde{\ttp}_x= 
			\sqrt{\frac{m_x}{\beta_x}} \left[\psi^0_x \tilde{\mathfrak{p}}_0+  \sum_{k=1}
			^{n-1} \psi^k_x\Big(i
		\sqrt{\frac{\gamma_k}{2}}(\tilde{\mathfrak{b}}_k^*-\tilde{\mathfrak{b}}_k) \Big) 
		\right]= -\frac{m_x}{\beta_x}\lambda^2 \Pi_0 + \sum_{k=1}
			^{n-1} \psi^k_x\Big(i
		\sqrt{\frac{\gamma_k}{2}}(\tilde{\mathfrak{b}}_k^*-\tilde{\mathfrak{b}}_k) \Big)
		,\\
		&\tilde{\ttr}_x =\frac{1}{\sqrt{\beta_x}} \sum_{k=1}^{n-1} \tilde{\psi_x^k}\Big( 
		\sqrt{\frac{\gamma_k}{2}}(\tilde{\mathfrak{b}}_k^*+\tilde{\mathfrak{b}}_k)\Big).
	\end{split}
 	\end{equation}

  Since $\expval{.}_{\rho}$ is linear, thanks to identities $\expval{\tilde{\mathfrak{b}}_k^*}_{\rho}=\expval{\tilde{\mathfrak{b}}_k}_{\rho}=0$ \eqref{bosonicthermalaverage}, we get \eqref{momentumelongationaverage1}. In order to obtain  \eqref{momentumelongationaverage2} from \eqref{momentumelongationaverage1} it's enough to recall these definitions: $\tilde{\ttp}_x=p_x-\frac{m_x}{\bar{m}}\bar{p}_x, \quad \tilde{\ttr}_x=r_x-\bar{r}_x$. 
 
\end{proof}
	The average of momentum and elongation in our thermal state is understood in \eqref{momentumelongationaverage2}. Later we need their fluctuation as well. Hence, we define 
	the following operators, we may refer to them as thermal coordinate, since they correspond 
	to the thermal fluctuation:
	\begin{equation} \label{thermalcoordinate}
		\begin{split}
		&\tilde{p}_x:=p_x-\expval{p_x}_{\rho}=p_x-\bar{p}_x \frac{m_x}{\bar{m}}+ \mathscr{E}
		^x_n=\tilde{\ttp}_x+\mathscr{E}_n^x,  \qquad \forall x \in \mathbb{I}_n , \\
 &\tilde{r}_x:=r_x-\expval{r_x}_{\rho}=r_x-\bar{r}_x= \tilde{\ttr}_x,  \qquad \forall x \in \mathbb{I}_{n-1},
				\end{split}
	\end{equation}
	Notice that these coordinate are similar to $\tilde{\ttp}$, and $\tilde{\ttr}$
	up to a vanishing error. \\
	Let us compute the fluctuation of $p_x$ and $r_x$ in the state $\rho$:

	\begin{corollary}
	As another straightforward consequence of Lemma \ref{avgboslem} we have: 
	\begin{equation} \label{p2thermalaverage}
	  \frac{\expval{\tilde{p}_x^2}_{\rho}}{m_x} = 
	  \frac{1}{\beta_x}\sum_{k=1}^{n-1} (\psi_x^k)^2 
	  \frac{\gamma_k}{2}\coth(\frac{\gamma_k}{2}),
	\end{equation}
	\begin{equation} \label{r2thermalaverage}
		\expval{\tilde{r}_x^2}_{\rho} =\frac{1}{\beta_x}\sum_{k=1}^{n-1} (\tilde{\psi}^k_x)^2 \frac{\gamma_k}{2}\coth(\frac{\gamma_k}{2}).
	\end{equation}
	\end{corollary}
	\begin{proof}
		From \eqref{bogoliinvthermal1} and \eqref{bosonicthermalaverage} one can observe: 
		\begin{equation} \label{lemma22}
		\begin{split} 		
 		&\expval{\tilde{\mathfrak{p}}_i \tilde{\mathfrak{p}}_j}_{\rho} =\expval{\tilde{\mathfrak{r}}_i \tilde{\mathfrak{r}}_j}_{\rho} =\delta_{i,j}\gamma_i\Big(\frac{1}{e^{\gamma_i}-1}+\frac{1}{2}\Big)=\delta_{i,j}\frac{\gamma_i}{2}\coth(\frac{\gamma_i}{2}), \quad \forall i,j \in \mathbb{I}_{n-1},\\
 		&\expval{\tilde{\mathfrak{p}}_i \tilde{\mathfrak{p}}_0}_{\rho}= \delta_{i,0} 
 		\lambda^2 \Pi_0^2, \qquad \forall i \in \mathbb{I}_n,
 	\end{split}
	\end{equation}	
	where in the second expression we take advantage of the fact that $\tilde{\mathfrak{p}}_0$
	is a constant  given in \eqref{pzero}, and  $\expval{\tilde{\mathfrak{p}}_i}_{\rho}=0$ for $i>0$. \\	 
	 Thanks to \eqref{thermalcoordinate}, we have $\tilde{p}_x=\tilde{\ttp}_x+ \mathscr{E}_n^x$,
	 and $\tilde{r}_x=\tilde{\ttr}_x$. If one replace  $\tilde{\ttp}_x$, and $\tilde{\ttr}_x$
	 with corresponding expressions from \eqref{bogoliinvthermal} and use the definition of 
	 $\mathscr{E}_n^x$, and $\tilde{\mathfrak{p}}_0$ it is clear that:
	 \begin{equation} \label{bogoliinvthermal00}
	 \tilde{p}_x=\sqrt{\frac{m_x}{\beta_x}} \sum_{k=1}^{n-1} \psi^k_x \tilde{\mathfrak{p}}_k,
	 \qquad \tilde{r}_x = \frac{1}{\sqrt{\beta_x}} \sum_{k=1}^{n-1} \tilde{\psi}^k_x 
	 \tilde{\mathfrak{r}}_k.
	\end{equation}	  
	 	 We compute 
	  $\frac{\expval{\tilde{p}_x^2}}{m_x}_{\rho}$ and $\expval{\tilde{r}_x^2}_{\rho}$
	 by squaring these expressions, and then taking the ensemble average $\expval{}_{\rho}$. 
	 Therefore, the linearity of the trace and the identities in \eqref{lemma22} give us the 
	 result:
	\begin{equation} \label{p2thermalaveragecomp}
		\begin{split}			
			&\tilde{p}_x^2=\frac{m_x}{\beta_x} \sum_{k,k'=1}^{n-1} \psi^k_x \psi^{k'}_x
			\tilde{\mathfrak{p}}_k \tilde{\mathfrak{p}}_{k'}, \implies 
		\expval{\tilde{p}_x^2}_{\rho}=\frac{m_x}{\beta_x} \sum_{k,k'=1}^{n-1} \psi^k_x \psi^{k'}_x \expval{\tilde{\mathfrak{p}}_k \tilde{\mathfrak{p}}_{k'}}_{\rho}. \\
		&\frac{\expval{\tilde{p}_x^2}_{\rho}}{m_x}=	\frac{1}{\beta_x}\sum_{k,k'=1}^{n-1} \psi^k_x \psi^{k'}_x \delta_{k,k'} \frac{\gamma_k}{2} \coth(\frac{\gamma_k}{2})= 
				\frac{1}{\beta_x} \sum_{k=1}^{n-1} (\psi_x^k)^2 \frac{\gamma_k}{2}\coth(\frac{\gamma_k}{2}).
		\end{split}
	\end{equation}
	The exact same computation using \eqref{bogoliinvthermal} and \eqref{lemma22}, gives us 
	corresponding expression in \eqref{r2thermalaverage} for  $\expval{\tilde{r}_x}_{\rho}$.
	\end{proof}
	
	\begin{remark} \label{rmkthermaleqavg2}
		Notice that for a clean chain (all masses equal to $m$), in thermal  equilibrium at temperature $\beta_{eq}^{-1}$, with periodic boundary conditions, we can obtain $\psi^k$  by discrete Fourier transform. In this case, $\gamma_k=\omega_k=2|\sin(\pi(\frac{k}{n}))|$. Therefore, since boundary effects disappear in the limit as $n \to \infty$,  we get up to a vanishing error:  	
		\begin{equation} \label{rmkthermaleq}
		\expval{\tilde{r}_x^2}_{\rho}=\frac{1}{\beta_{eq}}\frac{1}{2n} \sum_{k=1}^{n-1} \frac{\beta_{eq}\omega_k}{2\sqrt{m}}\coth(\frac{\omega_k\beta_{eq}}{2\sqrt{m}})+\epsilon_n.
		\end{equation}
The same expression can be obtained for $\frac{\expval{\tilde{p}_x}_{\rho}}{m}$, for a clean chain in thermal equilibrium. Taking the limit of $n \to \infty$, we obtain the constant $\mathrm{f}_{\beta}$ in \eqref{thermaleqcte}.
	\end{remark}
	
	Notice the difference of \eqref{r2thermalaverage} and \eqref{p2thermalaverage} with the classical case, where these averages are simply equal to $\frac{1}{\beta_x}$. Moreover observe that since $\gamma_k$ and $\psi^k (\tilde{\psi}^k) $ are eigenvalues and eigenvectors of $A_p^{\beta}=M_{\beta}^{-\frac12}(-\nabla_-\beta^0 \nabla_+)M_{\beta}^{-\frac12} ( A_r^{\beta}= (\beta^o)^{\frac12}(-\nabla_+ M_{\beta}^{-1} \nabla_-)(\beta^o)^{\frac12}
)$, it is obvious that for each configuration of the masses the averages $\frac{\expval{\tilde{p}_x^2}_{\rho}}{m_x}$ and $\expval{\tilde{r}_x^2}_{\rho}$ depend on the whole configuration of the masses.\\
	
	 For our purposes, in particular, in Section \ref{SLLNsection},  it would be  useful to rewrite \eqref{p2thermalaverage},\eqref{r2thermalaverage} in the following form: Recall the definition of $\ket{x}_n$\footnote{Notice the difference between the notation $\pmb{\ket{\Psi}}$ which is used for denoting the member of the Hilbert space, and $\ket{\psi}$ which denotes the finite dimensional vector spaces.} for $x \in \mathbb{I}_n$, as the canonical basis of $\mathbb{R}^n$, i.e $\ket{x}=(0,\dots,0,1,0\dots,0)^{\dagger}$, where $1$ is at the $x$th position. So we can write $\psi_x^k= \langle x ,\psi^k \rangle_n$. Similarly, let $\ket{x}$ for $x \in \mathbb{I}_{n-1}$, denotes the canonical basis for $\mathbb{R}^{n-1}$. Then we have: 

	\begin{equation} \label{r2p2thermalaverage}
		\expval{\tilde{r}_x^2}_{\rho}= \frac{1}{\beta_x}\expval{x ,\frac{(A_r^{\beta})^\frac12}{2} \coth(\frac{(A_r^{\beta})^\frac12}{2}) x}_{n-1}, \quad \frac{\expval{\tilde{p}_x^2}_{\rho}}{m_x}= \frac{1}{\beta_x}\expval{x ,\frac{(A_p^{\beta})^\frac12}{2} \coth(\frac{(A_p^{\beta})^{\frac12}}{2}) x}_n.
	\end{equation}
Here by convention, formally we denote $0\coth(0)=0$.\footnote{We will modify this convention later.}\\	
Since $A_p^{\beta}$ ($A_r^{\beta}$) is positive semidefinite (definite) one can define by spectral theorem the following matrices 
	 $$\mathsf{A}_p:=\frac{(A_p^{\beta})^\frac12}{2} \coth(\frac{(A_p^{\beta})^{\frac12}}{2}), \quad \mathsf{A}_r=\frac{(A_r^{\beta})^{\frac12}}{2} \coth(\frac{(A_r^{\beta})^\frac12}{2}).$$ So if we expand \eqref{r2p2thermalaverage} in the basis of $\psi^k$, and use the identity $\sum_{k=0}^{n-1} \dyad{\psi^k}=I_n$,  we get the exact same expression as in \eqref{p2thermalaverage}: 
	\begin{equation} \label{cothjustification}
		\expval{x,\mathsf{A}_p x}= \sum_{k=0}^{n-1} \expval{x, \psi^k}\expval{\psi^k,\mathsf{A}_px}=\sum_{k=0}^{n-1} \frac{\gamma_k}{2} \coth(\frac{\gamma_k}{2}) \expval{x,\psi^k}\expval{\psi^k,x}= \sum_{k=1}^{n-1} \frac{\gamma_k}{2} \coth(\frac{\gamma_k}{2}) (\psi^k_x)^2.
	\end{equation}
Similarly, we can justify the expression in \eqref{r2p2thermalaverage} for $\expval{\tilde{r}_x^2}$.
Using spectral properties of $\mathsf{A}_r$.

Recall the canonical transformations  \eqref{bogolip}, \eqref{bogolir}, \eqref{bogoliinv1},
we define yet another set of operators: let $\hat{\tilde{p}}$ and $\hat{\tilde{r}}$ to be the transformed form of $\tilde{r}$ and $\tilde{p}$, respectively:
\begin{equation} \label{hattilde}
		\begin{split}	
	&\hat{\tilde{p}}:=O^{\dagger}M^{-\frac12}\tilde{p}, \qquad \hat{\tilde{p}}_k:= 
	\langle M^{-\frac12}\varphi^k,\tilde{p} \rangle_n,  \\
	&\hat{\tilde{r}}:=\tilde{O}^{\dagger}\tilde{r}, \qquad  \qquad \:  \hat{\tilde{r}}_k:=
	\langle \phi^k, \tilde{r} \rangle_{n-1}.
	\end{split}
\end{equation}
 Since we use the explicit solution of  equations of motion, terms like $\expval{(\hat{\tilde{p}}_k)^2}_{\rho}$ and $\expval{(\hat{\tilde{r}}_k)^2}_{\rho}$ arise in our calculations, and the following lemma permits us to deal with them. Note that this lemma reflects one of the technical differences of this model with its classical counterpart.
 \begin{lemma} \label{boundinitial}
	Considering the above definitions, there exists a deterministic constant $C>0$ independent of $n$, such that for any realization of the masses and any $k \in \mathbb{I}_n$:
	\begin{equation} \label{essbound}
		\expval{(\hat{\tilde{p}}_k)^2}_{\rho} < C, \quad	\expval{(\hat{\tilde{r}}_k)^2}_{\rho} <C, \qquad \expval{\hat{\tilde{p}}_k}_{\rho}=\expval{\hat{\tilde{r}}_k}_{\rho}=0.
	\end{equation}
 \end{lemma}
\begin{proof}
Plugging $\tilde{p}_x$ and $\tilde{r}_x$ from \eqref{bogoliinvthermal00} into the definition \eqref{hattilde}, we get the following linear relation between these two sets of transformed operators:
	\begin{equation} \label{bogoli1vsbogoli22}		
		\begin{split}		
		\hat{\tilde{p}}_k=\sum_{i=1}^{n-1} \langle \varphi^k, \beta^{-\frac12} \psi^i \rangle  \tilde{\mathfrak{p}}_i, \qquad \forall k \in \mathbb{I}_{n-1}^0, \\
		\hat{\tilde{r}}_k=\sum_{i=1}^{n-1} \langle \phi^k, \beta^{-\frac12} \tilde{\psi}^i \rangle  \tilde{\mathfrak{r}}_i, \qquad \forall k \in \mathbb{I}_{n-1}.
		\end{split}
	\end{equation} 	 
 Since  $\tilde{\mathfrak{p}}_i$ and $\tilde{\mathfrak{r}}_i$ are linear combinations of bosonic operators $\tilde{\mathfrak{b}}_i,\tilde{\mathfrak{b}}_i^*$; obviously we have $\expval{\tilde{\mathfrak{r}}_i}_{\rho}=\expval{\tilde{\mathfrak{p}}_i}_{\rho}=0$. Hence, by linearity of $\expval{.}_{\rho}$, we have $\expval{\hat{\tilde{p}}_k}_{\rho}=\expval{\hat{\tilde{r}}_k}_{\rho}=0$. 
	For the purpose of establishing the bounds in \eqref{essbound}, we square the expression \eqref{bogoli1vsbogoli22} and by using \eqref{lemma22}, we have:
	\begin{equation} \label{hattilde2}
		\begin{split}
		\expval{(\hat{\tilde{p}}_k)^2}_{\rho} &=\sum_{i,j=1}^{n-1}\langle \varphi^k,\beta^{-\frac12} \psi^i \rangle \langle \beta^{-\frac12}\psi^j,\varphi^k \rangle \expval{\mathfrak{p}_i \mathfrak{p}_j}_{\rho} = \sum_{i,j=1}^{n-1}\langle \varphi^k,\beta^{-\frac12} \psi^i \rangle \langle \beta^{-\frac12}\psi^j,\varphi^k \rangle \delta_{i,j}\frac{\gamma_i}{2}\coth(\frac{\gamma_i}{2}) 
		\\&= \sum_{i=1}^{n-1} \big|\langle \beta^{-\frac12} \varphi^k, \psi^i \rangle 
		\big|^2 \frac{\gamma_i}{2} \coth(\frac{\gamma_i}{2}).
		\end{split}	
	\end{equation}
Since $\gamma_i$ are the eigenvalues of $A_p^{\beta}=M_{\beta}^{-\frac12}(-\nabla_-\beta^0 \nabla_+)M_{\beta}^{-\frac12}$, $\forall i, \: \gamma_i \leq ||A_p^{\beta}||_2 $, where $||.||_2$ denotes the matrix norm induced by the Euclidean norm.\footnote{For every linear function $f: \mathbb{R}^n \to \mathbb{R}^m  $ with corresponding $m\times n$ matrix $A$, we define $||A||_2 =\sup_{|x|_n=1}\frac{|Ax|_m}{|x|_n}$, where $|.|$ denotes the Euclidean norm in $\mathbb{R}^n$: $|x|_n=(\sum_{i=1}^n x_i^2)^{\frac12}$.} However, the following bound is evident from the definition of the matrices appearing in $A_p^{\beta}$:  $||A_{p}^{\beta}||_2 \leq||M_{\beta}^{-\frac12}||_2^2||\beta^o||_2||\nabla_-||_2||\nabla_+||_2 \leq \frac{4\beta_{max}^2}{m_{min}} $. This bound holds uniformly in $n$, for any realization of the masses, since the distribution of the masses is bounded, and  $\beta$ is continuous. Therefore, we deduce that there is a deterministic $c>0$, independent of $n$ such that for any realization of the masses $||A_p^{\beta}||_2 \leq c$. Furthermore, since the function $f(x)=x\coth x$ is continuous  in $(0,c)$, the expression $\frac{\gamma_i}{2} \coth(\frac{\gamma_i}{2})$ is nonnegative and bounded by a constant $c'$, independent of $n$, so we have for any realization of the masses: 

\begin{equation}
	\expval{(\hat{\tilde{p}}_k)^2} \leq c'\sum_{i=1}^{n-1} |\langle \beta^{-\frac12} \varphi^k, \psi^i \rangle |^2 = c'|\beta^{-\frac12} \varphi^k|^2 \leq c' ||\beta^{-\frac12}||_2^2|\varphi^k|^2 \leq c' \beta_{min} \leq C. 
\end{equation}
The sum in this expression is the expansion of the vector $\beta^{-\frac12}\varphi^k$ in the basis of $\psi^i$. We also used the fact that $|\varphi^k|_2=1$, and  norm of $\beta^{-\frac12}$ is bounded by $\beta_{min}^{-\frac12}$. 
For $\expval{(\hat{\tilde{r}}_k)^2}_{\rho}$, we proceed similarly and get the following expression: 
\begin{equation}
	\expval{(\hat{\tilde{r}}_k)^2}_{\rho}=\sum_{i=1}^{n-1} |\langle \beta^{-\frac12} \phi^k, \tilde{\psi}^i \rangle |^2 \frac{\gamma_i}{2} \coth(\frac{\gamma_i}{2}),
\end{equation}    
which can be treated exactly similar to the previous bound.
\end{proof}
\begin{remark} \label{themalboundinitialremark}
	Recall the averages $\expval{\tilde{r}_x^2}_{\rho}$ and $\frac{\expval{\tilde{p}_x^2}_{\rho}}{m_x}$ from \eqref{r2p2thermalaverage}:
	 $$\expval{\tilde{r}_x^2}_{\rho}= \frac{1}{\beta_x}\expval{x ,\frac{(A_r^{\beta})^\frac12}{2} \coth(\frac{(A_r^{\beta})^\frac12}{2}) x}_{n-1}, \quad \frac{\expval{\tilde{p}_x^2}_{\rho}}{m_x}= \frac{1}{\beta_x}\expval{x ,\frac{(A_p^{\beta})^\frac12}{2} \coth(\frac{(A_p^{\beta})^{\frac12}}{2}) x}_n.$$ In the proof of lemma \ref{boundinitial}, we observed that the the norm of the matrices $\frac{(A_p^{\beta})^{\frac12}}{2} \coth(\frac{(A_p^{\beta})^{\frac12}}{2})$,  \\
	 and $\frac{(A_r^{\beta})^{\frac12}}{2} \coth(\frac{(A_r^{\beta})^{\frac12}}{2})$ are bounded by a constant $c'$, uniformly in $n$. Therefore, we can deduce that there exists a constant $C$ uniform in $n$ such that for any realization of the masses: 
	\begin{equation} \label{thermalboundinitial}
		\expval{\tilde{r}_x^2} \leq C, \quad\frac{\expval{\tilde{p}_x^2}}{m_x} \leq C,
	\end{equation}
 where we used the fact that $\beta$ is continuous, with $\beta_{min}\leq\beta(y)$ for all $y \in [0,1]$, with $\beta_{min}$ strictly positive.
\end{remark}

\section{Wave equation} \label{WE}

In this section, we are going to show the limits \eqref{rconv} and \eqref{pconv}. Since our system is harmonic, the dynamic is linear.
As we already observed our chain evolves in time according to the Heisenberg dynamics generated by the Hamiltonian  \eqref{hambosontime}. Recall the definition of the dynamic: $\forall a \in \mathcal{B}(\mathcal{H}_n)$ define $a(t)$ as follows: 
\begin{equation} \label{heisenbergdynamic0}
	a(t) := e^{iH_nt}ae^{-iH_nt}.
\end{equation} 
Since $H_n$ is self-adjoint, using Stone's theorem, \eqref{heisenbergdynamic0} is the continuous one parameter group of authomorphisms. Moreover, we can extend the definition of this evolution to certain unbounded operators, such as $b_k$ and $b_k^*$, where we have: 
\begin{equation}
b_k(t) =e^{-i \omega_k t}b_k, \qquad b_k^*(t)=e^{i \omega_k t}b_k^*.
\end{equation}
By using linearity, we obtain the explicit time evolution for elongation and momentum operators in \eqref{bogoliinv2}. One may use this explicit solutions in order to demonstrate the limits \eqref{pconv} and \eqref{rconv}. However, we proceed using the equation of motions and certain homogenization lemmas.  \\
Recall the definition of the thermal state $\rho^n$ \eqref{initalstate}, and the thermal average: $ \expval{a}_{\rho^n}:=\Tr(\rho^na)$ for an observable  $a$, such that $a\rho^n$ be trace class. Since $\rho^n= \exp(-H_{\beta}^n)$, thanks to the spectral theory we observed that $\forall x $, $\rho^n p_x$, $r_x \rho^n$, $\rho^n r_x^2$, and $\rho^n p_x^2$ are trace class. Since the solution in \eqref{bogoliinv2} is linear, we can deduce that $\forall x, \forall t, \in [0,T]$, $p_x(nt)$, $r_x(nt)$, $p_x^2(nt)$, and $r_x^2(nt)$ are trace class. Hence, we can introduce the following notation:  
\begin{equation} \label{averagenotation}
\bar{p}_x(nt):=\expval{p_x(nt)}_{\rho^n}, \quad \bar{r}_x(t):=\expval{r_x(t)}_{\rho^n}.
\end{equation} \\
 Recall that  according to \eqref{momentumelongationaverage2}, we have $\forall x \in \mathbb{I}_n$:

\begin{equation} \label{initialaverage}
\bar{p}_x(0)=\frac{m_x}{\bar{m}} \bar{p}_x-\mathscr{E}_n^x, \quad 
\bar{r}_x(0)=\bar{r}_x.
\end{equation} 
Moreover, the time evolution of $\bar{p}_x(nt)$ and $\bar{r}_x(nt)$ can be represented as a system of coupled ODEs. 
First, observe that the dynamic defined in \eqref{heisenbergdynamic0}, gives us the equations of motions as \eqref{equationofmotionr}, then by the following simple computation we have:  
\begin{equation} \label{averagetimeevolution}
\begin{split}	
	& \frac{d}{dt} \bar{p}_x(t) =  \expval{\dot{p}_x(t)}_{\rho^n}= \expval{r_x(t)-r_{x-1}(t)}_{\rho^n}=  (\bar{r}_x(t)-\bar{r}_{x-1}(t))=  (\nabla_-\bar{r}(t))_x, \\
	& \frac{d}{dt} \bar{r}_x(t)= \expval{\dot{r}_x(t)}_{\rho^n}=\expval{\frac{p_{x+1}(t)}{m_{x+1}}-\frac{p_x(t)}{m_x}}_{\rho^n}= \Big(\frac{\bar{p}_{x+1}(t)}{m_{x+1}}-\frac{\bar{p}_x(t)}{m_x}\Big)=(\nabla_+ M^{-1}\bar{p}(t))_x,
\end{split} 
\end{equation}
where, $\nabla_-$ and $\nabla_+$ were defined in \eqref{nabla}. The justification for this computation is as follows: the operator $\rho^n p_x(nt)$ is bounded, and the time derivative exists in the operator norm (for this bounded operator) and we can change the trace and derivative by a simple argument.\\

Comparing the functions $\bar{p}(nt): \mathbb{R}^n \to \mathbb{R}$, and $\bar{r}(nt): \mathbb{R}^{n-1} \to \mathbb{R}$ for $ t \in [0,T]$, with their classical counterpart in \cite{BHO}, it is evident that they satisfiy the same coupled system of linear ODEs, with the same initial conditions up to a vanishing purturbation $\mathscr{E}^x_n$ (for each realization of the masses).
Therefore, these functions are very similar, and we can adapt the method of section 3 of \cite{BHO}, and prove  \eqref{rconv} and \eqref{pconv} in theorem \ref{maintheorem}. This conclusion is obtained from the fact that in the harmonic systems, both the classical and quantum evolutions are linear. Although, the result of \cite{BHO} is applicable here, their proof is not optimal and has a certain gap\footnote{The gap is as follows: comparing the relations (3.11), (3.12) with (3.14),(3.15) is not sufficient to close the argument, since the derivative of $f$ and $g$ appears in the RHS instead of $f$ and $g$.}. Moreover, we need
to take care of $\mathscr{E}_n^x$ separately, hence we state a modified version of that proof here. In fact, we prove this theorem assuming the function $\beta$ is Lipschitz continuous, 
since this proof is shorter and better illustrate the idea, then we bring the proof of general case $\beta \in C^0([0,1])$ afterwards.  

\begin{proof} [Proof of \eqref{rconv}, and \eqref{pconv} with $\beta$ Lipschitz]
We divide the proof into two steps:\\
\textit{Step1. A priori bound:} \\
Define $\bar{H}_n(t)$ and $\bar{I}_n(t)$ as follows:
\begin{equation} \label{conserevedquantities}
\begin{split}
&\bar{H}_n(t) := \sum_{x=1}^{n} \frac{\bar{p}_x^2(t)}{2m_x} + \sum_{x=1}^{n-1} \frac{\bar{r}_x^2(t)}{2}= \frac{1}{2}\expval{\bar{p}(t),M^{-1} \bar{p}(t)}_n + \frac{1}{2}\expval{\bar{r}(t),\bar{r}(t)}_{n-1}, \\
&\bar{I}_n(t) :=\frac{1}{2} \sum_{x=1}^n \frac{\big(\bar{r}_x(t)-\bar{r}_{x-1}(t)\big)^2}{m_x} + \frac12 \sum_{x=1}^{n-1} \Big(\frac{\bar{p}_{x+1}(t)}{m_{x+1}}-\frac{\bar{p}_x(t)}{m_x}\Big)^2 = \frac12 \expval{\nabla_- \bar{r}(t),M^{-1}\nabla_- \bar{r}(t)}_n \\
&+\frac12\expval{\nabla_+ M^{-1} \bar{p}(t),\nabla_+M^{-1}\bar{p}(t)}_{n-1}.
\end{split}
\end{equation}
From the time evolution \eqref{averagetimeevolution}, it is evident that $\forall n \in \mathbb{N}$ the quantities in $\eqref{conserevedquantities}$ are conserved. The first quantity $\bar{H}_n(t)$, can be viewed as the mechanical energy. We will see later that the average of the energy $\expval{H_n}_{\rho^n}$, can be decomposed into the mechanical and thermal parts, where $\bar{H}_n$ is the mechanical part. The second quantity $\bar{I}_n$, shows us a typical way of constructing the other conserved quantities by taking further gradients (See Remark \ref{OCQ}).

The conservation of $\bar{I}_n(t)$ and $\bar{H}_n(t)$ in \eqref{conserevedquantities},  the regularity assumptions, where $\bar{r},\bar{p} \in C^1([0,1])$, and the properties of the masses, give us the following bounds: there exists a deterministic $C>0$, such that for every $n \in \mathbb{N}$ and $t \in [0,T]$,  we have:
\begin{equation} \label{boundswaveeq1}
 \sum_{x=1}^{n}\big(\bar{p}_x^2(nt) +\bar{r}_x^2 (nt) \big)\leq Cn, 
 \end{equation}
\begin{equation} \label{boundwaveeq2}
	\sum_{x=1}^n \big(\bar{r}_x(nt)-\bar{r}_{x-1}(nt)\big)^2\leq \frac{C}{n}, \quad \sum_{x=1}^{n-1} \Big(\frac{\bar{p}_{x+1}(nt)}{m_{x+1}}-\frac{\bar{p}_x(nt)}{m_x}\Big)^2 \leq \frac{C}{n}.
\end{equation}
Notice that these bounds hold for every time scale $n^{\alpha}t$, for $\alpha>0$. First, observe that \eqref{boundswaveeq1}, is bounded by $2(1+m_{max})\bar{H}_n(nt)$,  and
expressions in \eqref{boundwaveeq2} are bounded by 
 and $2(1+m_{max})\bar{I}_n(nt)$.
By the conservation of $\bar{I}_n(t)$ and $\bar{H}_n(t)$ in  \eqref{conserevedquantities}, it is enough to show $\bar{H}_n(nt) \leq Cn$, and  $\bar{I}_n(nt) \leq \frac{C}{n}$ for $t=0$. But this is obvious: Since $\bar{r}, \bar{p} \in C^1([0,1])$, $\bar{p}_x^2$ and $\bar{r}_x^2$ are bounded for every $x$, which gives \eqref{boundswaveeq1}. Moreover, $(\bar{r}_x-\bar{r}_{x-1})^2=(\bar{r}(\frac{x}{n})-\bar{r}(\frac{x-1}{n}))^2 \leq \frac{c_2}{n^2}$, since $\bar{r} \in C^1([0,1])$ (choose $c_2 = ||\bar{r}'||_{L^{\infty}}$). Moreover, thanks to \eqref{initialaverage}  we have

\begin{equation} \label{lippp}
\Big(\frac{\bar{p}_{x+1}(0)}{m_{x+1}}-\frac{\bar{p}_x(0)}{m_x}\Big)^2 \leq 
\frac{2}{\bar{m}^2}\left(\bar{p}\big(\frac{x+1}{n}\big)-\bar{p}\big(\frac{x}{n}\big) \right)^2 +
2\big(\frac{\mathscr{E}_n^{x+1}}{m_{x+1}}-\frac{\mathscr{E}_n^x}{m_x}\big)^2 \leq \frac{c_3}{n}.
\end{equation}
Note that in \eqref{lippp}, first, we apply the property $\bar{p} \in C^1([0,1])$. Then we take advantage  of the identity  $\frac{\mathscr{E}^x_n}{m_x}=(\beta(\frac{x}{n}))^{-1}(\sum_{x=1}^n \frac{m_x}{\beta_x})^{-1}(\sum_{x=1}^n \bar{p}(\frac{x}{n})\frac{m_x}{\bar{m}})$ and we bounded $|\frac{1}{\beta_x}-\frac{1}{\beta_{x-1}}| \leq \frac{c'_3}{n}$ by using the assumption that $\beta$ is Lipschitz and $0< \beta_{min}<\beta(y)< \beta_{max}$. Finally, we bounded the rest by a constant thanks to the properties of $\bar{p}$ and $m_x$. (Note that this is the only place where we use the assumption that $\beta$ is Lipschitz.

From \eqref{boundswaveeq1}, \eqref{boundwaveeq2} by Cauchy-Schwartz inequality we can deduce that $\bar{r}_x(nt), \bar{p}_x(nt)$ are H\"{o}lder regular and bounded, in the following sense:  there exists a deterministic constant $C>0$, such that for every $n$ and every $x,x' \in \mathbb{I}_n$ we have:
\begin{equation} \label{holderreg}
	|\bar{r}_x(nt)-\bar{r}_{x'}(nt)| \leq \frac{C|x-x'|^{\frac12}}{\sqrt{n}},  \qquad \Big|\frac{\bar{p}_x(nt)}{m_x}-\frac{\bar{p}_{x'}(nt)}{m_{x'}}\Big| \leq \frac{C|x-x'|^{\frac12}}{\sqrt{n}}. 
\end{equation}  
 Moreover thanks to \eqref{boundswaveeq1} and \eqref{holderreg}, there exists $C'>0$, such that $\forall n$ and $\forall x \in \mathbb{I}_n$, we have: 
 \begin{equation} \label{boundwaveeq3}
 	|\bar{r}_x(nt)|\leq C', \qquad |\bar{p}_x(nt)| \leq C',
 \end{equation}

  \textit{Step2.  Mass Homogenization } \label{masshom} \\
 For every $f \in C^0([0,1])$ and $t \in [0,T]$, as 
 $N \to \infty$ we have:
 \begin{equation} \label{masshomo1}
 \frac{1}{N} \sum_{x=1}^{N} f(\frac{x}{N}) \frac{\bar{p}_x(Nt)}{m_x}(m_x-\bar{m}) \to 0,
 \end{equation}
	\begin{equation} \label{masshomo2}
		\frac{1}{N} \sum_{x=1}^{N} f(\frac{x}{N}) \Big(\frac{\bar{p}_x(Nt)}{m_x}\Big)^2(m_x-\bar{m}) \to 0,
	\end{equation}
 almost surely with respect to the distribution of the masses, where $\bar{m}=\mathbb{E}(m_x)$. This step permits us to deal with the randomness of the masses by homogenizing them. The second limit \eqref{masshomo2} will be used in the next section. 
  \begin{proof}
 For the proof of this step, one can see Lemma 2 in \cite{BHO}. Let us emphasize the fact that 
 this lemma's proof only need estimates \eqref{holderreg}, and \eqref{boundswaveeq1}. 
 Therefore, we can use Lemma 2 of \cite{BHO} here. 
 \end{proof}
\textit{Step3. Weak Convergence to the solution of the wave equation with a $C^2$ test function.}\\
In this step, we prove the convergences \eqref{rconv}, \eqref{pconv}, for a special class of test functions. In the next step, we complete the proof by using the H\"{o}lder bounds in \eqref{holderreg}. Notice that here, we follow a different path in comparison to \cite{BHO}. 
Let $\forall t \in [0,T]$, and $f,g \in C^2([0,1])$, such that $f(0)=f(1)=0$, $g(0)=g(1)$. Moreover, $f$ and $g$ are continuously differentiable at $0$ and $1$, with  $g'(0)=g'(1)=0$ and $f'(0)=f'(1)$ ($f,g$ are periodic, with Dirichlet boundary condition for $f$, and Neumann boundary condition for $g$). Then, we have: 

\begin{align}
	\frac{1}{n} \sum_{x=1}^n f(\frac{x}{n}) \expval{r_x(nt)}_{\rho} = \frac{1}{n} \sum_{x=1}^nf(\frac{x}{n})\bar{r}_x(nt) \to \int_0^1 f(y)\mathrm{r}(y,t)dy, \label{rconv2}\\
			\frac{1}{n} \sum_{x=1}^n g(\frac{x}{n}) \expval{p_x(nt)}_{\rho}=\frac{1}{n} \sum_{x=1}^{n} g(\frac{x}{n})\bar{p}_x(nt) \to \int_0^1 g(y)\mathrm{p}(y,t)dy, \label{pconv2}
\end{align}

almost surely, with respect to the distribution of the masses as $n \to \infty$, where $\mathrm{r}(y,t)$ and $\mathrm{p}(y,t)$ are the unique strong solutions to the following system of conservation laws \eqref{pde1}, \eqref{bc1}, and \eqref{bc2}. 

\begin{proof} [Proof of Step 3]
First, for every $f,g \in C^0([0,1])$, such that $f(0)=f(1)=0$, $g(0)=g(1)$,    define the following kernels: 
\begin{equation} \label{kerneldef}
R(f,t):= \int_0^1 \mathrm{r}(y,t)f(y)dy, \quad \: P(g,t):= \int_0^1 \mathrm{p}(y,t)g(y)dy.
\end{equation}
In particular, if we assume $f,g \in C^2([0,1])$ satisfying the assumption of this Step, we can argue as follows:
 Since the solutions of \eqref{pde1} are explicit, and the unique weak solution coincide with the strong solution (See e.g. \cite{evans}), by using the explicit form of the solution, we have a uniform bound on the time derivative of $\mathrm{r}$ and $\mathrm{p}$. Therefore, by using the dominated convergence theorem we have: 
 \begin{equation} \label{kernelder1}
		\begin{split} 	
 	\partial_s R(f,s) &=\int_0^1 \partial_s \mathrm{r}(y,t) f(y)dy= \frac{1}{\bar{m}}\int_0^1 \partial_y \mathrm{p}(y,t) f(y)dy= -\frac{1}{\bar{m}}\int_0^1\mathrm{p}(y,t)f'(y)dy \\
 	&= -\frac{1}{m}P(f',t),
	\end{split} 
 \end{equation}
	where we integrated by parts, and used the property $f(0)=f(1)=0$. Similarly, using the boundary condition $\mathrm{r}(1,s)=\mathrm{r}(0,s)$, implies: 
	\begin{equation} \label{kernelder2}
\begin{split}		
		\partial_s P(g,s) &= \int_0^1 \partial_s \mathrm{p}(y,s)g(y)=\int_0^1 \partial_y\mathrm{r}(y,t)g(y)dy = -\int_0^1 \mathrm{r}(y,t)g'(y)dy \\
                          &=R(g',t).
\end{split}	
	\end{equation}	 
Therefore, by using \eqref{kernelder1}, \eqref{kernelder2}, we can characterize $R(f,t)$ and $P(g,t)$, for the aforementioned $f,g \in C^2([0,1])$, with proper boundary conditions, as follows: 
\begin{equation} \label{kernelchar}
\begin{split}
	&R(f,t)=R(f,0)-\frac{1}{\bar{m}} \int_0^t P(f',s)ds, \\
	&P(g,t)=P(g,0)-\int_0^t R(g',s)ds.
	\end{split}
\end{equation} 
Notice that we have: 
\begin{equation}
R(f,0)= \int_0^1 \bar{r}(y)f(y)dy, \quad P(g,0)=\int_0^1 \bar{p}(y)g(y)dy.
\end{equation}
In the microscopic level, for $n \in \mathbb{N}$, we define the kernels $R_n(f,t)$ and $P_n(g,t)$, for $t \in [0,T]$ and $f,g \in C^0([0,1])$: 
\begin{equation} \label{kerneln} 
\begin{split}
R_n(f,t)= \frac{1}{n} \sum_{x=1}^n f(\frac{x}{n}) \expval{r_x(nt)}_{\rho}=\frac{1}{n}\sum_{x=1}^{n}f(\frac{x}{n})\bar{r}_x(nt), \\
P_n(g,t)= \frac{1}{n} \sum_{x=1}^n g(\frac{x}{n}) \expval{p_x(nt)}_{\rho}=\frac{1}{n}\sum_{x=1}^{n}g(\frac{x}{n})\bar{p}_x(nt).
\end{split}
\end{equation}
In particular, for $f,g \in C^2([0,1])$, satisfying the assumptions of this step, we can characterize $R_n(f,t), P_n(f,t)$ as follows: 
\begin{equation} \label{kernelnchar}
\begin{split}
	&R_n(f,t)= R_n(f,0)+ \int_0^t \partial_s R_n(f,s)ds = R_n(f,0) + \int_0^t \frac{1}{n} \sum_{x=1}^n f(\frac{x}{n}) \partial_s\big(\bar{r}_x(ns)\big)ds= \\ & R_n(f,0)+\int_0^t  \sum_{x=1}^n f(\frac{x}{n})\Big(\frac{\bar{p}_{x+1}(ns)}{m_{x+1}}-\frac{\bar{p}_x(ns)}{m_x}\Big)ds = R_n(f,0)- \int_0^t\sum_{x=1}^n \Big(f(\frac{x}{n})-f(\frac{x-1}{n})\Big) \frac{\bar{p}_x(ns)}{m_x}ds,
\end{split}
\end{equation}
where we used the time evolution of $\bar{r}(t)$ from \eqref{averagetimeevolution}  in the second line, then we performed a summation by parts, using the assumption $f(0)=f(1)=0$. Since $f \in C^2([0,1])$, \\$f(\frac{x}{n})-f(\frac{x-1}{n})=\frac{1}{n}f'(\frac{x}{n})+\frac{\epsilon_n^x}{n^2}$, where $\epsilon_n^x$ is bounded by a constant $C>0$, uniformly in $x$ and $n$. Hence, we have: 
\begin{equation} 
R_n(f,t)=R_n(f,0)-\int_0^t \frac{1}{n} \sum_{x=1}^n f'(\frac{x}{n})\frac{\bar{p}_x(ns)}{m_x}ds + \epsilon_n,
\end{equation}
where $\epsilon_n$ is the remainder term and one can observe $\epsilon_n \to 0$ in a deterministic way, as $n \to \infty$.   
Moreover, we can use the result of Step 2, namely \eqref{masshomo1}, in order to replace $m_x$ by $\bar{m}$ and get  $\frac{1}{n}\sum_{x=1}^n f'(\frac{x}{n})\frac{\bar{p}_x(ns)}{m_x} - \frac{1}{n}\sum_{x=1}^n f'(\frac{x}{n})\frac{\bar{p}_x(ns)}{\bar{m}}=\epsilon'_n(s)$, where $\epsilon'_n \to 0$, almost surely as $n \to \infty$. Note that $\epsilon'(s)$ is bounded thanks to \eqref{boundwaveeq3}, and by dominated convergence theorem, $\bar{\epsilon}_n =\int_0^t \epsilon'(s)ds  \to 0$, almost surely. Hence, by using the definition of $P_n(g,s)$, we have: 
\begin{equation} \label{kernelncharr}
R_n(f,t)= R_n(f,0)- \frac{1}{\bar{m}} \int_0^t P_n(f',s)ds +\bar{\epsilon}_n,
\end{equation} 
where $\bar{\epsilon}_n \to 0$, as $n \to \infty$, almost surely. 
We can proceed similarly in order to obtain the counterpart of \eqref{kernelncharr} for $P_n(f,t)$. Notice that for $P_n(g,t)$, homogenization over the masses is not necessary.
\begin{equation}
\begin{split}
P_n(g,t)= &P_n(g,0)+\int_0^t \partial_s P_n(g,s)ds = P_n(g,0)+ \int_0^t \frac{1}{n} \sum_{x=1}^n  g(\frac{x}{n})\big(\partial_s \bar{p}_x(ns)\big)ds=  P_n(g,0)+ \\ &\int_0^t  \sum_{x=1}^n g(\frac{x}{n})\big(\bar{r}_x(ns)-\bar{r}_{x-1}(ns)\big)ds= P_n(g,0)- \int_0^t  \sum_{x=1}^n \bar{r}_x(ns)\Big(g\big(\frac{x+1}{n}\big)-g\big(\frac{x}{n}\big)\Big)ds, 
\end{split}
\end{equation}
where we advanced similar to \eqref{kernelchar}, using \eqref{averagetimeevolution} and the microscopic boundary condition $\bar{r}_0(s)=\bar{r}_n(s)=0$, for the summation by parts. Again, since $g \in C^2([0,1])$, we shall write $g(\frac{x+1}{n})-g(\frac{x}{n})=\frac{1}{n}g'(\frac{x}{n})+\tilde{\epsilon}_x^n$, where $|\tilde{\epsilon}_x^n| \leq \frac{C'}{n^2} $, for a $C'$, uniform in $x$ and $n$. Therefore, by using the bound $|\bar{r}_x(ns)| \leq C$, and the definition of $R_n(g',s)$, we have:
\begin{equation} \label{kernelncharp}
P_n(g,t)=P_n(g,0)-\int_0^t \frac{1}{n} \sum_{x=1}^n \bar{r}_x(ns)g'(\frac{x}{n}) + \tilde{\epsilon}_n
=P_n(g,0)-\int_0^t R_n(g',s)ds + \tilde{\epsilon}_n,
\end{equation}
such that $\tilde{\epsilon}_n \to 0$ as $n \to \infty$. 

 For every continuous $f,g$ (therefore, for $f,g$ satisfying our assumptions), we have the following observation: One can obtain the solution to the system of ODEs \eqref{averagetimeevolution} explicitly \footnote{These solutions are exactly  similar to \eqref{bogoliinv2}, where the operators at time zero replaced by the averaged function at time zero.}  and observe that for any fixed continuous $(f,g)$: $\forall n,  R_n(f,t), P_n(g,t)$ are smooth in time (at least $C^1$). Moreover, by computing their derivatives,  we obtain a uniform bound $\bar{C}>0$ (uniform in $t$ and $n$):
 \begin{equation}
	\begin{split} 	
 	&|\partial_t R_n(f,t)|=\Big| \sum_{x=1}^{n-1} f(\frac{x}{n}) \Big(\frac{\bar{p}_{x+1}(nt)}{m_{x+1}}-\frac{\bar{p}_x(nt)}{m_x}\Big) \Big| \leq C_1 \sum_{x=1}^n \Big|\frac{\bar{p}_{x+1}(nt)}{m_{x+1}}-\frac{\bar{p}_x(nt)}{m_x} \Big| \leq \\
 	&C_1 (\sum_{x=1}^n \Big|\frac{\bar{p}_{x+1}(nt)}{m_{x+1}}-\frac{\bar{p}_x(nt)}{m_x}\Big|^2)^{\frac12}(n)^{\frac12}
\leq C_1n^{\frac12}(\frac{C}{n})^{\frac12} \leq \bar{C},
\end{split}
  \end{equation}
  where we used the Cauchy Schwartz inequality and the bound in \eqref{boundwaveeq2}. Similarly, using the other inequality in \eqref{boundwaveeq2}, we have the similar uniform bound  $\bar{C}$ for $|\partial_tP_n(g,t)|$.\\
  
   For proving this step, we show that $R_n(f,t) \to R(f,t)$ and $P_n(g,t) \to P(g,t)$, almost surely as $n \to \infty$, for every $f$ and $g$, satisfying the smoothness and boundary condition assumption, and every $t\in [0,T]$. First, notice that for $t=0$, we have $R_n(f,0)=\frac{1}{n}\sum_{x=1}^nf(\frac{x}{n})\bar{r}(\frac{x}{n})$, hence, the convergence to $R(f,0)=\int_0^1 f(y)\bar{r}(y)dy$  is evident.\\
    For  $P_n(g,0)= \frac{1}{n} \sum_{x=1}^n g(\frac{x}{n})\bar{p}_x(0)=\frac{1}{n} \sum_{x=1}^n g(\frac{x}{n})\frac{m_x}{\bar{m}}(\bar{p}_x-\mathscr{E}^x_n)$, where we used the definition of $\bar{p}_x(0)$ from \eqref{initialaverage}, we can deduce  $\frac{1}{n} \sum_{x=1}^n g(\frac{x}{n})\frac{m_x}{\bar{m}}(\bar{p}_x) \to P(g,0)=\int_0^1 g(y)\bar{p}(y)$ almost surely, by the Strong Law of Large Numbers. Moreover, 
    $$\Big|\frac{1}{n}\sum_{x=1}^n g(\frac{x}{n})\mathscr{E}^x_n \Big| \leq \frac{C}{n} \sum_{x=1}^n |\mathscr{E}^x_n|=\frac{C}{n}\Big| \sum_{x=1}^n \bar{p}(\frac{x}{n})\frac{m_x}{\bar{m}}\Big| \to 0,$$ 
    
    almost surely, where we use the definition of $\mathscr{E}_n^x$ \eqref{ERROR!}, and 
    the fact that $\frac{\Pi_0}{n} \to 0$ almost surely \eqref{ERRORZero}. 
    Combining the last two convergences we obtain $P_n(g,0) \to P(g,0)$ almost surely for any 
    continuous $g$.
    
    We use the characterizations \eqref{kernelchar}, \eqref{kernelncharr}, and \eqref{kernelncharp} in order to prove the result for $t \in (0,T]$. 

 Fix proper $f$ and $g$, and consider the families of functions $ \{ (R_n(f,.)\}_n, \{ P_n(g,.)) \}_n$. These families are equicontinuous, since we established  a uniform bound on their derivatives\footnote{For every $\epsilon>0$, take $\delta=\frac{\epsilon}{\bar{C}}$, then for every $\phi$ in this family, and $t,t' \in [0,T]$, if $|t-t'|<\delta$, we have $|\phi(t)-\phi(t')| \leq |t-t'||\phi'(t^*)| \leq \bar{C}|t-t'| \leq \epsilon$, for $t<t^*<t'$, since all the functions in this family are smooth, and their derivative are uniformly bounded. }. Hence, by Arzel\`{a}-Ascoli theorem, there exist continuous functions $\tilde{\varphi}^f_r(t), \tilde{\varphi}^g_p(t)$ on $[0,T]$, such that a subsequence of  $R_n(f,t)$  and $P_n(g,t)$ converges to these functions, respectively. \\ 
  
   In particular, we can take $f_k(y)=\sin(k\pi y)$ and $g_k(y)=\cos(k\pi y)$, for every $k \in \mathbb{N}_0$. Denote $R(f_k,t)$ and $P(f_k,t)$ by $\varphi_r^k(t)$ and $\varphi_p^k(t)$, respectively. Since $R$ and $P$ are linear in their first arguments, $f_k'=\pi kg_k $ and $g_k'= -\pi k f_k$,  by using the characterization \eqref{kernelchar}, we have $\varphi_p^k$ and $\varphi_r^k$ for every $k \in \mathbb{N}_0$, satisfy the following system of ODEs:
   \begin{equation} \label{ODE1}
   		\varphi_r^k(t)= \varphi_r^k(0)-\frac{\pi k}{\bar{m}}\int_0^t \varphi^k_p(s)ds, \quad \varphi_p^k(t)= \varphi_p^k(0)+\pi k\int_0^t \varphi^k_r(s)ds. 
	\end{equation}     
	
	Recall the continuous functions $\tilde{\varphi}^{f_k}_r$ and $\tilde{\varphi}^{g_k}_p$ as the limit of a subsequence of $R_n(f_k,t)$ and $P_n(f_k,t)$ and denote them by $\tilde{\varphi}^{k}_r$ and $\tilde{\varphi}^{k}_p$, respectively. Recall the characterizations \eqref{kernelncharr} and \eqref{kernelncharp}, observe that $P_n$ and $R_n$ are linear in their first argument and take the limit of these characterizations for the subsequences converging to $\tilde{\varphi}^{k}_r$ and $\tilde{\varphi}^{k}_p$. By using the dominated convergence theorem (since these functions are bounded on a compact domain), we deduce that $\tilde{\varphi}^{k}_r$ and $\tilde{\varphi}^{k}_p$ satisfy the exact same system of ODEs as \eqref{ODE1}, almost surely. Moreover, we observed earlier the convergence at time zero i.e $\tilde{\varphi}^{k}_r(0)=\varphi^k_r(0)$ and $\tilde{\varphi}^k_p(0)=\varphi^{k}_p(0)$ almost surely. Therefore, by a uniqueness argument, we have $\tilde{\varphi}^{k}_r(t)=\varphi^k_r(t)$ and $\tilde{\varphi}^k_p(t)=\varphi^{k}_p(t)$, for every $t$ in $[0,T]$ almost surely. \\
 Notice that this argument is valid for any limiting point of $R_n(f_k,t)$  and $P_n(f_k,t)$. Hence, $\forall k \in \mathbb{N}_0$, and $\forall t \in [0,T]$, we have $R_n(f_k,t) \to R(f_k,t)$ and $P_n(g_k,t) \to P(g_k,t)$, as $n \to \infty$, almost surely. \\
   
   Finally, using the fact that the sets $\{ f_k \}_{k=0}^{\infty}$ and $\{ g_k \}_{k=0}^{\infty}$ are orthonormal (Fourier) basis of $L^2([0,1])$, we can finish the proof with an $\frac{\epsilon}{3}$ argument, thanks to the fact that $R_n(.,t)$ and $P_n(.,t)$ are linear in their first argument.    
\end{proof}

\textit{Step4: The pointwise convergence of $\bar{r}$ and $\bar{p}$:}\\
In this step, we  prove the pointwise convergence $\bar{r}_{[ny]}(nt) \to \mathrm{r}(y,t)$ and $\frac{\bar{p}_{[ny]}}{m_{[ny]}} \to \frac{\mathrm{p}(y,t)}{\bar{m}}$, almost surely, exploiting the "H\"{o}lder" bounds in \eqref{holderreg}. Concretely,  $\forall y \in (0,1)$ and $t \in [0,T]$ we have: 
\begin{equation} \label{pointwiseconvergence}
 \bar{r}_{[ny]}(nt) \to \mathrm{r}(y,t), \quad \frac{\bar{p}_{[ny]}(nt)}{m_{[ny]}} \to \frac{\mathrm{p}(y,t)}{\bar{m}},  
\end{equation}
 almost surely, with respect to the distribution of the masses.
 \begin{proof}
		Fix $y \in (0,1)$, and take $\epsilon_1>0$ such that $y \in (2\epsilon_1,1-2\epsilon_1)$. Let $\zeta$ be a positive symmetric mollifier i.e. $\zeta \in C^{\infty}_c (\mathbb{R})$ (infinitely differentible, compactly supported, with $supp(\zeta)=[-1,1]$), $\int_{\mathbb{R}} \zeta(y')dy'=1$, $\zeta(y) \geq 0$, and $\zeta(y)=\zeta(-y)$. Let $\zeta_{\epsilon}:= \frac{1}{\epsilon}\zeta(\frac{y}{\epsilon})$, for $0<\epsilon< \epsilon_1$, be a regularizing family, notice that we have $\zeta_{\epsilon} \in C^{\infty}_c(\mathbb{R})$, $supp(\zeta_{\epsilon})=[-\epsilon,\epsilon]$, and $\int_{\mathbb{R}} \zeta_{\epsilon}(y')dy'=1$, as well as $\zeta_{\epsilon}(y)\geq 0$. Since $\int_{\mathbb{R}}\zeta_{\epsilon}(y')dy'=1$, multiplying by $\bar{r}_{[ny]}(nt)$, we have $\forall \epsilon_1>\epsilon>0$: 
		\begin{equation} \label{pointwise1}
			\begin{split}
					&\bar{r}_{[ny]}(nt)= \bar{r}_{[ny]}(nt) \int_{\mathbb{R}} \zeta_{\epsilon}(y-y')dy'	= \int_{\mathbb{R}}	\zeta_{\epsilon}(y-y')\bar{r}_{[ny]}(nt)dy' =\\ &\int_{\mathbb{R}}\zeta_{\epsilon}(y-y')\big(\bar{r}_{[ny]}(nt)-\bar{r}_{[ny']}(nt)\big)dy' + \int_{\mathbb{R}} \zeta_{\epsilon}(y-y') \bar{r}_{[ny']}(nt)dy'.
			\end{split}
		\end{equation}
		Since $\zeta_{\epsilon}(y-y')$ is supported on $(y-\epsilon,y+\epsilon)$, we have:
		\begin{equation} \label{pointwise11}
\begin{split}		
		&\Big|\int_{\mathbb{R}}\zeta_{\epsilon}(y-y')(\bar{r}_{[ny]}(nt)-\bar{r}_{[ny']}(nt))dy'\Big| \leq \sup_{y' \in (y-\epsilon,y+\epsilon)}|\bar{r}_{[ny]}(nt)-\bar{r}_{[ny']}(nt)| \int_{\mathbb{R}} \zeta_{\epsilon}(y')dy' = \\ & \sup_{y' \in (y-\epsilon,y+\epsilon)}|\bar{r}_{[ny]}(nt)-\bar{r}_{[ny']}(nt)| \leq C \sqrt{\epsilon}, 
\end{split}		
		\end{equation}
		where the last bound is deduced from \eqref{holderreg}, since $|y-y'|\leq \epsilon$. In order to deal with the second term in \eqref{pointwise1}, notice that by the choice of $\epsilon_1$, $supp(\zeta_{\epsilon}) \subset (0,1)$, hence we have 
		\begin{equation}
		\begin{split}
			& \int_{\mathbb{R}} \zeta_{\epsilon}(y-y') \bar{r}_{[ny']}(nt)dy' = \int_0^1 \zeta_{\epsilon}(y-y') \bar{r}_{[ny']}(nt)dy' = \sum_{x=0}^{n-1} \int_{\frac{x}{n}}^{\frac{x+1}{n}} \zeta_{\epsilon}(y-y') \bar{r}_{[ny']}(nt)dy' \\ 
			&=\sum_{x=1}^n \bar{r}_x(nt) \int_{\frac{x}{n}}^{\frac{x+1}{n}} \zeta_{\epsilon}(y-y') dy'	=\frac{1}{n} \sum_{x=1}^n \bar{r}_x(nt) \zeta_{\epsilon}(y-\frac{x}{n}) 
			\\ &+\sum_{x=1}^n \bar{r}_x(nt)\int_{\frac{x}{n}}^{\frac{x+1}{n}} \big(\zeta_{\epsilon}(y-y')-\zeta_{\epsilon}(y-\frac{x}{n})\big)dy',
		\end{split}
		\end{equation}
where we used the fact that $\zeta_{\epsilon}$ is smooth, and $\bar{r}_{[ny']}(nt)$ is a step function. Since $\zeta_{\epsilon}$ is smooth ($C^{\infty}$) and compactly supported, $|\int_{\frac{x}{n}}^{\frac{x+1}{n}} (\zeta_{\epsilon}(y-y')-\zeta_{\epsilon}(y-\frac{x}{n}))dy'| \leq \frac{M}{n^2}$, where $M$ is a constant independent of $n$ (for example, one can take $M$ as $\sup |\zeta_{\epsilon}'|$). Therefore, thanks to the bound on $\bar{r}_x(nt)$ in \eqref{boundwaveeq3}, the last term is bounded by $\frac{c'}{n}$, where $c'$ is a constant uniform in $y$ and $n$ \footnote{Notice that we misuse the bound of the sums in the last expression thanks to the support of $\zeta_{\epsilon}$.}.
After all, we have $\forall \: 0<\epsilon < \epsilon_1$: 
\begin{equation} \label{pointwise2}
\bar{r}_{[ny]}(nt)=\frac{1}{n}\sum_{x=1}^n\bar{r}_x(nt)\zeta_{\epsilon}(y-\frac{x}{n}) +\epsilon'(\epsilon) +\epsilon''(n),
\end{equation} 
 where $|\epsilon'(\epsilon)|$ is bounded by $C\sqrt{\epsilon}$, and $|\epsilon''(n)|$ is bounded by $\frac{c'}{n}$.  
		 
 However, observe that thanks to the choice of $\epsilon$, $\forall \epsilon$, $\zeta_{\epsilon}(y-.)$ satisfies the properties of the test function $f$, in the  step 3. 
 Therefore, by using \eqref{rconv2}, and \eqref{pconv2}, for $f(.)=\zeta_{\epsilon}(y-.)$ and taking the limit $n \to \infty$ in \eqref{pointwise2}, we get $\forall \epsilon >0$, as $n \to \infty$:
 \begin{equation} \label{pointwise3}
   \bar{r}_{[ny]}(nt) \to \int_0^1 \zeta_{\epsilon}(y-y') \mathrm{r}(y',t) +\epsilon'(\epsilon),
\end{equation}
almost surely.
Taking the limit $\epsilon \to 0$, in \eqref{pointwise3}, since the left hand side is independent of $\epsilon$, thanks to the continuity of $\mathrm{r}(y,t)$, and properties of $\zeta_{\epsilon}$, the first term converges to $\mathrm{r}(y,t)$. The second term converges to zero, thanks to the bound $\epsilon'(\epsilon) \leq C\sqrt{\epsilon}$.  This finishes the proof of $\bar{r}_{[ny]}(nt) \to \mathrm{r}(y,t)$. \\
In order to deal with $\bar{p}_{[ny]}(nt)$, we can proceed similarly. First, for the sake of obtaining the counterpart of \eqref{pointwise11}, we may use the second bound in \eqref{holderreg}, where $|\frac{\bar{p}_x(nt)}{m_x} -\frac{\bar{p}_{x'}(nt)}{m_{x'}}|\leq C \frac{|x-x'|^{\frac12}}{\sqrt{n}}$. Then we will get the following expression, similar to \eqref{pointwise2} (where we used the bound $\frac{\bar{p}_x}{m_x} \leq C$, as well):
\begin{equation} \label{pointwise4}
 \frac{\bar{p}_{[ny]}(nt)}{m_{[ny]}}=\frac{1}{n} \sum_{x=1}^n \frac{\bar{p}_x(nt)}{m_x} \zeta_{\epsilon}(y-\frac{x}{n}) +\tilde{\epsilon}'(\epsilon) + \tilde{\epsilon}''(n),
\end{equation}
 where similar to the previous case, $|\tilde{\epsilon}'(\epsilon)| \leq C \sqrt{\epsilon}$, and $|\tilde{\epsilon}''(n)| \leq \frac{c'}{n}$. Here, since $$\frac{1}{n} \sum_{x=1}^n \zeta_{\epsilon}(y-\frac{x}{n})\frac{\bar{p}_x(nt)}{m_x}=\frac{1}{n \bar{m}} \sum_{x=1}^n \zeta_{\epsilon}(y-\frac{x}{n})\frac{\bar{p}_x(nt)}{m_x}(\bar{m}-m_x) +\frac{1}{n}\sum_{x=1}^n \zeta_{\epsilon}(y-\frac{x}{n})\frac{p_x(nt)}{\bar{m}},$$
we can replace $m_x$ with $\bar{m}$ in \eqref{pointwise4}, with the cost of an error term (the first term in the last relation), that goes to zero almost surely, thanks to the result \eqref{masshomo1} from Step 2.  
Since $\zeta_{\epsilon}(y-.)$ satisfies the criteria of $g$ in Step 3, by using the result of this step and taking $n \to \infty$ in \eqref{pointwise4}, we have: 
\begin{equation}
\frac{\bar{p}_{[ny]}}{m_{[ny]}} \to \int_0^1 \frac{\mathrm{p}(y',t)}{\bar{m}} \zeta_{\epsilon}(y-y')dy' +\tilde{\epsilon}'(\epsilon), 
\end{equation}
almost surely.
Taking $\epsilon \to 0 $ similar to the previous case, the continuity of $\mathrm{p}(y,t)$, and the fact that $\tilde{\epsilon}'(\epsilon) \leq C \sqrt{\epsilon}$, finish the proof of this step. 
 \end{proof}
\textit{Step 5. Finishing the proof }\\ In this step, we finish the proof of the first part of \ref{maintheorem}, namely the convergences \eqref{rconv}, \eqref{pconv}. 
Take $f \in C^0([0,1])$, then as $n \to \infty$ we have: 
\begin{equation} \label{rconv3}
\frac{1}{n} \sum_{x=1}^n f(\frac{x}{n})\expval{r_x(nt)}_{\rho}=\frac{1}{n} \sum_{x=1}^n f(\frac{x}{n})\bar{r}_x(nt) \to \int_0^1 f(y) \mathrm{r}(y,t)dy,
\end{equation}
\begin{equation} \label{pconv3}
\frac{1}{n} \sum_{x=1}^n f(\frac{x}{n})\expval{p_x(nt)}_{\rho}=\frac{1}{n} \sum_{x=1}^n f(\frac{x}{n})\bar{p}_x(nt) \to \int_0^1 f(y) \mathrm{p}(y,t)dy,
\end{equation}
almost surely, with respect to the distribution of the masses.	
	\begin{proof}
		In order order to prove \eqref{rconv3}, notice that we have:
		$$ \frac{1}{n} \sum_{x=1}^{n} \bar{r}_x(nt) f\left(\frac{x}{n}\right) = \int_0^1 f\big(\frac{[ny]}{n}\big) \bar{r}_{[ny]}(nt)dy.$$	
		Using the pointwise convergence result \eqref{pointwiseconvergence}, from the previous step, and the continuity of $f$, we have the pointwise convergence: $f(\frac{[ny]}{n})\bar{r}_{[ny]}(nt) \to f(y)\mathrm{r}(y,t),$ almost surely. Thanks to the bound $|\bar{r}_{[ny]}(nt)| \leq C,$ in \eqref{boundwaveeq3}, we deduce the result i.e. \eqref{rconv3}, by dominated convergence theorem. 
		Finally, to prove \eqref{pconv3}, we write the right hand side as: 
		\begin{equation} \label{step51}
			\frac{1}{n} \sum_{x=1}^{n} f(\frac{x}{n})\bar{p}_x(nt)= \frac{1}{n} \sum_{x=1}^n f(\frac{x}{n})\frac{\bar{p}_x(nt)}{m_x}(m_x-\bar{m}) +\frac{\bar{m}}{n} \sum_{x=1}^n f(\frac{x}{n}) \frac{\bar{p}_x(nt)}{m_x}.
		\end{equation}			 
	The first term in the latter goes to zero almost surely, thanks to \eqref{masshomo1}. For the second term  we can argue similar to the term corresponding to $r$: it converges to $\int_0^1f(y)\mathrm{p}(y,t)$, almost surely, thanks to the  pointwise convergence: $f(\frac{[ny]}{n})\frac{\bar{p}_{[ny]}(nt)}{m_{[ny]}} \to f(y)\frac{\mathrm{p}(y,t)}{\bar{m}}$ in \eqref{pointwiseconvergence}, the bound $|\bar{p}_x(nt)| \leq C$ in \eqref{boundwaveeq3}, and the dominated convergence theorem. This finishes the proof.
	\end{proof}

\end{proof}
\begin{remark} \label{lippprem}
Notice that we only use the fact that $\beta: \mathbb{R} \to [0,1]$ is Lipschitz in the bound 
\eqref{lippp}. In particular, if we have $\mathscr{E}^x_n=0$ for all $x$, the same result \eqref{rconv3}, \eqref{pconv3} holds with $\beta \in C^0([0,1])$.  In fact, if $\bar{p}_x(nt)$ and $\bar{r}_x(nt)$ denote the solution of time evolution \eqref{averagetimeevolution} with initial condition $\bar{p}_x(0)=\frac{m_x}{\bar{m}}$ and $\bar{r}_x(0)=\bar{r}_x$ then we can deduce \eqref{rconv3},  \eqref{pconv3}, and 
\eqref{pointwiseconvergence} with $\beta \in C^0([0,1])$. 
\end{remark}
 \begin{proof} [Proof of \eqref{rconv} and \eqref{pconv} in general case]

As we observed in Remark \ref{lippprem}, in case $\mathscr{E}^x_n=0$ for all $x$, we have \eqref{rconv3}, and \eqref{pconv3}. Since the evolution of $\bar{p}(nt)$, $\bar{r}(nt)$ \eqref{averagetimeevolution} is linear, to prove \eqref{rconv} and \eqref{pconv} it is sufficient to prove the following: 
For any $n$, let $\pi(nt) \in \mathbb{R}^n$ $\varrho(nt) \in \mathbb{R}^{n-1}$ be the solution of \eqref{averagetimeevolution} with initial datum 
$$ \pi_x(0)=-\mathscr{E}^n_x, \qquad \varrho_x(0)=0.  $$
Then for any test function $f \in C^0([0,1])$, we have: 
\begin{equation}\label{errorconv}
\begin{split}
&\frac{1}{n}\sum_{x=1}^n f\big(\frac{x}{n}\big) \pi_x(nt)  \to 0,\\ 
&\frac{1}{n} \sum_{x=1}^n f\big( \frac{x}{n} \big) \varrho_x(nt) \to 0,
\end{split}
\end{equation} 
almost surely as $n \to \infty$. 
Since $\pi_x(nt)$, $\rho_x(nt)$ solves \eqref{averagetimeevolution} we have the conservation of the mechanical energy
$\bar{H}_n(t)$: 
\begin{equation} \nonumber
\bar{H}_n(t) = \frac12 \sum_{x=1}^n \left( \frac{(\pi_x(nt))^2}{m_x}+ (\varrho_x(nt))^2 \right).
\end{equation}
By a Cauchy-Schwartz inequality we get: 
\begin{equation}
\begin{split}
&\Big|\frac{1}{n}\sum_{x=1}^n f\big(\frac{x}{n}\big) \pi_x(nt) \Big| \leq 
\left(\frac{1}{n} \sum_{x=1}^n f^2\big(\frac{x}{n}\big) \right)^{\frac12}
\left(\frac{1}{n}\sum_{x=1}^n \pi^2_x(nt) \right)^{\frac12} \leq \\
 &\Big|\frac{C}{n} \bar{H}_n(t) \Big|^{\frac12}= \Big|\frac{C}{n} \bar{H}_n(0) \Big|^{\frac12}
 =\Big|\frac{C}{n} \sum_{x=1}^n(\mathscr{E}_n^x)^2 \Big|^{\frac12} \to 0,
\end{split}
\end{equation}
almost surely, where we used the initial datum, as well as the fact that $\mathscr{E}_n^{[ny]} \to 0$ almost surely thanks to its definition \eqref{ERROR!}, and \eqref{ERRORZero}. We can proceed similarly for $\varrho(nt)$, and
this finishes the proof.  
 \end{proof}

\section{Energy at time zero}
\subsection{Strong Law of Large Numbers for Energy at time zero} \label{SLLNsection}

In order to prove the convergence of the distribution of the energy to the solution of the Euler equation \eqref{econv}, first, we  need to show this convergence at time zero, for any test function $g \in C^0([0,1])$, almost surely i.e.  
\begin{equation} \label{e0conv}
	\frac1n \sum_{x=1}^{n} g(\frac{x}{n}) \expval{e_x}_{\rho} \to \int_0^1 g(y)\Big(\frac{\bar{p}(y)^2}{2\bar{m}}+\frac{\bar{r}(y)^2}{2} + \mathrm{f}_{\beta}^{\mu}(y)\Big)dy.
\end{equation}   

First, we decompose the energy into the mechanical and thermal (fluctuation) part: Recall the definition of $\tilde{p}_x$ and $\tilde{r}_x$ as $\tilde{p}_x=p_x-\expval{p_x}_{\rho}= p_x-\frac{m_x}{\bar{m}}\bar{p}_x+ \mathscr{E}_n^x$, and $\tilde{r}_x=r_x-\expval{r_x}_{\rho}=r_x-\bar{r}_x$, respectively. Then we have: 
\begin{equation} \label{decompose0}
\begin{split}
\expval{e_x}_{\rho}=\expval{\frac{p_x^2}{2m_x}+\frac{r_x^2}{2}}_{\rho}= 
\frac{1}{2}(\frac{\expval{p_x}_{\rho}^2}{m_x} + \expval{r_x}_{\rho}^2 +  \frac{\expval{\tilde{p}_x^2}_{\rho}}{m_x}+ \expval{\tilde{r}_x^2}_{\rho}).
\end{split}
\end{equation}
Moreover, observe that 
 \begin{equation} \label{edit1}
 \frac{1}{n}\sum_{x=1}^n \frac{1}{2}g(\frac{x}{n})\Big(\frac{\expval{p_x}_{\rho}^2}{m_x} + \expval{r_x}_{\rho}^2\Big) \to  \int_0^1 g(y)\Big(\frac{\bar{p}(y)^2}{2\bar{m}}+\frac{\bar{r}(y)^2}{2}\Big)dy,
 \end{equation}
  almost surely, with respect to the distribution of the masses. Notice that \eqref{edit1} is a direct consequence of Corollary \eqref{meavg1}, where we have $\expval{r_x}_{\rho}=\bar{r}_x=\bar{r}(\frac{x}{n})$ and $\expval{p_x}_{\rho}=\frac{m_x}{\bar{m}}\bar{p}_x- \mathscr{E}_x^n$. Then we applied the Strong Law of Large Numbers for $\{m_x \}$ and take advantage of the fact that $p,r,g$ are continuous. Moreover, we use the  fact that terms corresponding to 
  $\mathscr{E}_n^x$ are  vanishing \eqref{ERROR!}, \eqref{ERRORZero}.
	Therefore, in order to deduce \eqref{e0conv}, we shall show the following convergence: 
	\begin{equation}
	\frac{1}{n} \sum_{x=1}^{n}g(\frac{x}{n})\frac12\Big(\frac{\expval{\tilde{p}_x^2}_{\rho}}{m_x}+ \expval{\tilde{r}_x^2}_{\rho}\Big) \to \int_0^1 g(y)\mathrm{f}_{\beta}^{\mu}(y),
	\end{equation}
	 almost surely. Before proceeding, we define the function $\mathrm{f}_{\beta}^{\mu}(y)$ as follows: denote $\tilde{e}_{[ny]}:=\frac12(\frac{\tilde{p}_{[ny]}^2}{m_x}+ \tilde{r}_{[ny]}^2)$ and define: 
	 \begin{equation} \label{fdef}
		\mathrm{f}_{\beta}^{\mu}(y):=\lim_{n \to \infty} \mathbb{E}\big(\expval{\tilde{e}_{[ny]}}_{\rho}\big).
	 \end{equation}
 
 We prove the existence of this limit in the Appendix \ref{app1}. Moreover, we show that $\mathrm{f}^{\mu}_{\beta}$ is continuous. 

In pursuance of establishing the limit \eqref{e0conv}, it is sufficient to prove a sufficient decay of the following covariance: $\mathbb{E}(\expval{\tilde{e}_x}_{\rho} \expval{\tilde{e}_{x'}}_{\rho})-\mathbb{E}(\expval{\tilde{e}_x}_{\rho})\mathbb{E}(\expval{\tilde{e}_{x'}}_{\rho})$. The rest will be the proof of SLLN for weakly correlated random variables, where we will follow the line of \cite{rlyons}. Precisely, we express this decay in the following lemma: First, for every random variables $X,Y$, define $\Cov(X,Y):= \mathbb{E}(XY)-\mathbb{E}(X)\mathbb{E}(Y)$, 
\begin{lemma} \label{decaylemma}
	There exists $\: 0<c,C, \mathcal{C}<\infty$, independent of $n$, such that for every $n$ we have  $\forall \: x,x' \in \mathbb{I}_n$:
	\begin{equation}	\label{decay} 
		|\Cov(\expval{\tilde{e}_x}_{\rho},\expval{\tilde{e}_y}_{\rho})|= \big|\mathbb{E}\big(\expval{\tilde{e}_x}_{\rho} \expval{\tilde{e}_{x'}}_{\rho}\big)-\mathbb{E}\big(\expval{\tilde{e}_x}_{\rho}\big)\mathbb{E}\big(\expval{\tilde{e}_{x'}}_{\rho}\big)\big| <C\exp(-c|x-x'|) +\frac{\mathcal{C}}{n}.
	\end{equation}
\end{lemma} 
In order to proof \eqref{decaylemma}, first, we  rewrite $\expval{\tilde{e}_x}_{\rho}$ as: $$\expval{\tilde{e}_x}_{\rho}=\frac{1}{2}\Big(\expval{\tilde{r}_x^2}_{\rho}+\frac{\expval{\tilde{p}_x^2}_{\rho}}{m_x}\Big)= \frac{1}{2\beta_x}\Bigg(\expval{x ,\frac{\sqrt{A_r^{\beta}}}{2} \coth(\frac{\sqrt{A_r^{\beta}}}{2}) x}_{n-1}  + \expval{x ,\frac{(A_p^{\beta})^\frac12}{2} \coth(\frac{(A_p^{\beta})^{\frac12}}{2}) x}_n\Bigg),$$ thanks to \eqref{p2thermalaverage}. Then, we use the analyticity of the function $f(y)=\sqrt{y}\coth(\sqrt{y})$, and expand the matrix $\frac{\sqrt{A_p^{\beta}}}{2}\coth(\frac{\sqrt{A_p^{\beta}}}{2})$, in terms of its Taylor series around an appropriate point. By using the fact that $A_p^{\beta}$ is tridiagonal, and mass terms appear locally in this matrix, we observe that first $|x-x'|$ terms in $\mathbb{E}(\expval{e_x}_{\rho}\expval{e_x'}_{\rho})$, can be factorized,  and the rest of the expansion is exponentially small. This proves \eqref{decay}. In the rest of this  section, first, we make this argument rigorous, and then, we prove the SLLN, by using the results form \cite{rlyons}. 

  In the succeeding lemma we observe that $((A_p^{\beta})^k)_{xx}= \langle x , (A_p^{\beta})^k x \rangle$, only depends on the masses in the interval $[x-[\frac{k}{2}],x+[\frac{k}{2}]]\cap \mathbb{N}$.

\begin{lemma} \label{powerofA}
Fix $n$, and recall the definition of the matrix $A_p^{\beta}=M_{\beta}^{-\frac12}(-\nabla_-\beta^0 \nabla_+)M_{\beta}^{-\frac12}$, where $M_{\beta}=M\beta^{-1}$, $M=\diag(m_1,\dots,m_n)$, and $\{ m_x \}$  are i.i.d random variables with smooth (w.r.t Lebesgue) density, $d\mu=\mu(x)dx$. Moreover, recall the notation of the canonical basis of $\mathbb{R}^n$ i.e. $\ket{x}$ for $x \in \mathbb{I}_n$. Take $x \in \mathbb{I}_n$, then for any $k \in \mathbb{I}_n$, denote the $x$th diagonal element of $(A_p^{\beta})^k$ by $\vartheta^k(x):=(A^p_{\beta})^k_{xx}= \langle x, (A_p^{\beta})^k x \rangle$, as a function of masses $m_1 ,\dots m_n$. Then, $\vartheta^k(x)$ only depends on $m_i$, for $i \in I(x,k)$, where $I(x,k)$ is defined as follows:
\begin{equation} \label{dependinterval}
I(x,k)=\big[\min \{x-[\frac{k}{2}] ,1 \},\max \{x+[\frac{k}{2}],n \}\big] \cap \mathbb{N}.
\end{equation}
 In other words, for a fixed realization of the masses, the function $\tilde{\vartheta}(y):=\frac{\partial \vartheta^k(x)}{\partial m_y}$ is supported on $I(x,k)$. 
\end{lemma}
Before proving lemma \ref{powerofA}, we deduce the following corollary: 
\begin{corollary} \label{corollary1}
For $x,y \in \mathbb{I}_n$, and $k+k' <2|x-y|$, we have: 		
\begin{equation} \label{decayofcorrelation0}
\begin{split}
  \mathbb{E}\big(\langle x, (A_p^{\beta})^k x \rangle \langle y, (A_p^{\beta})^{k'} y \rangle\big)=\mathbb{E}\big(\langle x, (A_p^{\beta})^k x \rangle\big)\mathbb{E}\big(\langle y, (A_p^{\beta})^{k'} y \rangle\big).
	\end{split}	
	\end{equation}
\end{corollary}
\begin{proof} 
We deduce this corollary directly from \eqref{dependinterval}, since the assumption $k+k'<2|x-y|$, implies that $I(x,k)\cap I(y,k')= \emptyset$. Therefore, $\langle x, (A_p^{\beta})^{k} x \rangle$, and $\langle y, (A_p^{\beta})^{k'} y \rangle$, are functions of two disjoint set of random variables, and we get the result \eqref{decayofcorrelation0}.   
\end{proof}

	Before proceeding, we state the proof of Lemma \ref{powerofA}:
	\begin{proof} [Proof of Lemma \ref{powerofA}]
		Recall the definition of $A_p^{\beta}=M_{\beta}^{-\frac12}(-\nabla_-\beta^0 \nabla_+)M_{\beta}^{-\frac12}$, and observe that it's a symmetric matrix, which can be expressed in the following explicit way (Here $\beta_{n+1}=0$): 
		\begin{equation} \label{aprep}
			\begin{split}			
			(A_p^{\beta})_{11}=\frac{\beta_1^2}{m_1}, \qquad (A_p^{\beta})_{xx}=\frac{\beta_x(\beta_x+\beta_{x+1})}{m_x}, \quad 1<x\leq n, \\
			(A_p^{\beta})_{x(x+1)}=-\frac{\beta_x \sqrt{\beta_x\beta_{x+1}}}{\sqrt{m_xm_{x+1}}}, \qquad 1 \leq x \leq n-1.  
\end{split}		
		\end{equation}
	Now, consider the expression $\langle x, (A_p^{\beta})^k x \rangle$, we rewrite this expression by multiplying the identity  matrix for $k-1$ times. Denote these $k-1$ matrices by $\sum_{x_j=1}^n \dyad{x_j}$, for $j=1,\dots, k-1$: 
	\begin{equation} \label{rwrep}
		\begin{split}
				\langle x, (A_p^{\beta})^k x \rangle=\sum_{x_1,\dots x_{k-1}=1}^n \langle x,A_p^{\beta} x_1 \rangle \langle x_1 ,A_p^{\beta} x_2 \rangle \dots \langle x_{k-1} ,A_p^{\beta} x \rangle.
		\end{split}
	\end{equation}
	Since $A^{\beta}_p$ is tridiagonal, each term of the form $\langle x_j, A_p^{\beta} x_{j+1} \rangle$ is zero, unless $x_j=x_{j+1}$, or $x_j=x_{j+1}\pm 1$. Hence, each non-zero term in the sum \eqref{rwrep}, is corresponding to a sequence $(x,x_1,\dots,x_{k-1},x)$. This sequence can be interpreted as a discrete random walk path from point $x$, at time zero, to the same point $x$, at time $k$, where at each  time-step, one can choose to go right  i.e., $z \to z+1$, or left, i.e., $z \to z-1$, or stay at the same position, i.e., $z \to z$. However, since the path should return  to point $x$, it is supported on the interval $I(x,k)$. Furthermore, using the explicit form of $A_p^{\beta}$ from \eqref{aprep}, each term $\langle x_j, A_p^{\beta} x_{j'} \rangle$, depends on $\frac{1}{\sqrt{m_jm_{j'}}}$, for $j=j'$, or $j= j'  \pm 1$. Hence, for each path, the corresponding contribution to \eqref{rwrep} only depends on the masses of the points where the path is crossing. Therefore, $\langle x  ,(A_p^{\beta})^k x \rangle$ only depends on the following set of masses: $\{m_i | i \in I(x,k) \}$. 
	\end{proof}
		 Note that the same line of reasoning can be done for $A_r^{\beta}$. Since $A_r^{\beta}$ is symmetric, and can be expressed as follows: 
		\begin{equation} \label{arrep}
	\begin{split}		
		&(A_r^{\beta})_{xx}=\beta_x\Big(\frac{\beta_x}{m_x}+\frac{\beta{x+1}}{m_{x+1}}\Big),
		\\ 
		&(A_r^{\beta})_{xx+1}=\frac{\beta_{x+1}}{m_{x+1}}\sqrt{\beta_{x}\beta_{x+1}}.
		\end{split}		
		\end{equation}
		By using the similar argument as in the previous lemma, we deduce that $\forall x \in \mathbb{I}_{n-1}$, and $k<n-1$, $\langle x , (A_r^{\beta})^k x \rangle$, only depends on the masses $m_i$, for $i \in \tilde{I}(x,k)$, where we have:
		\begin{equation}
			\tilde{I}(x,k)=\big[\min\{1,x-[\frac{k}{2}] \},\max \{ n ,x+[\frac{k}{2}]+1 \}\big].
		\end{equation}
	Similar to the corollary \ref{corollary1}, with the exact same argument we deduce the following:
	\begin{corollary} \label{corollary2}
		For $x,y \in \mathbb{I}_n$, and $k+k'<2|x-y|-2$, we have:
		\begin{equation} \label{decay1}
			\begin{split}		
		\Cov\big(\langle x, (A_r^{\beta})^k x \rangle ,\langle y, (A_r^{\beta})^{k'} y \rangle\big)=0,\\
		\Cov \big(\langle x, (A_p^{\beta})^{k} x \rangle ,\langle y, (A_r^{\beta})^{k'} y \rangle\big)=0.
			\end{split}		
		\end{equation}			
	\end{corollary}

\begin{remark} \label{covfirstterms}
One can observe that in \eqref{decay1}, and \eqref{decayofcorrelation0}, we can substitute $A_p^{\beta}$ and $A_r^{\beta}$, respectively with $A_p^{\beta}-cI_n$ and $A_r^{\beta}-cI_{n-1}$, for some constant $c$. This is straightforward, since adding a constant to the diagonal elements does not change the support of $\vartheta(y)$ and $\tilde{\vartheta}(y)$. Therefore, we can repeat the same argument with these new set of matrices, and obtain for any $x,y \in \mathbb{I}_n$, and $k+k'<|x-y|+1$:
\begin{equation} \label{covrk}
\begin{split}
&\Cov\big(\langle x, (A_p^{\beta}-cI_n)^k x \rangle, \langle y, (A_r^{\beta}-cI_{n-1})^{k'} y \rangle\big)=0, \\
&\Cov\big(\langle x, (A_p^{\beta}-cI_n)^k x \rangle ,\langle y, (A_p^{\beta}-cI_n)^{k'} y \rangle\big)=0,\\
&\Cov\big(\langle x, (A_r^{\beta}-cI_{n-1})^k x \rangle ,\langle y, (A_r^{\beta}-cI_{n-1})^{k'} y \rangle\big)=0.
\end{split}
\end{equation}
\end{remark}	
	
	In order to proof lemma \ref{decaylemma}, define the function $\mathfrak{f}: \mathbb{C}  \to \mathbb{C}$ as follows: 
	\begin{equation} \label{xcothx}		
			\mathfrak{f}(z)=
			\begin{cases}
				z^{\frac12}\coth(z^{\frac12}), \quad z \neq  0, \\
				1, \quad z=0.
			\end{cases}
	\end{equation}

One can easily observe that the poles of the function $z^{\frac12}\coth(z^{\frac12})$ is the following set: $\{z \in \mathbb{C} | z= -k^2\pi^2, k \in \mathbb{Z} \}$, and this function is analytic on the rest of the complex plane. However, the point zero is a removable pole, and  by redefining the function at zero, we can remove this pole: It is well known that the function $\coth(z)$ has the following Taylor series expression for $0<|z|<\pi$: $\coth(z)=z^{-1}+\sum_{n=1}^{\infty}a_n z^{2n-1}$, where $a_n=\frac{2^{2n}B_{2n}}{(2n)!}$, and $B_{2n}$ 
are Bernoulli numbers. Hence, we have $z\coth(z)=1+\sum_{n=1}^{\infty} a_nz^{2n}$, and   $z^{\frac12}\coth(z^{\frac12})$ is given by the following Taylor series: $1+\sum_{n=1}^{\infty}a_nz^n$ for $0<|z|<\pi^2$.  Hence, the pole of $\mathfrak{f}$ is given by the set $\{z \in \mathbb{C} | z= -k^2\pi^2, k \in \mathbb{N}, k>0 \}.$ 

	Finally, we can state the proof of Lemma \ref{decaylemma}: 
	\begin{proof} [Proof of Lemma \ref{decaylemma}]
		First, from Section \ref{ensemble average}, recall that there is a constant $c_0>0$, uniform in $n$\footnote{This constant can be taken equal to $4\frac{\beta_{max}^2}{m_{min}}$.}, such that for any configuration of the masses, we have $||A_p^{\beta}||_2 ,||A_r^{\beta}||_2 \leq c_0$. Define $\alpha:=\frac12(c_0+1)$, let $\mathcal{R}:=\alpha+\pi^2$, by the above argument $\mathfrak{f}(z)$ is analytic in the open disk $|z-\alpha|<\mathcal{R}$, and $\mathcal{R}$ is the radius of convergence for the Taylor expansion of $\mathfrak{f}$,  $\mathfrak{f}(z)=\sum_{k=0}^{\infty}a_k(z-\alpha)^k$. Moreover, by the choice of $\alpha$ and $c_0$, one can easily observe that all the eigenvalues of $A_p^{\beta}$ and $A_r^{\beta}$ lies in the disk $|z-\alpha|< \mathcal{R}$. Explicitly, $\forall k \in \mathbb{I}_{n-1}$, we have $|\gamma_k^2-\alpha|<\mathcal{R}$. Hence, we can write the following Taylor expansions for $\mathfrak{f}(A_p^{\beta})$ and $\mathfrak{f}(A_r^{\beta})$ (For the proof of this fact one can see Theorem 4.7 of \cite{Higham}). 
		\begin{equation}\label{Taylorexpansionmatrix}
			\mathfrak{f}(A_p^{\beta})=\sum_{k=0}^{\infty} a_k (A_p^{\beta}-\alpha I_n)^k, \qquad \mathfrak{f}(A_r^{\beta})=\sum_{k=0}^{\infty} a_k (A_r^{\beta}-\alpha I_{n-1})^k. 			
		\end{equation}
	Comparing the definition of $\mathfrak{f}$ in \eqref{xcothx}, where $\mathfrak{f}(0)=1$, with the expression  \eqref{r2p2thermalaverage}, where we had $0\coth(0)=0$ by convention, we deduce the following expression for $\expval{\tilde{e}_x}_{\rho}$:
	\begin{equation} \label{averageintermoff}
		\begin{split}		
		\expval{\tilde{e}_x}_{\rho} &= \frac{1}{\beta_x}\big(\langle x,\mathfrak{f}(A_r^{\beta}) x \rangle_{n-1} + \langle x, \mathfrak{f}(A_p^{\beta}) x \rangle_n - (\psi^0_x)^2\big) \\ &= \frac{1}{\beta_x}\big(\langle x,\mathfrak{f}(A_r^{\beta}) x \rangle_{n-1} + \langle x, \mathfrak{f}(A_p^{\beta}) x \rangle_n - \frac{\frac{m_x}{\beta_x}}{\sum_{x=1}^{n}\frac{m_x}{\beta_x}}\big),
		\end{split}
	\end{equation}	 
	where we used the equality $\psi^0=(\sum_{x=1}^{n} \frac{m_x}{\beta_x})^{-\frac12}M_{\beta}^{\frac12}\ket{\textbf{1}}$. We denote $\frac{\frac{m_x}{\beta_x}}{\sum_{x=1}^{n}\frac{m_x}{\beta_x}}$ by $\epsilon^x_n$. Notice that $|\epsilon_n^x|$ is bounded by $\frac{\mathcal{C}_0}{n}$, where $\mathcal{C}_0$ is independent from $n$ \footnote{Precisely, one can choose $\mathcal{C}_0 $ to be equal to $\frac{m_{max}\beta_{max}}{m_{min}\beta_{min}}$.}, thanks to the assumptions on the distribution of the masses, and temperature profile $\beta(y)$. Moreover, since  $||A_p^{\beta}||_2$ and $||A_r^{\beta}||_2$ are bounded by $c_0$, uniformly in $n$ for any realization of the masses. By using the fact that $\mathfrak{f}$ is continuous and increasing in the interval $[0,c_0]$, $||\mathfrak{f}(A_r^{\beta})||_2$ and $||\mathfrak{f}(A_p^{\beta})||_2$ are bounded by a constant $c_1$ \footnote{One can choose $c_1=\mathfrak{f}(c_0)$.}, independent of $n$, for any realization of the masses. Therefore, $|\langle x,\mathfrak{f}(A_r^{\beta}) x \rangle_{n-1} + \langle x, \mathfrak{f}(A_p^{\beta}) x \rangle_n|$ is bounded by $2c_1$. 
	Taking advantage of the aforementioned bounds, we can deduce the following inequality:
	\begin{equation}  \label{killepsilon}
		\begin{split}
			&\big|\Cov(\expval{\tilde{e}_x}_{\rho} ,\expval{\tilde{e}_{x'}}_{\rho})\big| = \\ &\frac{1}{\beta_x\beta_{x'}}\big|\Cov\big((\langle x,\mathfrak{f}(A_r^{\beta}) x \rangle_{n-1} + \langle x, \mathfrak{f}(A_p^{\beta}) x \rangle_n -\epsilon_x^n),(\langle x',\mathfrak{f}(A_r^{\beta}) x' \rangle_{n-1} \langle x', \mathfrak{f}(A_p^{\beta}) x' \rangle_n -\epsilon_{x'}^n)\big)\big| \leq \\ &  \frac{1}{\beta_{min}^2} \big[\big|\Cov\big((\langle x,\mathfrak{f}(A_r^{\beta}) x \rangle_{n-1} + \langle x, \mathfrak{f}(A_p^{\beta}) x \rangle_n ),(\langle x',\mathfrak{f}(A_r^{\beta}) x' \rangle_{n-1} +\langle x', \mathfrak{f}(A_p^{\beta}) x' \rangle_n)\big)\big| + 
			\\ & \big|\Cov\big(\epsilon_x^n,(\langle x',\mathfrak{f}(A_r^{\beta}) x' \rangle_{n-1} +\langle x', \mathfrak{f}(A_p^{\beta}) x' \rangle_n\big)\big|+ \big|\Cov\big(\epsilon_{x'}^n,(\langle x,\mathfrak{f}(A_r^{\beta}) x \rangle_{n-1} +\langle x, \mathfrak{f}(A_p^{\beta}) x \rangle_n\big)\big|+\big|\epsilon^n_{x,x'}\big|\big],
		\end{split}
	\end{equation}
	where we denote $\epsilon^n_{x,x'}:=\Cov(\epsilon_x^n,\epsilon_{x'}^n)$. Since $|\epsilon_x^n| ,|\epsilon_{x'}^n| \leq \frac{\mathcal{C}_0}{n}$, and $|\langle x,\mathfrak{f}(A_r^{\beta}) x \rangle_{n-1} + \langle x, \mathfrak{f}(A_p^{\beta}) x \rangle_n|$, $|\langle x,\mathfrak{f}(A_r^{\beta}) x' \rangle_{n-1} + \langle x', \mathfrak{f}(A_p^{\beta}) x' \rangle_n| \leq 2c_1$, $|\epsilon^n_{x,x'}|$ is bounded by $\frac{1}{\beta_{min}^2}8c_1\frac{\mathcal{C}_0}{n}+o(\frac{1}{n^2})$. Therefore, there exists $\mathcal{C}$ independent of $n$, such that 
	\begin{equation} \label{killedepsilon}
		\begin{split}	
		&\big|\Cov(\expval{\tilde{e}_x}_{\rho} ,\expval{\tilde{e}_{x'}}_{\rho})\big| \leq \\
		& \frac{1}{\beta_{min}^2}\big|\Cov\big((\langle x,\mathfrak{f}(A_r^{\beta}) x \rangle_{n-1} + \langle x, \mathfrak{f}(A_p^{\beta}) x \rangle_n ),(\langle x',\mathfrak{f}(A_r^{\beta}) x' \rangle_{n-1} +\langle x', \mathfrak{f}(A_p^{\beta}) x' \rangle_n)\big)\big| + \frac{\mathcal{C}}{n}.
		\end{split}	
	\end{equation}
	 
	First term can be written as the sum of the following terms: 
	\begin{equation} \label{4terms}
	\begin{split}
				&\big|\Cov\big((\langle x,\mathfrak{f}(A_r^{\beta}) x \rangle_{n-1} + \langle x, \mathfrak{f}(A_p^{\beta}) x \rangle_n ),(\langle x',\mathfrak{f}(A_r^{\beta}) x' \rangle_{n-1} +\langle x', \mathfrak{f}(A_p^{\beta}) x' \rangle_n)\big)\big|	= \\&
				\big|\Cov\big((\langle x,\mathfrak{f}(A_r^{\beta}) x \rangle_{n-1}),(\langle x',\mathfrak{f}(A_r^{\beta}) x' \rangle_{n-1})\big)+\Cov\big((\langle x,\mathfrak{f}(A_r^{\beta}) x \rangle_{n-1}),(\langle x',\mathfrak{f}(A_p^{\beta}) x' \rangle_{n})\big)+
				\\
				&\Cov\big((\langle x,\mathfrak{f}(A_p^{\beta}) x \rangle_{n}),(\langle x',\mathfrak{f}(A_r^{\beta}) x' \rangle_{n-1})\big)+\Cov\big((\langle x,\mathfrak{f}(A_p^{\beta}) x \rangle_{n}),(\langle x',\mathfrak{f}(A_p^{\beta}) x' \rangle_{n})\big)\big|	 \leq \\ &
					\big|\Cov\big((\langle x,\mathfrak{f}(A_r^{\beta}) x \rangle_{n-1}),(\langle x',\mathfrak{f}(A_r^{\beta}) x' \rangle_{n-1})\big)\big|+\big|\Cov\big((\langle x,\mathfrak{f}(A_r^{\beta}) x \rangle_{n-1}),(\langle x',\mathfrak{f}(A_p^{\beta}) x' \rangle_{n})\big)\big|+
				\\
				&\big|\Cov\big((\langle x,\mathfrak{f}(A_p^{\beta}) x \rangle_{n}),(\langle x',\mathfrak{f}(A_r^{\beta}) x' \rangle_{n-1})\big)\big|+\big|\Cov\big((\langle x,\mathfrak{f}(A_p^{\beta}) x \rangle_{n}),(\langle x',\mathfrak{f}(A_p^{\beta}) x' \rangle_{n})\big)\big|.
			\end{split}
\end{equation}		
	
	In order to complete the proof, we observe that each of the terms in \eqref{4terms} are exponentially small. We show this fact for one of these terms, and the rest can be treated exactly in the same way. We do this task, using the Taylor series \eqref{Taylorexpansionmatrix}. We divide the series into two parts: The first $|x-x'|$ terms, and the rest, which is exponentially small. Let us define $\mathfrak{f}_{\prec}$ and $\mathfrak{f}_{\succ}$, as follows \footnote{Note that these definition depends on the $n,x,x'$.}:
	\begin{equation}\label{taylorcut}
		\begin{split}		
		&\mathfrak{f}_{\prec}(A_r^{\beta}):= \sum_{k=0}^{|x-x'|-1} a_k(A_r^{\beta}-\alpha I_{n-1})^k, \qquad \mathfrak{f}_{\prec}(A_p^{\beta}):= \sum_{k=0}^{|x-x'|-1} a_k(A_p^{\beta}-\alpha I_{n})^k, \\
		&\mathfrak{f}_{\succ}(A_r^{\beta}):= \sum_{k>|x-x'|-1} a_k(A_r^{\beta}-\alpha I_{n-1})^k, \qquad \mathfrak{f}_{\succ}(A_p^{\beta}):= \sum_{k>|x-x'|-1} a_k(A_p^{\beta}-\alpha I_{n})^k. 
		\end{split}
	\end{equation}

Notice that $\mathfrak{f}(A_r^{\beta})=\mathfrak{f}_{\prec}(A_r^{\beta})+\mathfrak{f}_{\succ}(A_r^{\beta})$, and $\mathfrak{f}(A_p^{\beta})=\mathfrak{f}_{\prec}(A_p^{\beta})+\mathfrak{f}_{\succ}(A_p^{\beta})$, we substitute $\mathfrak{f}(A_r^{\beta})$  in the first term of \eqref{4terms} with this expression:
\begin{equation} \label{fdivided}
\begin{split}
&\big|\Cov\big((\langle x,\mathfrak{f}(A_r^{\beta}) x \rangle_{n-1}),(\langle x',\mathfrak{f}(A_r^{\beta}) x' \rangle_{n-1})\big)\big|= \\&\big|\Cov\big((\langle x,(\mathfrak{f}_{\prec}+\mathfrak{f}_{\succ})(A_r^{\beta}) x \rangle_{n-1}),(\langle x',(\mathfrak{f}_{\prec}+\mathfrak{f}_{\succ})(A_r^{\beta}) x' \rangle_{n-1})\big)\big|= \\
&\big|\Cov\big((\langle x,\mathfrak{f}_{\prec}(A_r^{\beta}) x \rangle_{n-1}),(\langle x',\mathfrak{f}_{\prec}(A_r^{\beta}) x'\rangle_{n-1})\big)+\Cov\big((\langle x,\mathfrak{f}_{\succ}(A_r^{\beta}) x \rangle_{n-1}),(\langle x',\mathfrak{f}_{\prec}(A_r^{\beta}) x'\rangle_{n-1})\big)+ \\ &\Cov\big((\langle x,\mathfrak{f}_{\prec}(A_r^{\beta}) x \rangle_{n-1}),(\langle x',\mathfrak{f}_{\succ}(A_r^{\beta}) x'\rangle_{n-1})\big)+ \Cov\big((\langle x,\mathfrak{f}_{\prec}(A_r^{\beta}) x \rangle_{n-1}),(\langle x',\mathfrak{f}_{\prec}(A_r^{\beta}) x'\rangle_{n-1})\big) \big|.
\end{split}
\end{equation}  
The first term in \eqref{fdivided}	is equal to zero, thanks to the third equality in \eqref{covrk} in the remark \ref{covfirstterms}: 
\begin{equation} \label{covfterms}
\begin{split}
&\Cov\big((\langle x,\mathfrak{f}_{\prec}(A_r^{\beta}) x \rangle_{n-1}),(\langle x',\mathfrak{f}_{\prec}(A_r^{\beta}) x'\rangle_{n-1})\big) = \\ &\sum_{k,k'<|x-x'|-1} a_ka_{k'} \Cov\big((\langle x, (A_r^{\beta}-\alpha I_{n-1})^k x \rangle) ,(\langle x', (A_r^{\beta}-\alpha I_{n-1})^{k'} x' \rangle)\big)=0,
\end{split}
\end{equation} 		
	where each term in the sum is equal to zero, thanks to the remark \ref{covfirstterms}.
	
	We take care of the remaining terms, by using the properties of Taylor series. First, observe that by the choice of  $\alpha$, we have $||A_r^{\beta}-\alpha I_{n-1}||_2 \leq \alpha$. Moreover, $\alpha +1 < \mathcal{R}$ so the series $\sum_{k=0}^{\infty} a_k (\alpha+1)^k$ is convergent, and there exist a constant $M>0$ (independent of $x,x'$ and $n$), such that $\forall k$, $|a_k(\alpha+1)^k| \leq M$, and by denoting $\varrho=\frac{\alpha}{\alpha+1}$, we get: 
	\begin{equation} \label{taylorremainder}
		\begin{split}
			&||\mathfrak{f}_{\succ}(A_r^{\beta})||_2 = ||\sum_{k>|x-x'|-1} a_k(A_r^{\beta}-\alpha I_{n-1})^k||_2		\leq \sum_{k>|x-x'|-1}|a_k|||A_r^{\beta}-\alpha I_{n-1}||_2^k \leq \sum_{k>|x-x'|-1}|a_k| \alpha^k  \\ & =  \sum_{k>|x-x'|-1} |a_k| (\alpha+1)^k \varrho^k \leq M  \sum_{k>|x-x'|-1} \varrho^k \leq \mathcal{C}_1 \varrho^{(|x-x'|)}.
					\end{split}
	\end{equation}	  
	Considering the fact that $||\mathfrak{f}(A_r^{\beta}))||_2 \leq c_1$, and $\mathfrak{f}(A_r^{\beta})=\mathfrak{f}_{\prec}(A_r^{\beta})+\mathfrak{f}_{\succ}(A_r^{\beta})$, as a direct consequence of \eqref{taylorremainder}, we have: $||\mathfrak{f}_{\prec}(A_r^{\beta})||_2 \leq c_2$, where $c_2$ is a constant uniform in $n$. Therefore, we have:
	\begin{equation} \label{remainderbound}
	\begin{split}		
		&\big|\Cov\big((\langle x,\mathfrak{f}_{\succ}(A_r^{\beta}) x \rangle_{n-1}),(\langle x',\mathfrak{f}_{\prec}(A_r^{\beta}) x'\rangle_{n-1}))  + \Cov((\langle x,\mathfrak{f}_{\prec}(A_r^{\beta}) x \rangle_{n-1}),(\langle x',\mathfrak{f}_{\succ}(A_r^{\beta}) x'\rangle_{n-1})\big)\big| \leq \\ &
		4||\mathfrak{f}_{\succ}(A_r^{\beta}) ||_2||\mathfrak{f}_{\prec}(A_r^{\beta})||_2 \leq 4c_2\mathcal{C}_1 \varrho^{(|x-x'|)}, \\
		&\big|\Cov\big((\langle x,\mathfrak{f}_{\succ}(A_r^{\beta}) x \rangle_{n-1}),(\langle x',\mathfrak{f}_{\succ}(A_r^{\beta}) x'\rangle_{n-1})\big)\big| \leq 2||\mathfrak{f}_{\succ}(A_r^{\beta}) ||_2||\mathfrak{f}_{\succ}(A_r^{\beta})||_2 \leq 2\mathcal{C}_1^2 \varrho^{2|x-x'|}.
 	\end{split}	
	\end{equation}
	Notice that all the bounds here are deterministic and independent of the realization of the masses.  
	Combining \eqref{fdivided},  \eqref{covfterms}  and \eqref{remainderbound}, there exists a deterministic constant independent of $n$ and realization of the masses, $\mathcal{C}_2$, such that we have:
	\begin{equation} \label{bound1}
		\big|\Cov\big((\langle x,\mathfrak{f}(A_r^{\beta}) x \rangle_{n-1}),(\langle x',\mathfrak{f}(A_r^{\beta}) x' \rangle_{n-1})\big)\big| \leq \mathcal{C}_2 \varrho^{(x-x')}.
	\end{equation}
		Recall \eqref{4terms}-the other terms can be treated exactly similar to \eqref{bound1}- $\mathfrak{f}$ can be divided as in \eqref{taylorcut}. Then, we obtain the same expression as in \eqref{fdivided}, where the first term is equal to zero, thanks to the second and third equality in \eqref{covrk}, and the remainder can be bounded with the exact same bound. Hence, there exist a constant $C>0$ uniform in $n$, such that:
		\begin{equation} \label{bound2}
			\frac{1}{\beta_{min}^2} \big|\Cov\big((\langle x,\mathfrak{f}(A_r^{\beta}) x \rangle_{n-1} + \langle x, \mathfrak{f}(A_p^{\beta}) x \rangle_n ),(\langle x',\mathfrak{f}(A_r^{\beta}) x' \rangle_{n-1} +\langle x', \mathfrak{f}(A_p^{\beta}) x' \rangle_n)\big)\big| \leq C \varrho^{(|x-x'|)}.
		\end{equation}
	Comparing \eqref{bound2} and \eqref{killedepsilon}, and recalling the fact that $0<\varrho =\frac{\alpha}{\alpha+1}<1$, where $\alpha$ is independent of $n$, we get the constants $0<c,\mathcal{C},C<\infty$, independent of $n$, such that:
	\begin{equation}
	|\Cov(\expval{\tilde{e}_x}_{\rho},\expval{\tilde{e}_{x'}}_{\rho})| \leq C\exp(-c|x-x'|) + \frac{\mathcal{C}}{n}.
	\end{equation}
	
	\end{proof}
	
		Thanks to the the  exponential decay of covariances \eqref{decay}, we have the SLLN \eqref{e0conv}. For proving the SLLN from this decay, we follow the lines of \cite{rlyons}: 
		\begin{theorem} \label{SLLN}
			Recall the definition of density state $\rho^n$ from \eqref{initalstate}, and the definition of the ensemble average with respect to this state by $\expval{.}_{\rho^n}$ from \eqref{mixedaverage}. Let $g \in C^0([0,1])$, be a test function. Then we have the following convergence, almost surely with respect to the distribution of the masses.
			\begin{equation} \label{e0conv1}
			 \lim_{n \to \infty} \frac{1}{n} \sum_{x=1}^{n}g(\frac{x}{n})\expval{\tilde{e}_x}_{\rho^n} \to \int_0^1 g(y)\mathrm{f}_{\mu}^{\beta}(y)dy, 
			\end{equation}
	where $\mathrm{f}_{\mu}^{\beta}(y)$ is
 defined in \eqref{fdef}. 		
		\end{theorem}      
	\begin{proof}
		Define the random variable $Y^n_x:=g(\frac{x}{n})\big(\expval{\tilde{e}_x}_{\rho^n}-\mathbb{E}(\expval{\tilde{e}_x}_{\rho^n}) \big)$. First, notice that $\mathbb{E}(Y^n_x)=0$, and by the definition of $g$, and the fact that $\expval{e_x}_{\rho^n}$ is bounded (we established this fact in Lemma \ref{ensemble average}), $|Y^n_x|$ is uniformly bounded by a constant $C_0$. Moreover, thanks to the Lemma \ref{decaylemma}, and inequality \eqref{decay}, and the fact that $g$ is bounded we have: 
		\begin{equation} \label{boundY}
		\mathbb{E}(Y^n_xY^n_y)| \leq C_1 \exp(-c|x-y|)+\frac{\mathcal{C}}{n},
		\end{equation}
		for some constant $C_1,c, \mathcal{C}$, uniform in $n$.   Let $S_n=\frac{1}{n}\sum_{x=1}^n Y_x^n$, by using \eqref{boundY}, we have $\mathbb{E}(S_n^2) \leq \frac{\mathcal{C}_2}{n}$. Hence, $\forall \epsilon >0$, $\sum_{n=1}^{\infty} \frac{1}{n} \frac{\mathbb{E}(S_n^2)}{\epsilon^2} < \infty$. On the other hand, by Cauchy condensation lemma, we know if $\sum_{n=1}^{\infty} \frac{b_n}{n} < \infty$, with $b_n \geq 0$, then there exists a sequence $n_k$ of integers, such that $\sum_{k=1}^{\infty} b_{n_k} < \infty$, and $\lim_{k \to \infty} \frac{n_{k+1}}{n_k}=1$, (For the proof of this fact one can see Lemma 3 in \cite{d49}). Therefore, there exists a subsequence $S_{n_k}$,  such that $\lim_{k \to \infty}\frac{n_{k+1}}{n_k} \to 1$, and $\forall \epsilon > 0 $, $\sum_{k=1}^{\infty} \frac{\mathbb{E}(S_{n_k}^2)}{\epsilon^2} < \infty$.   
	Hence, by Borel-Contelli lemma, since $\forall \epsilon >0$, $\sum_{k=1}^{\infty} \mathbb{P}(|S_{n_k}|>\epsilon) \leq \sum_{k=1}^{\infty} \frac{\mathbb{E}(S_{n_k}^2)}{\epsilon^2} $, we have $\lim_{k \to \infty}S_{n_k} \to 0$, almost surely. Now take $n$ such that $n_k \leq n < n_{k+1}$; then, by using the fact that $\forall n,$ $\forall x \in \mathbb{I}_n$, $|Y_x|\leq C_0$, we have: 
	\begin{equation} \label{edit2}
	|S_n-S_{n_k}| \leq \frac{1}{n_k}\sum_{x=1}^{n_k}|Y^{n}_x-Y^{n_k}_x| + \frac{1}{n_k}\sum_{x>n_k}^{n}|Y_x^n| \leq \frac{1}{n_k}\sum_{x=1}^{n_k}|Y^{n}_x-Y^{n_k}_x| + C_0\frac{n-n_k}{n_k}.
	\end{equation}
	Since $\frac{n_{k+1}}{n_k} \to 1$, for any $\varepsilon > 0$, there exits $N_*$, such that if $n_k>N_*$, then for the second term in \eqref{edit2}, we have: $C_0\frac{n-n_k}{n_k}< \frac{\varepsilon}{4}$. Moreover, by the Lemma \ref{app2} in Appendix \ref{app1}, for $n_k$ sufficiently large, we have $\forall x \in \mathbb{I}_{n_k}$, with $n_*<x<n-n_*$: $|Y^n_x-Y^{n_k}_x| \leq \frac{\varepsilon}{2}$, where $n_*$ is independent of $n_k$. Moreover, the terms corresponding to $1\leq x \leq n_*$ can be bounded by $\frac{\varepsilon}{4}$, since $n_*$ does not depend on $n_k$, and $Y_x^n$ is uniformly bounded. Hence, the first term in  \eqref{edit2} is bounded by $\frac{3\varepsilon}{4}$, for $n_k$ sufficiently large.  Therefore, for every $\varepsilon>0$, there exist $N$, such that for $n_k>N$, $|S_n-S_{n_k}|<\varepsilon$ \footnote{This bound is true for every configuration of masses.}. Since $\lim_{k \to \infty}S_{n_k} \to 0$, almost surely, we deduce that $S_n \to 0$ almost surely.
	Hence, \begin{equation} \label{convtemp}
	\frac{1}{n}\sum_{x=1}^n g(\frac{x}{n})\Big(\expval{\tilde{e}_x}_{\rho^n}-\mathbb{E}(\expval{\tilde{e}_x}_{\rho^n})\Big) \to 0,
	\end{equation}  almost surely. 
This gives us the result \eqref{e0conv1}, thanks to the definition of  $\mathrm{f}_{\mu}^{\beta}$ in \eqref{fdef},  Corollary \ref{fdeflimit}, and dominated convergence theorem.	
	 
\end{proof}	
\subsection{Thermal Equilibrium}
	Recall the definition \eqref{initalstate} of density operator $\rho^{n}_{\beta,\bar{p},\bar{r}}$ corresponding to our locally Gibbs state. We denote the density operator corresponding to thermal equilibrium at inverse temperature $\beta_{eq} \in (0,\infty)$, by $\rho^{n,\beta_{eq}}_{\bar{p},\bar{r}}$. Recall that by thermal equilibrium, we refer to the case where the temperature profile $\beta(.)$ is constant, i.e., $\forall y \in [0,1], \: \beta(y)=\beta_{eq}$. In this case, the matrices $A_p^{\beta}$, and $A_r^{\beta}$ have the following form: 
	$$A_p^{\beta_{eq}}=(\beta_{eq})^2 M^{-\frac12}(-\Delta)M^{-\frac12}=\beta_{eq}^2A_p^0, \quad \quad A_r^{\beta_{eq}}=(\beta_{eq})^2\nabla_+M^{-1}\nabla_-=\beta_{eq}^2A_r^0.  $$ 
	Therefore, in thermal equilibrium, the average of the fluctuation part of the kinetic and potential energy ($\expval{\tilde{r}_x^2}_{\rho}$ and  $\frac{\expval{\tilde{p}_x^2}_{\rho}}{m_x}$) is given as follows: 
	\begin{equation} \label{r2p2thermalaveragetheq}
		\begin{aligned}		
		\expval{\tilde{r}_x^2}_{\rho}= \expval{x ,\frac{(A_r^{0})^\frac12}{2} \coth(\frac{\beta_{eq}(A_r^0)^\frac12}{2}) x}_{n-1}, \quad  \frac{\expval{\tilde{p}_x^2}_{\rho}}{m_x}= \expval{x ,\frac{(A_p^{0})^\frac12}{2} \coth(\frac{\beta_{eq}(A_p^{0})^{\frac12}}{2}) x}_n.
		\end{aligned}	
	\end{equation}
 In the finite system, one can see from these expressions (by analyzing their Taylor expansions) that $\mathbb{E}(\expval{\tilde{e}_{[ny]}}) \neq \mathbb{E}(\expval{\tilde{e}_{[ny']}}) $, for $y,y' \in (0,1)$. However, if we take the limit as $n \to \infty$, thanks to Corollary \ref{feqdef} in Appendix \ref{app1}, we recover the space homogeneity  in the bulk, i.e., denoting the $\lim_{n \to \infty} \mathbb{E}(\expval{\tilde{e}_{[ny]}}_{\rho^{n,\beta_{eq}}})$ in thermal equilibrium by $\mathrm{f}^{\mu}_{\beta_{eq}}$ we have:
	$$\forall y,y' \in (0,1), \quad \mathrm{f}^{\mu}_{\beta_{eq}}(y)=\mathrm{f}^{\mu}_{\beta_{eq}}(y').  $$
Take $y \in (0,1)$, since $\mathrm{f}^{\mu}_{\beta_{eq}}(y)$ is independent of $y$, we define the function $\mathrm{f}^{\mu}(\beta_{eq})$ to be the thermal equilibrium average at inverse temperature $\beta_{eq} \in (0, \infty)$:
\begin{equation} \label{thermaleqfunc}
	\mathrm{f}^{\mu}(\beta_{eq}):= \mathrm{f}^{\mu}_{\beta_{eq}}(y) = \lim_{n \to \infty} \mathbb{E}\big(\expval{\tilde{e}_{[ny]}}_{\rho^{n,\beta_{eq}}}\big). 
\end{equation} 	
		 Thanks to Proposition \ref{spro}, out of thermal equilibrium with a proper $\beta \in C^0([0,1])$ satisfying the assumption of definition \eqref{initalstate}, we can express the function $\mathrm{f}^{\mu}_{\beta(.)}$ \eqref{fdef}, in terms of $\mathrm{f}^{\mu}(\beta)$ as follows: 
		\begin{equation} \label{stefano}
		\forall y \in (0,1), \quad \mathrm{f}^{\mu}_{\beta(.)}(y)=\mathrm{f}^{\mu}(\beta(y)),
\end{equation}				
		where in the second expression, the equilibrium average is computed at inverse temperature $\beta_{eq}=\beta(y)$. \\
	
		\section{Energy Evolution} \label{EE}

		In this section, we finish the proof of Theorem \ref{maintheorem} by proving \eqref{econv}. The idea is as follows: We can decompose the energy into the  mechanical and thermal parts in both microscopic ($\expval{e_x(nt)}_{\rho}$) and macroscopic ($\mathrm{e}(y,t)$) scale. The contribution of the mechanical part in the left hand side (LHS) of \eqref{econv} converges to the mechanical part in the right hand side, thanks to \eqref{masshomo2} and \eqref{pointwiseconvergence}. The contribution of the thermal energy in the LHS of \eqref{econv} converges to the thermal part of the RHS, thanks to  \eqref{e0conv}, at time zero. Finally, the thermal energy in the LHS of \eqref{econv} remains constant in the limit as $n \to \infty$, thanks to the localization phenomena, similar to the RHS of \eqref{econv}, where the contribution of the thermal part is given by a function constant in time. This constant function can be obtained by solving \eqref{pde1}, and finding the explicit solution for $\mathrm{e}(y,t)$. We make this heuristic rigorous in this section.\\ 
		\subsection{Mechanical Energy}
		
		  Before proceeding, we state the following lemma, in order to deal with the mechanical part. 
		\begin{lemma} \label{kinticlemma}
			For any test function $g \in C^0([0,1])$, by recalling the notation $\bar{p}_x(nt)=\expval{p_x(nt)}_{\rho^n}$, and $\bar{r}_x(nt)=\expval{r_x(nt)}_{\rho^n}$, we have: 
			\begin{equation} \label{kinetic1}
				\lim_{n \to \infty} \frac{1}{n} \sum_{x=1}^{n} g(\frac{x}{n})\Big(\frac{\expval{p_x(nt)}_{\rho^n}^2}{2m_x}+\frac{\expval{r_x(nt)}^2_{\rho^n}}{2}\Big) \to \int_0^1 g(y) \Big(\frac{\mathrm{p}(y,t)^2}{2\bar{m}} +\frac{\mathrm{r(y,t)}^2}{2}\Big)dy,
			\end{equation}
almost surely, w.r.t the distribution of the masses.		
		\end{lemma}
		\begin{proof}
		 	Denote the solution of the evolution equation \eqref{averagetimeevolution}, with initial datum $\bar{p}_x(0)=\bar{p}_x\frac{m_x}{\bar{m}}$, and $\bar{r}_x(0)=\bar{r}_x$ with $\bar{\pi}_x(nt)$ and $\bar{\varrho}
		 	_x(nt)$.
		 	Moreover, we denote the solution of \eqref{averagetimeevolution} with initial datum $\pi_x(0)=-\mathscr{E}_n^x$, and $\varrho_x(0)=0$ by $\pi_x(nt)$, 
		 	and $\varrho_x(nt)$, respectively. Thanks to linearity of the evolution equation we have for any $x$: 
			\begin{equation} \label{ME00}
			\expval{p_x(nt)}_{\rho}= \bar{\pi}_x(nt)+\pi_x(nt),\qquad 
		 	\expval{r_x(nt)}_{\rho}=\bar{\varrho}_x(nt)+ \varrho_x(nt).
\end{equation}

		 	 First,  since  $\bar{\varrho}_{[ny]}(nt) \to \mathrm{r}(y,t)$, a.s. by \eqref{pointwiseconvergence}, we have $(\bar{\varrho}_{[ny]}(nt))^2 \to \mathrm{r}^2(y,t)$. By using the fact that $g(\frac{[ny]}{n}) \to g(y)$, we get: 
		 	 \begin{equation} \label{ME001}
		 	 \frac{1}{2n} \sum_{x=1}^n g(\frac{x}{n})(\bar{\varrho}_{x}(nt))^2 = \frac12 \int_0^1 g(\frac{[ny]}{n}) (\bar{r}_{[ny]}(nt))^2 dy \to \int_0^1 g(y)\mathrm{r}^2(y,t)dy, 
		 	 \end{equation}
		 	 almost surely, by dominated convergence theorem. The fact that $\bar{\varrho}_{[ny]}(nt)^2$ is bounded and hence integrable is obvious from the conservation of the energy bounds in \eqref{boundswaveeq1}. \\ The momentum part can be treated by \eqref{masshomo2} as follows: 
		\begin{equation} \label{ee1}
		\begin{aligned}
			&\frac{1}{2n}\sum_{x=1}^n g(\frac{x}{n})\frac{(\bar{\pi}_x(nt))^2}{m_x} = \frac{1}{2n}\sum_{x=1}^n g(\frac{x}{n})\frac{(\bar{\pi}_x(nt))^2}{m_x^2}m_x= \frac{1}{2n}\sum_{x=1}^{n} g(\frac{x}{n}) (\frac{\bar{\pi}_x(nt)}{m_x})^2\bar{m} + \\
&\frac{1}{2n} \sum_{x=1}^{n} g(\frac{x}{n}) \Big(\frac{\bar{\pi}_x(nt)}{m_x}\Big)^2(m_x -\bar{m})= \bar{m}\int_0^1 g(\frac{[ny]}{n}) \Big(\frac{\bar{\pi}_{[ny]}}{m_{[ny]}}\Big)^2 dy + \frac{1}{2n} \sum_{x=1}^{n} g(\frac{x}{n}) \Big(\frac{\bar{\pi}_x(nt)}{m_x}\Big)^2(m_x -\bar{m})\\
&\to \int_0^1 g(y) \frac{\mathrm{p}(y,t)^2}{2\bar{m}},	
		\end{aligned}
		\end{equation} 
almost surely as $n \to \infty$, where the last  sum converges to $0$, almost surely, thanks to \eqref{masshomo2}. The last  integral in the second line  converges to $\int_0^1 g(y) \frac{\mathrm{p}(y,t)^2}{2\bar{m}}$, using the convergence $\frac{\bar{\pi}_{[ny]}}{m_{[ny]}} \to \frac{\mathrm{p}(y,t)}{\bar{m}}$ a.s, in \eqref{pointwiseconvergence}, and dominated convergence theorem.  	\\
On the other hand, for $\pi_x(nt)$ we have: 
\begin{equation} \label{ME002}
\Big|\frac{1}{n} \sum_{x=1}^n g\big(\frac{x}{n}\big)\frac{\pi_x^2(nt)}{m_x} \Big|
\leq  \frac{C}{n} \sum_{x=1}^n \frac{\pi_x^2(nt)}{m_x} \leq \frac{C'}{n} \sum_{x=1}^n 
(\mathscr{E}_n^x)^2 \to 0,
\end{equation}	
where we used the conservation of the energy as well as the properties of $\mathscr{E}_n^x$ \eqref{ERRORZero}. We can argue similarly and obtain
\begin{equation} \label{ME003}
 \Big|\frac{1}{n} \sum_{x=1}^n g\big(\frac{x}{n}\big)\frac{\varrho^2(nt)}{m_x} \Big|
 \to 0.
\end{equation}
almost surely, as $n \to \infty$. Using the decomposition \eqref{ME00}, and combining \eqref{ME001} and \eqref{ee1}, with \eqref{ME002} and \eqref{ME003}, and taking advantage of a Cauchy-Schwartz inequality yield the result \eqref{kinetic1}.  
\end{proof}				

One can think of $\frac{\expval{p_x(nt)}_{\rho^n}^2}{2m_x}+\frac{\expval{r_x(nt)}^2_{\rho^n}}{2}$ as the Mechanical energy of the particle $x \in \mathbb{I}_n$, and $\frac{\mathrm{p}^2(y,t)}{2\bar{m}}+\frac{\mathrm{r}^2(y,t)}{2}$ as the Mechanical energy of the macroscopic material coordinate  $y \in \mathbb{I}$. This lemma is basically saying that the microscopic Mechanical energy converges to the macroscopic one, almost surely, in the sense of \eqref{kinetic1}. 
\subsection{Thermal Energy and Localization}
	
	In this section, we provide the necessary tool in order to deal with the thermal energy, i.e., localization of the "high modes" of the chain, which enables us to close the equation in \eqref{econv}. We state the localization in the sense of the following lemma.  This lemma is a consequence of the well known locaization phenomena in the disordered chain of harmonic ocillators  (see e.g. \cite{Kunz}, \cite{FA}, \cite{A98},\cite{Theo},\cite{BHO}), and we bring it here directly from (\cite{BHO} Lemma 3, Section 5) without a proof. 
	\begin{lemma} \label{localizationlemma}
		Recall the definition of the random matrix $A_p^0=M^{-\frac12}(-\Delta)M^{-\frac12}$, from Section \ref{pb}, where $M$ is the diagonal matrix of the masses, and $\Delta$ is the matrix of discrete gradient \eqref{productofgrad}. Moreover, recall the ordered eigenvalues of $A_p^0: 0=\omega_0<\omega_1 \dots <\omega_{n-1}$, and their corresponding eigenvectors: $\{\varphi^k \}_{k=0}^{n-1}$. Denote $\tilde{\varphi}^k:=M^{-\frac12}\varphi^k$.  \\
			Fix $\alpha ,\eta>0$, such that $0<2\alpha<\eta <1$. Recall the distribution of the masses $\mathbb{P}$, then there exists almost surely, $n_0 \in \mathbb{N}$ such that $\forall \: n>n_0$, and $\forall \: k \in I(\alpha):=(n^{(1-\alpha)},n-1] \cap \mathbb{Z}$, there exists an interval $J(k) \subset [0,n]$ with $|J(k)| \leq 2n^{\eta}$, such that: 
			\begin{equation} \label{localizationestimate}
					\forall x \notin J(k) , \quad |\tilde{\varphi}^k_x| \leq n^{-\frac{1}{\eta}}. 
			\end{equation}			 
	Equivalently, $|\varphi^k_x| \leq n^{-\frac{1}{\eta}}\sqrt{m_{max}}$.
	\end{lemma}		
		 Exploiting this lemma, one can deal with the contribution of the momentum to the thermal energy.  In order to deal with the contribution of the elongation, one needs to establish the localization of the eigenvectors of $A_r^0$. However, since the eigenvectors of $A_r^0$ i.e. $\phi^k$ are related to $\varphi^k$ by the following identity $\phi^k=\frac{1}{\omega_k}\nabla_+M^{-\frac12}\varphi^k$, we do not establish the localization directly. Instead, we control the contribution of the elongation to the thermal energy, by using lemma \ref{localizationlemma} as well as the following lemma (Notice that the contribution of the elongation to the thermal energy has not been discussed throughly in \cite{BHO}):    
		
\begin{lemma} \label{localizationlemma2}
 Recall the set-up of Lemma \ref{localizationlemma}, for every $\tilde{\varphi}^k$ satisfying  \eqref{localizationestimate}, i.e. there exists an interval $J(k)$ with $|J(k)| \leq 2n^{\eta}$, such that $\forall \: x \notin J(k)$, $|\tilde{\varphi}^k_x| \leq n^{-\frac{1}{\eta}}$, there exists a constant $c$ independent of $n$, with $\frac{1}{\omega_k} \leq cn^{\frac{3\eta}{2}}$. In particular, we  have:
 \begin{equation} \label{localizationestimate2}
 		\forall x \in \tilde{J}(k), \quad  |\phi^k_x|=\Big|\frac{1}{\omega_k}\nabla_+ \tilde{\varphi}^k_x \Big| \leq 2c n^{-\frac{1}{\eta}+\frac{3\eta}{2}},
 \end{equation}
where $\tilde{J}(k)$ is the interval: $  [\min\{J(k)\}+1,\max\{J(k)\}-1] $.
\end{lemma}		
 	\begin{proof}
 		First, recall  $\sum_{x=1}^n |\varphi^k_x|^2=1$, therefore $\sum_{x=1}^n|\tilde{\varphi}_x^k|^2 \geq \frac{1}{m_{max}}$. On the other hand, using the assumption \eqref{localizationestimate}, since $|J(k)^c|<n$, we have $\sum_{x \notin J(k)} |\tilde{\varphi}_x^k|^2 \leq n^{-\frac{2}{\eta}+1}$. Combining last two inequalities, we have $\sum_{x\in J(k)}|\tilde{\varphi}_x^k|^2 \geq \frac{1}{m_{max}}-\frac{1}{n^{\frac{2}{\eta}-1}} \geq \frac{c_0}{m_{max}}$, where $c_0$ is a constant independent of $n$, and the last inequality is deduced from the fact that $0<\eta<1$, and $m_{max}$ is bounded. Consequently, since $|J(k)|=2n^{\eta}$, we can choose $x_0 \in J(k)$ such that: 
 		
 	\begin{equation} \label{52lemma}
 			\frac{c_1}{n^{\frac{\eta}{2}}}	 \leq |\tilde{\varphi}_{x_0}^k|,
		\end{equation} 		  
		where $c_1$ is a constant independent of $n$. Now, choose $x_1 \notin J(k)$	to be the closest member of $J(k)^c$ to $x_0$. Using the assumption $|\tilde{\varphi}_x^k| \leq n^{-\frac{1}{\eta}}$, and the inequality \eqref{52lemma}, we have(assume $x_0>x_1$, the other situation will be exactly similar):
		\begin{equation} \label{521lemma}
			\begin{aligned}
			 	&c_2n^{-\frac{\eta}{2}} \leq c_1 n^{-\frac{\eta}{2}}-n^{-\frac{1}{\eta}} \leq ||\tilde{\varphi}_{x_0}^k|-|\tilde{\varphi}_{x_1}^k|| \leq |\tilde{\varphi}_{x_0}^k-\tilde{\varphi}_{x_1}^k| = |\sum_{j=x_1}^{x_0-1} \nabla_+ \tilde{\varphi}^k_j| \leq \sum_{j=x_1}^{x_0-1} |\nabla_+ \tilde{\varphi}^k_j| \\
			 	& \leq |x_1-x_0| \max_{j \in [x_0,x_1]} |\nabla_+ \tilde{\varphi}^k_j| \leq 2n^{\eta} \max_{j \in [x_0,x_1]} |\nabla_+ \tilde{\varphi}^k_j|,
			\end{aligned}
		\end{equation}		 
	where $\nabla_+\tilde{\varphi}^k_j=\tilde{\varphi}^k_{j+1}-\tilde{\varphi}^k_{j}$. Here, $c_2$ is a constant independent of $n$, and  we used the  choice of $x_1$: since $x_0 \in J(k)$, $|x_0-x_1| \leq |J(k)|=2n^{\eta}$.
Therefore, there exists  $j_0 \in \mathbb{I}_{n-1}$, and a constant $c_3$,  such that:
	\begin{equation} \label{522lemma}
		|\nabla_+\tilde{\varphi}^k_{j_0}| \geq c_3 n^{-\frac{3\eta}{2}}.
	\end{equation} 	
	Finally, we use the fact  $\sum_{x=1}^{n-1} |\phi^k_x|^2= \sum_{x=1}^{n-1}\frac{|\nabla_+\tilde{\varphi}^k_x|^2}{\omega_k^2}=1$, and thanks to \eqref{522lemma}, we obtain: 
	\begin{equation}
			\omega_k^2 \geq |\nabla_+\tilde{\varphi}^k_{j_0}|^2 \geq cn^{-3\eta},
	\end{equation}	 
where, this finishes the proof of the bound $\frac{1}{\omega_k} \leq cn^{\frac{3\eta}{2}}$. Since we assumed $\forall x \notin J(k)$, $|\tilde{\varphi}_x^k| \leq n^{-\frac{1}{\eta}}$, hence, $\forall x \notin \tilde{J}(k)$, $|\tilde{\varphi}_{x+1}^k-\tilde{\varphi}_x^k| \leq 2n^{-\frac{1}{\eta}}$. Now, using the bound $\frac{1}{\omega_k^2} \leq cn^{3\eta}$, and the definition $\phi^k_x= \frac{1}{\omega_k}(\tilde{\varphi}_{x+1}^k-\tilde{\varphi}_x^k)$, give us the estimate \eqref{localizationestimate2} as well.	
	\end{proof} 				
		
	We finish this section, by expressing the following lemma: Recall the "thermal" operators $\tilde{p}_x$ and $\tilde{r}_x$, defined in \eqref{thermalmatrices}. These operators can be defined at any time $t$:
	\begin{equation} \label{thermalcoordinatet}
		\tilde{p}_x(nt) =p_x(nt) -\expval{p_x(nt)}_{\rho}, \quad \tilde{r}_x(nt)=r_x(nt)-\expval{r_x(nt)}_{\rho}.
	\end{equation}
		Notice that here $\expval{p_x(nt)}_{\rho}$, should be understood as the constant times the identity operator. Then we have:
		\begin{lemma} \label{Thermallimitlemma}
			For any test function $g \in C^0([0,1])$, define $T_N^g(t)$ as follows: 
			\begin{equation} \label{Tn}
				T_n^g(t):= \frac{1}{n} \sum_{x=1}^n g(\frac{x}{n})\Big(\frac{\expval{\tilde{p}_x^2(nt)}_{\rho}}{2m_x}+\frac{\expval{\tilde{r}_x^2(nt)}_{\rho}}{2}\Big).
			\end{equation}
			Then, $\forall g \in C^1([0,1])$:
			\begin{equation} \label{Thermallimitmain}
				\lim_{n \to \infty} T_n^g(t)-T_n^g(0) \to 0,
			\end{equation}			  		
		almost surely. 		
		\end{lemma}
		One can see $T^g_n(t)$ as the contribution of the thermal energy in the LHS of \eqref{econv}. Observe that $\lim_{n \to \infty} T_n^g(0) = \int_0^1 g(y)\mathrm{f}_{\beta}^{\mu}(y)dy$,  $\forall g \in C^0([0,1])$, by \eqref{e0conv}. \\
		  
		\begin{proof}
		We denoted the average with respect to $\rho$ with $\expval{.}_{\rho}$, and the inner product in $\mathbb{R}^n$ with $\expval{.,.}_n$. Only in this proof, for the convenience and in order to  prevent any confusion, we will denote the average with respect to $\rho$ with $\pmb{\expval{.}_{\rho}}$, whenever these two appear in the same expression. \\		
			We use the explicit solution of the evolution equation, since $(p_x(nt),r_x(nt))$, and $(\expval{p_x(nt)}_{\rho}$, $\expval{r_x(nt)}_{\rho})$ are respectively solutions to the similar  linear equations \eqref{equationofmotionr}, and \eqref{averagetimeevolution}, by linearity, $\tilde{p}(nt)$ and $\tilde{r}(nt)$ can be obtained directly from \eqref{bogoliinv2}:    
			\begin{equation} \label{tildesolution}
				\begin{aligned}
					&\tilde{p}(nt)=\sum_{k=0}^{n-1}\big(\cos(\omega_knt)\hat{\tilde{p}}_k(0)- \sin(\omega_knt)\hat{\tilde{r}}_k(0))M^{\frac12}\varphi^k =:\sum_{k=0}^{n-1}(\hat{\tilde{p}}_k(nt)\big)M^{\frac{1}{2}} \varphi^k,
					\\
					&\tilde{r}(nt)=\sum_{k=1}^{n-1}\big(\cos(\omega_knt)\hat{\tilde{r}}_k(0)+\sin(\omega_knt)\hat{\tilde{p}}_k(0)\big) \phi^k =: \sum_{k=1}^{n-1} \hat{\tilde{r}}_k(nt)\phi^k,
				\end{aligned}
			\end{equation}					
where  $\hat{\tilde{r}}_k(0)= \expval{\phi^k,\tilde{r}(0)}_{n-1}=\expval{\phi^k,r(0)}_{n-1}-\expval{\phi^k,\bar{r}(0)}_{n-1}$, and $ \hat{\tilde{p}}_k(0)=\expval{M^{\frac12}\varphi^k,\tilde{p}(0)}_{n}=\expval{M^{\frac12}\varphi^k,p(0)}_{n}-\expval{M^{\frac12}\varphi^k,\bar{p}(0)}_{n}$, 	were defined in \eqref{hattilde}. Moreover, the definition of $\hat{\tilde{r}}_k(nt)$, and $\hat{\tilde{p}}_k(nt)$	are implicit in this expression.

			We prove this lemma in the following steps: \\
						\textit{Step1. Contribution of the low modes tends to zero.} \\
		 Define $\tilde{p}^{o}$ and $\tilde{r}^o$ to be the low mode portion of $\tilde{p}$ and $\tilde{r}$, respectively, for proper $0<\alpha<1$. The choice of $\alpha$ will become clear later: 
		 \begin{equation} \label{lowmodeportion}
		  \tilde{p}^o_x(nt):= \sum_{k \in \mathbb{Z} \cap [0,n^{(1-\alpha)}]} \hat{\tilde{p}}_k(nt) \sqrt{m_x}\varphi^k_x, \qquad    \tilde{r}^o_x(nt):= \sum_{k \in \mathbb{Z} \cap [1,n^{(1-\alpha)}]} \hat{\tilde{r}}_k(nt) \phi^k_x.
		 \end{equation}
			Then we have for any fixed $t \in [0,T]$: 
			\begin{equation} \label{lowmodelimit}
			\mathcal{L}_n^g(t) := \frac{1}{n}\sum_{x=1}^n g(\frac{x}{n})\Big(\frac{\expval{(\tilde{p}_x^o(nt))^2}_{\rho}}{2m_x}+\frac{\expval{(\tilde{r}_x^o(nt))^2}_{\rho}}{2}\Big) \to 0, 
			\end{equation}
			as $n \to \infty$. 			
		
		First, observe that $\frac{\expval{(\tilde{p}_x^o)^2}_{\rho}}{2m_x}$, and $\frac{\expval{(\tilde{r}_x^o)^2}_{\rho}}{2}$ are positive. This is elementary since $\tilde{p}_x^o$ and $\tilde{r}_x^o$ are self-adjoint. 
		Therefore, since $g$ is bounded, we proceed as follows (Notice that $\tilde{r}_n^o$ is zero by boundary condition): 
		\begin{equation} \label{lowmodesbound1}
			\begin{aligned}
				&|\mathcal{L}_n^g (t) | \leq \frac{||g||_{\infty}}{2n} \sum_{x=1}^n	\expval{\frac{(\tilde{p}_x^o)^2}{2m_x}+\frac{(\tilde{r}_x^o)^2}{2}}_{\rho} \leq \frac{C}{2n}\Bigg( \expval{\sum_{x=1}^n \frac{1}{m_x}\sum_{k,k'=0}^{[n^{1-\alpha}]+1} \hat{\tilde{p}}_k(nt)\hat{\tilde{p}}_{k'}(nt)\sqrt{m_x}\varphi_x^k  \sqrt{m_x}\varphi_x^{k'} }_{\rho_n}	 
			\\	& + \expval{\sum_{x=1}^{n-1}\sum_{k,k'=1}^{[n^{1-\alpha}]+1} \hat{\tilde{r}}_k(nt)\hat{\tilde{r}}_{k'}(nt)\phi_x^k \phi_x^{k'}}_{\rho}\Bigg) =\frac{C}{2n}\Bigg(\expval{\sum_{k,k'=0}^{[n^{1-\alpha}]+1}\hat{\tilde{p}}_k(nt)\hat{\tilde{p}}_{k'}(nt) \sum_{x=1}^n \varphi_x^k \varphi_x^{k'}}_{\rho}  \\& +\expval{\sum_{k,k'=1}^{[n^{1-\alpha}]+1}\hat{\tilde{r}}_k(nt)\hat{\tilde{r}}_{k'}(nt) \sum_{x=1}^{n-1} \phi_x^k \phi_x^{k'}}_{\rho}\Bigg)=\frac{C}{2n}\expval{\sum_{k,k'=0}^{[n^{1-\alpha}]+1} (\hat{\tilde{p}}_k(nt)\hat{\tilde{p}}_{k'}(nt)+\hat{\tilde{r}}_k(nt)\hat{\tilde{r}}_{k'}(nt))(\delta_{k,k'})}_{\rho}  \\
			& +\frac{C}{2n} \expval{\sum_{k=0}^{[n^{1-\alpha}]+1} (\hat{\tilde{p}}_k(nt))^2+(\hat{\tilde{r}}_k(nt)^2)}_{\rho},
		\end{aligned}
		\end{equation}
		where we substitute $\tilde{p}_x^o$ and $\tilde{r}_x^o$ by their definitions in \eqref{lowmodelimit},  and obtain the double sum, then we benefited from the linearity of $\tr$: $\expval{.}_{\rho}$, and the fact that $\{ \varphi^k \} $	and $\{ \phi^k \}$ are orthonormal basis for $\mathbb{R}^n$ and $\mathbb{R}^{n-1}$, respectively, hence $\sum_{x=1}^n \varphi_x^k \varphi_x^{k'}= \sum_{x=1}^n \phi_x^k \phi_x^{k'}=\delta_{k,k'}$. Notice that by abusing the notation, we start the last two sums from $k=0$, in spite of the fact that $\hat{\tilde{r}}_0$ has not been defined and by convention one can take $\hat{\tilde{r}}_0 \equiv 0$, at any time.\\
		 On the other hand, one can see that $\hat{\tilde{p}}_k^2 + \hat{\tilde{r}}_k^2 $ is conserved in time, $\forall k \in \mathbb{I}_n$, by the direct computation from the definition  \eqref{tildesolution}: $(\hat{\tilde{p}}_k(nt))^2 + (\hat{\tilde{r}}_k(nt))^2=(\hat{\tilde{p}}_k(0))^2(\sin^2(\omega_knt)+ \cos^2(\omega_knt) +(\hat{\tilde{r}}_k(0))^2(\sin^2(\omega_knt)+ \cos^2(\omega_knt) + \hat{\tilde{p}}_k \hat{\tilde{r}}_k(\cos(\omega_knt)\sin(\omega_knt)-\sin(\omega_knt)\cos(\omega_knt)) +\hat{\tilde{r}}_k \hat{\tilde{p}}_k (\sin(\omega_knt)\cos(\omega_knt)-\cos(\omega_knt)\sin(\omega_knt))=(\hat{\tilde{p}}_k(0))^2+\hat{\tilde{r}}_k(0))^2) $. Hence, using the bounds in \eqref{essbound} i.e. $\expval{(\hat{\tilde{p}}_k(0))^2}_{\rho} < C, \quad	\expval{(\hat{\tilde{r}}_k(0))^2}_{\rho} <C, $ from Lemma \ref{boundinitial}, we obtain:
		 \begin{equation} \label{lowmodecint}
		 	|\mathcal{L}_n^g(t) | \leq \frac{C}{2n} \sum_{k=0}^{[n^{1-\alpha}]+1} \expval{(\hat{\tilde{p}}_k(nt))^2+(\hat{\tilde{r}}_k(nt))^2}_{\rho}= \frac{C}{2n} \sum_{k=0}^{[n^{1-\alpha}]+1} \expval{(\hat{\tilde{p}}_k(0))^2+(\hat{\tilde{r}}_k(0))^2}_{\rho} \leq C' \frac{n^{(1-\alpha)}}{n},
		 \end{equation} 
		which clearly goes to zero as $n \to \infty$, by the choice of $0<\alpha<1$. Hence, we get \eqref{lowmodelimit}.\\
		\textit{Step2. Localization and freezing of the high modes.} \\
		In this step, we prove that the part of thermal energy coming from high modes is frozen in time, thanks to the localization Lemmas \ref{localizationlemma} and \ref{localizationlemma2}. In the same spirit of the previous step \eqref{lowmodeportion}, recall $I(\alpha)=(n^{(1-\alpha)},n-1] \cap \mathbb{Z}$, and define $\tilde{p}^\bullet_x(nt)$ and $\tilde{r}^{\bullet}_x(nt)$ as:
		\begin{equation} \label{highmodedef}
			\tilde{p}_x^{\bullet}(nt):= \tilde{p}_x(nt)-\tilde{p}_x^{o}(nt)=\sum_{k \in I(\alpha)} \hat{\tilde{p}}_k(nt) \sqrt{m_x}\varphi_x^k, \quad \tilde{r}_x^{\bullet}:= \tilde{r}_x(nt)-\tilde{r}_x^{o}(nt)=\sum_{k \in I(\alpha)} \hat{\tilde{r}}_k(nt) \phi_x^k.
		\end{equation}
		Moreover, define $\mathcal{U}_n^g(t)$ as: 
		\begin{equation} \label{highmodeportion} 
			\mathcal{U}_n^g(t) := \frac{1}{n}\sum_{x=1}^n g(\frac{x}{n})\Big(\frac{\expval{(\tilde{p}_x^{\bullet}(nt))^2}_{\rho}}{2m_x}+\frac{\expval{(\tilde{r}_x^{\bullet}(nt))^2}_{\rho}}{2}\Big).
		\end{equation}				
		Then we have for any $t \in [0,T]$, and $g \in C^1[0,1]$:
		\begin{equation} \label{highmodelimit}				
			\mathcal{U}_n^g(t) - \mathcal{U}_n^g(0) \to 0,
		\end{equation}
			almost surely. 
		
	In order to prove \eqref{highmodelimit}, we decompose $\mathcal{U}_n^g(t)$ into two parts, one which is constant in time, and the other which is small.  \\
	For a fixed $n$,  a function $g:[0,1] \to \mathbb{R}$, and a vector of $n$ operators $p$, let $g.p$ denotes the following  vector of operators: $g.p(x)= g(\frac{x}{n})p_x, \: \forall x \in \mathbb{I}_n$. Moreover, for a vector of $n-1$ operators $r$, $g.r$ is the following vector of operators: $g.r(x)=g(\frac{x}{n})r_x, \: \forall x \in \mathbb{I}_{n-1}$. Since $n$ is fixed in our computation, this notation does not cause any confusion.   Using this notation, and linearity of the trace, one can rewrite $\mathcal{U}_n^g(t)$ as follows:
	\begin{equation}
			\mathcal{U}_n^g(t)= \frac{1}{2n} \pmb{\langle} \sum_{x=1}^n g(\frac{x}{n})(\frac{\tilde{p}^{\bullet}_x(nt)^2}{m_x} + \tilde{r}^{\bullet}_x(nt)^2) \pmb{\rangle_{\rho}} = \frac{1}{2n} \pmb{\langle}\expval{g.\tilde{p}^{\bullet}(nt),M^{-1}\tilde{p}^{\bullet}(nt)}_n + \expval{g.\tilde{r}^{\bullet}(nt),\tilde{r}^{\bullet}(nt)}_{n-1} \pmb{\rangle_{\rho}}.
	\end{equation}
		By the resolution of the identity i.e. $I_n = \sum_{k=0}^{n-1} \ket{\varphi^k}\bra{\varphi^k}$, and $I_{n-1} = \sum_{k=1}^{n-1} \ket{\phi^k}\bra{\phi^k}$ in $\mathbb{R}^n$, and $\mathbb{R}^{n-1}$, we expand the later in the basis of $\phi^k$ and $\varphi^k$. We also split  $M^{-1}$ and recall the definition $\tilde{\varphi}^k=M^{-\frac12}\varphi^k$:
		\begin{equation} \label{highmodes2}
			\begin{aligned}
				 \mathcal{U}_n^g(t) &= \frac{1}{2n} \pmb{\Bigg\langle} \sum_{k=1}^n  \expval{g.\tilde{p}^{\bullet}(nt),\tilde{\varphi}^k}_n \expval{\tilde{\varphi}^k,\tilde{p}^{\bullet}(nt)}_n + \sum_{k=1}^{n-1} \expval{g.\tilde{r}^{\bullet},\phi^k}_{n-1}\expval{\phi^k,\tilde{r}^{\bullet}(nt)}_{n-1} \pmb{\Bigg\rangle_{\rho}}	\\ &= \frac{1}{2n} \pmb{\Bigg\langle} \sum_{k \in I(\alpha)} \Big(\expval{g.\tilde{p}^{\bullet}(nt),\tilde{\varphi}^k}_n \hat{\tilde{p}}_k(nt) + \expval{g.\tilde{r}^{\bullet}(nt),\phi^k}_{n-1} \hat{\tilde{r}}_k(nt) \Big)\pmb{\Bigg\rangle_\rho}.
			\end{aligned}
		\end{equation}
	Notice that in order to obtain the second line, we used the identities $\expval{\tilde{\varphi}^k,\tilde{p}^{\bullet}(nt)}_n=\hat{\tilde{p}}_k(nt)$ for $k \in I(\alpha)$, and $\expval{\tilde{\varphi}^k,\tilde{p}^{\bullet}(nt)}_n=0$ for $k \notin I(\alpha)$,  and their counterparts for $\tilde{r}^{\bullet}(nt)$, thanks to the definition of $\tilde{p}^{\bullet}(nt)$ and $\tilde{r}^{\bullet}(nt)$ in \eqref{highmodedef}.\\
	Now, let us split $g$ for each  $k$. For each $k \in I(\alpha)$, recall the interval $J(k)$ given by the Lemma \ref{localizationlemma}, and let $x_k$ be the center of this interval. Then, let $g_k(x) := g(\frac{x_k}{n})$, $\forall x \in \mathbb{I}_n$, be the constant vector for each $k \in I(\alpha)$,  and define $\tilde{g}_k(x)=g(x)-g_k(x)$, $\forall x \in \mathbb{I}_n$. We simply have: $\forall k \in I(\alpha)$, $g= \tilde{g}_k+g_k$. By linearity of $g.p$, and $\expval{.}_n$, we rewrite \eqref{highmodes2} as follows:
	\begin{equation} \label{Useperation}
\begin{aligned}		
	&\mathcal{U}_n^g(t)= \frac{1}{2n}\pmb{\Bigg\langle} \sum_{k \in I(\alpha)}  \Big(\expval{g_k.\tilde{p}^{\bullet}(nt),\tilde{\varphi}^k}_n \hat{\tilde{p}}_k(nt) + \expval{g_k.\tilde{r}^{\bullet}(nt),\phi^k}_{n-1} \hat{\tilde{r}}_k(nt) \Big)  \pmb{\Bigg\rangle_{\rho}} + \\ 
&  \frac{1}{2n}\pmb{\Bigg\langle} \sum_{k \in I(\alpha)}  \Big(\expval{\tilde{g}_k.\tilde{p}^{\bullet}(nt),\tilde{\varphi}^k}_n \hat{\tilde{p}}_k(nt) + \expval{\tilde{g}_k.\tilde{r}^{\bullet}(nt),\phi^k}_{n-1} \hat{\tilde{r}}_k(nt) \Big) \pmb{\Bigg\rangle_{\rho}}.
\end{aligned}	
	\end{equation}
		 In the later decomposition, the first line is constant in time, and the second line vanishes as $n \to \infty$. Let $\mathcal{U}_n^g(t)=\bar{\mathcal{U}}_n^g(t)+\tilde{\mathcal{U}}_n^g(t)$, where $\bar{\mathcal{U}}_n^g(t)$ is the first line in \eqref{Useperation}, and  $\tilde{\mathcal{U}}_n^g(t)$ is the second line. Since  for each $k$, $g_k$ is  constant in $x$, we can factor it and observe: 
		\begin{equation} \label{constanintimeU}
			\begin{aligned}			
			\bar{\mathcal{U}}_n^g(t) &:= \frac{1}{2n} \pmb{\Bigg\langle} \sum_{k \in I(\alpha)} g(\frac{x_k}{n}) \Big(\expval{\tilde{p}^{\bullet}(nt),\tilde{\varphi}^k}_n \hat{\tilde{p}}_k(nt) + \expval{\tilde{r}^{\bullet}(nt),\phi^k}_{n-1} \hat{\tilde{r}}_k(nt) \Big)  \pmb{\Bigg\rangle_{\rho}} \\
			&= \frac{1}{2n} \pmb{\Big\langle} \sum_{k \in I(\alpha)} g(\frac{x_k}{n}) \big(\hat{\tilde{p}}_k(nt)^2+\hat{\tilde{r}}_k(nt)^2\big)  \pmb{\Big\rangle_{\rho}} = \bar{\mathcal{U}}_n^g(0),
			\end{aligned}		
		\end{equation}
	where we take advantage of the identity $\expval{\tilde{\varphi}^k,\tilde{p}^{\bullet}(nt)}_n=\hat{\tilde{p}}_k(nt)$, for $k \in I(\alpha)$,  and the similar identity for $\tilde{r}^{\bullet}(nt)$, as in \eqref{highmodes2}. Moreover, we already observed in \eqref{lowmodecint}, that the expression $\hat{\tilde{p}}_k(nt)^2+\hat{\tilde{r}}_k(nt)^2$  for each $k$ -which represents the thermal energy of the $k$th mode- is conserved by the dynamics. Hence, $\bar{\mathcal{U}}_n^g(t)=\bar{\mathcal{U}}_n^g(0)$ is constant in time. \\
	In the rest of this step, we prove that $\tilde{\mathcal{U}}_n^g \to 0$, almost surely as $n \to \infty$. In preparation of this proof, we need to use the following form of the Cauchy Schwartz inequality. For certain operators $a$ and $b$, we have: 
	\begin{equation} \label{CSineq}
	 	|\expval{ab^*}_{\rho}|^2 \leq \expval{aa^*}_{\rho} \expval{bb^*}_{\rho}. 
	\end{equation}

In order to deal with $\tilde{\mathcal{U}}_n^g(t)$, first, we bound $\pmb{\langle} \expval{\tilde{g}_k.\tilde{p}^{\bullet}(nt),\tilde{\varphi}^k}_n^2 \pmb{\rangle_{\rho}}$ as follows: Recall Lemma \ref{localizationlemma}, since this lemma is valid for any choice of $0<2\alpha<\eta<1$, we take $\eta<\frac23$. Then, there exists a constant $c>0$, independent of $n$, such that for any $k \in I(\alpha)$, and $n>n_0$, where $n_0$ is given by \ref{localizationlemma}, we have:  
\begin{equation} \label{gkbound}
	\pmb{\Big\langle} \expval{\tilde{g}_k.\tilde{p}^{\bullet}(nt),\tilde{\varphi}^k}_n^2 \pmb{\Big\rangle_{\rho}} \leq \frac{c}{n^{1-3\eta}},  \qquad  \pmb{\Big\langle} \expval{\tilde{g}_k.\tilde{r}^{\bullet}(nt),\phi^k}_{n-1}^2 \pmb{\Big\rangle_{\rho}} \leq \frac{c}{n^{1-3\eta}},
\end{equation}  
almost surely. These bounds can be achieved by expanding the inner product, and performing the following computation:
\begin{equation} \label{gkbound1}
	\begin{aligned}
		&\pmb{\Big\langle} \expval{\tilde{g}_k.\tilde{p}^{\bullet}(nt),\tilde{\varphi}^k}_n^2 \pmb{\Big\rangle_{\rho}} = \pmb{\Big\langle} (\sum_{x=1}^n\tilde{g}_k(\frac{x}{n})\tilde{\varphi}^k_x\tilde{p}^{\bullet}_x(nt))^2	 \pmb{\Big\rangle_{\rho}} = \sum_{x,y=1}^n \tilde{g}_k(\frac{x}{n})\tilde{g}(\frac{y}{n}) \tilde{\varphi}^k_x\tilde{\varphi}^k_y \pmb{\langle}\tilde{p}^{\bullet}_x(nt) \tilde{p}^{\bullet}_y(nt) \pmb{\rangle_{\rho}} \leq \\ & \sum_{x,y=1}^n |\tilde{g}_k(\frac{x}{n})\tilde{g}(\frac{y}{n}) \tilde{\varphi}^k_x\tilde{\varphi}^k_y| |\pmb{\langle}\tilde{p}^{\bullet}_x(nt) \tilde{p}^{\bullet}_y(nt) \pmb{\rangle_{\rho}}| \leq \sum_{x,y=1}^n  |\tilde{g}_k(\frac{x}{n})\tilde{g}(\frac{y}{n}) \tilde{\varphi}^k_x\tilde{\varphi}^k_y| \big(\pmb{\langle} \tilde{p}_x^{\bullet}(nt))^2 \pmb{\rangle_{\rho}}\big)^{\frac12} \big(\pmb{\langle} \tilde{p}_y^{\bullet}(nt))^2 \pmb{\rangle_{\rho}}\big)^{\frac12} = \\&
		\Big(\sum_{x=1}^n |\tilde{g}_k(\frac{x}{n}) \tilde{\varphi}^k_x| \big(\pmb{\langle} \tilde{p}_x^{\bullet}(nt))^2 \pmb{\rangle_{\rho}}\big)^{\frac12}\Big)^2 \leq \Big(\sum_{x=1}^n|\tilde{g}_k(\frac{x}{n})\tilde{\varphi}^k_x|^2\Big)\Big(\sum_{x=1}^n \pmb{\langle} (\tilde{p}_x^{\bullet}(nt))^2 \pmb{\rangle_{\rho}}\Big).
	\end{aligned}
\end{equation}
	 
	 The second inequality obtained using the aforementioned form of Cauchy Schwartz inequality in \eqref{CSineq},  and the last inequality is evident, using the Cauchy Schwartz inequality for finite dimensional vectors.\\ 
	 In the following computation, using the definition of $\tilde{p}^{\bullet}_x(nt)$ \eqref{highmodedef}, and getting a double sum, we bound the second term in \eqref{gkbound1} by $c_0n$, where $c_0>0$ is a constant independent of $n$:
	\begin{equation} \label{gkbound2}
		\begin{aligned}
			&\sum_{x=1}^n \pmb{\langle}\tilde{p}_x^{\bullet}(nt)^2\pmb{\rangle_{\rho}} \leq \sum_{x=1}^n \frac{m_{max}}{m_x} \sum_{k,k'\in I(\alpha)} m_x \varphi_x^k \varphi_x^{k'} \pmb{\langle} \hat{\tilde{p}}_k(nt) \hat{\tilde{p}}_{k'}(nt)  \pmb{\rangle_{\rho}} \leq \\ & c_1\sum_{k,k' \in I(\alpha)} \pmb{\langle} \hat{\tilde{p}}_k(nt) \hat{\tilde{p}}_{k'}(nt)  \pmb{\rangle_{\rho}} \sum_{x=1}^n \varphi_x^k \varphi_x^{k'} 	  = \sum_{k \in I(\alpha)} \pmb{\langle} (\hat{\tilde{p}}_k(nt))^2 \pmb{\rangle_{\rho}} \leq c_14C|I(\alpha)|\leq c_0n.
		\end{aligned}
	\end{equation}
	In the last line, we bounded $\pmb{\langle} (\hat{\tilde{p}}_k(nt))^2 \pmb{\rangle_{\rho}} \leq 4C$ (2C works as well, with conservation argument), by using the explicit form of $\hat{\tilde{p}}_k(nt)= \cos(\omega_knt)\hat{\tilde{p}}_k-\sin(\omega_knt)\hat{\tilde{r}}_k$:
	\begin{equation} \label{boundpk}
		\begin{aligned}		
		&\pmb{\langle} (\hat{\tilde{p}}_k(nt))^2 \pmb{\rangle_{\rho}} = \\ &\pmb{\langle} \hat{\tilde{p}}_k(0)^2 \cos^2(\omega_knt) + \hat{\tilde{r}}_k(0)^2 \sin^2(\omega_k(nt)) -\sin(\omega_knt)\cos(\omega_knt)(\hat{\tilde{r}}_k(0)\hat{\tilde{p}}_k(0)+\hat{\tilde{p}}_k(0)\hat{\tilde{r}}_k(0)) \pmb{\rangle_{\rho}} \leq \\
		&  \pmb{\langle} (\hat{\tilde{p}}_k(0))^2 \pmb{\rangle_{\rho}} + \pmb{\langle} (\hat{\tilde{r}}_k(0))^2 \pmb{\rangle_{\rho}} + |\pmb{\langle}\hat{\tilde{p}}_k(0) \hat{\tilde{r}}_k(0) \pmb{\rangle_{\rho}}|	+|\pmb{\langle}\hat{\tilde{r}}_k(0) \hat{\tilde{p}}_k(0) \pmb{\rangle_{\rho}}| \leq		
		 \\& \pmb{\langle} \hat{\tilde{p}}_k(0))^2 \pmb{\rangle_{\rho}} + \pmb{\langle} (\hat{\tilde{r}}_k(0))^2 \pmb{\rangle_{\rho}} + 2 \pmb{\langle} (\hat{\tilde{p}}_k(0))^2 \pmb{\rangle_{\rho}}^{\frac12} \pmb{\langle} \hat{\tilde{r}}_k(0))^2 \pmb{\rangle_{\rho}} ^{\frac12} \leq 4C,
		\end{aligned}	
	\end{equation}
	where we bounded $|\sin(.)|$ and $|\cos(.)|$ by one, used Cauchy Schwartz \eqref{CSineq} in the second line, and used the appropriate bound from Lemma \ref{boundinitial} in the last line. \\
	Moreover, we can bound the first term in the RHS of \eqref{gkbound} by $\frac{c_1}{n^{2-3\eta}}$, almost surely, thanks to \eqref{localizationestimate}, and regularity of $g$. Recall the definition of $J(k)$ from lemma \ref{localizationlemma}, then we can rewrite the later as:
	\begin{equation} \label{gkbound3}
\begin{aligned}	
	&\sum_{x=1}^n |\tilde{g}_k(\frac{x}{n})\tilde{\varphi}^k_x|^2=\sum_{x\in J(k)}|\tilde{g}_k(\frac{x}{n})\tilde{\varphi}^k_x|^2+\sum_{x \notin J(k)}|\tilde{g}_k(\frac{x}{n})\tilde{\varphi}^k_x|^2 \leq c_1' \sum_{x \in J(k)} |\tilde{g}_k(\frac{x}{n})|^2+ c_2'\sum_{x \notin J(k) } |\tilde{\varphi}^k_x|^2 \\
	& \leq c_1'C_1\sum_{x \in J(k)} \frac{|x-x_k|^2}{n^2} + c_2' n^{-\frac{2}{\eta}}|J(k)^c| \leq 2c_1'C_1 \sum_{j=0}^{n^{\eta}}j^2+ c_2'n^{-\frac{2}{\eta}+1} \leq C_1' \frac{n^{3\eta}}{n^2}+c_2'n^{-\frac{2}{\eta}+1}  \\ & \leq \frac{c_1}{n^{2-3\eta}},
\end{aligned}	
	\end{equation} 
where in the second inequality, we bound $|\tilde{\varphi}^k_x|$, in the second term by $n^{-\frac{1}{\eta}}$, almost surely, for $x \notin J(k)$, thanks to the bound \eqref{localizationestimate} in Lemma \eqref{localizationlemma}. In the second inequality, for the first term, we exploited the definition of $\tilde{g}_k(\frac{x}{n})=g(\frac{x}{n})-g(\frac{x_k}{n})$, as well as the fact that $g \in C^1([0,1])$, and therefore, there exists a constant $C_1>0$ \footnote{One can simply take $C_1=||g'||_{\infty}$.}, such that $|g(\frac{x}{n})-g(\frac{x_k}{n})| \leq C_1|\frac{x}{n}-\frac{x_k}{n}|$.  In the third inequality, we applied the definition of $x_k$ as the center of the interval $J(k)$, we also used the fact that $|J(k)|=2n^{\eta}$, and $|J(k)^c|=n-2n^{\eta} <n$.   \\

By inserting the bounds $\sum_{x=1}^n \pmb{\langle} \tilde{p}^{\bullet}_x(nt)^2   \pmb{\rangle_{\rho}}\leq c_0n$, from \eqref{gkbound2}, and $\sum_{x=1}^n |\tilde{g}_k(\frac{x}{n})\tilde{\varphi}^k_x|^2 \leq \frac{c_1}{n^{2-3\eta}}$, from   \eqref{gkbound3}, in the expression \eqref{gkbound1}, we obtain the first bound in \eqref{gkbound}, which is the product of these two bounds, namely:
 $$\pmb{\Big\langle} \expval{\tilde{g}_k.\tilde{p}^{\bullet}(nt),\tilde{\varphi}^k}_n^2 \pmb{\Big\rangle_{\rho}}\leq \Big(\sum_{x=1}^n \pmb{\langle} \tilde{p}^{\bullet}_x(nt)^2   \pmb{\rangle_{\rho}}\Big)\Big(\sum_{x=1}^n |\tilde{g}_k(\frac{x}{n})\tilde{\varphi}^k_x|^2 \Big)\leq c_0n \times \frac{c_1}{n^{2-3\eta}} \leq \frac{c}{n^{1-3\eta}}.$$

	The second bound in \eqref{gkbound}, corresponding to the elongation operator $r$ can be treated similarly, with a small modification (using Lemma \ref{localizationlemma2} instead of Lemma \ref{localizationlemma}). \\
	 Performing the exact same computation as in  \eqref{gkbound1}, one can obtain the following bound using the Cauchy Shwartz inequality twice:
	\begin{equation} \label{rkbound1}
		\pmb{\Big\langle} \expval{\tilde{g}_k.\tilde{r}^{\bullet}(nt),\phi^k}_{n-1}^2\pmb{\Big\rangle_{\rho}} \leq \Big(\sum_{x=1}^{n-1}|\tilde{g}_k(\frac{x}{n})\phi_x^k|^2\Big)\Big(\sum_{x=1}^{n-1} \pmb{\langle} (\tilde{r}^{\bullet}_x(nt))^2 \pmb{\rangle_{\rho}}\Big),	
	\end{equation} 
	Again, similar to \eqref{gkbound2}, we have: 
	\begin{equation} \label{rkbound2}
			\begin{aligned}		
		& \sum_{x=1}^n \pmb{\langle} (\tilde{r}^{\bullet}_x(nt))^2 \pmb{\rangle_{\rho}}= \sum_{x=1}^{n-1} \sum_{k,k'\in I(\alpha)}\phi_x^k \phi_x^{k'} \pmb{\langle} \hat{\tilde{r}}_k(nt) \hat{\tilde{r}}_{k'}(nt) \pmb{\rangle_{\rho}}= \sum_{k,k' \in I(\alpha)} \pmb{\langle} \hat{\tilde{r}}_k(nt) \hat{\tilde{r}}_{k'}(nt) \pmb{\rangle_{\rho}} \sum_{x=1}^{n-1} \phi_x^k \phi_x^{k'}= \\
		& \sum_{k \in I(\alpha)} \pmb{\langle} \hat{\tilde{r}}_k(nt)^2  \pmb{\rangle_{\rho}} \leq 4C |I(\alpha)| \leq c_0'n,
			\end{aligned}
	\end{equation}
	
	where we bounded $\pmb{\langle} \hat{\tilde{r}}_k(nt)^2  \pmb{\rangle_{\rho}} \leq 4C$, exactly similar to \eqref{boundpk}, using the explicit form of $\hat{\tilde{r}}_k(nt)= \cos(\omega_knt)\hat{\tilde{r}}_k(0) + \sin(\omega_knt) \hat{\tilde{r}}_k(0)$, and the bounds in Lemma \ref{boundinitial}.\\
	Dealing with the first term in the RHS of \eqref{rkbound1} requires more attention (this problem have not been addressed in \cite{BHO}). We can proceed similar to  \eqref{gkbound3}, recalling $\tilde{J}(k)$ from Lemma \ref{localizationlemma2}: 
	\begin{equation}\label{rkbound3}
\begin{aligned}
	&\sum_{x=1}^{n-1} |\tilde{g}_k(\frac{x}{n})\phi^k_x|^2=\sum_{x \in \tilde{J}(k)}|\tilde{g}_k(\frac{x}{n})\phi^k_x|^2+\sum_{x \notin \tilde{J}(k)}|\tilde{g}_k(\frac{x}{n})\phi^k_x|^2 \leq c_1'' \sum_{x \in \tilde{J}(k)} |\tilde{g}_k(\frac{x}{n})|^2+ c_2'' \sum_{x \notin \tilde{J}(k) } |\phi^k_x|^2 \\
	& \leq (c_1''C_1\sum_{x \in \tilde{J}(k)} \frac{|x-x_k|^2}{n^2}) + (c_3 n^{-\frac{2}{\eta}+3\eta}|\tilde{J}(k)^c|) \leq (2c_1''C_1 \sum_{j=0}^{n^{\eta}}j^2)+( c_3n^{-\frac{2}{\eta}+3\eta+1}) 
	\\ &\leq C_1' \frac{n^{3\eta}}{n^2}+c_3n^{-\frac{2}{\eta}+3\eta+1}   \leq \frac{c_1}{n^{2-3\eta}},
	\end{aligned}
	\end{equation}
	where these computation can be justified similar to \eqref{gkbound3}, except from the fact that on the second inequality we exploited the bound \eqref{localizationestimate2}: $|\phi^k_x| \leq 2cn^{-\frac{1}{\eta}+\frac{3\eta}{2}}$ from Lemma \ref{localizationlemma2}. Moreover, in the last inequality, by using the assumption $\eta \in (0,\frac23)$, we deduced that $3\eta-2>-\frac{2}{\eta}+3\eta+1$, and $n^{3\eta-2}> n^{-\frac{2}{\eta}+3\eta+1}$. Hence, we have the last bound by choosing the proper constant. Finally, notice that here we can replace $J(k)$ by $\tilde{J}(k)$, since $|J(k)| \geq |\tilde{J}(k)|$.\\
	By inserting \eqref{rkbound2} and \eqref{rkbound3}  into \eqref{rkbound1}, we obtain the second bound in \eqref{gkbound}:
	\begin{equation}
	\pmb{\Big\langle} \expval{\tilde{g}_k.\tilde{r}^{\bullet}(nt),\phi^k}_{n-1}^2\pmb{\Big\rangle_{\rho}} \leq \Big(\sum_{x=1}^{n-1}|\tilde{g}_k(\frac{x}{n})\phi_x^k|^2\Big)\Big(\sum_{x=1}^{n-1} \pmb{\langle} (\tilde{r}^{\bullet}_x(nt))^2 \pmb{\rangle_{\rho}}\Big) \leq  \frac{c_1}{n^{2-3\eta}} \times c_0' n \leq \frac{c}{n^{1-3\eta}}.
	\end{equation}
		
	Recall the definition of  $\tilde{\mathcal{U}}_n^g(t) $
\begin{equation}
\tilde{\mathcal{U}}_n^g(t) := \frac{1}{2n} \sum_{k \in I(\alpha)} \pmb{\Big\langle} \expval{\tilde{g}_k.\tilde{p}^{\bullet}(nt),\tilde{\varphi}^k}_n \hat{\tilde{p}}_k(nt) + \expval{\tilde{g}_k.\tilde{r}^{\bullet}(nt),\phi^k}_{n-1} \hat{\tilde{r}}_k(nt)  \pmb{\Big\rangle_{\rho}},
\end{equation} 
by using the Cauchy Schwartz inequality \eqref{CSineq}, we have\footnote{We already discussed the positivity of the terms under the square root in the second line, which can be obtained by the fact that these operators are linear combination of bosonic operators $\tilde{\mathfrak{b}}_k$, and their adjoints, and they are self adjoint, hence, by using Lemma \ref{bosonicthermalaverage}, we obtain the positivity (more abstract proof is of course possible).}:
\begin{equation} \label{tildelimit}
\begin{aligned}
		&|\tilde{\mathcal{U}}_n^g(t) | \leq \frac{1}{2n} \sum_{k \in I(\alpha)} \Big|\pmb{\Big\langle} \expval{\tilde{g}_k.\tilde{p}^{\bullet}(nt),\tilde{\varphi}^k}_n \hat{\tilde{p}}_k(nt) \pmb{\Big\rangle_{\rho}}\Big| + \Big|\pmb{\Big\langle} \expval{\tilde{g}_k.\tilde{r}^{\bullet}(nt),\phi^k}_{n-1} \hat{\tilde{r}}_k(nt) \pmb{\Big\rangle_{\rho}}\Big| \\
		& \leq \frac{1}{2n} \sum_{k \in I(\alpha)} \pmb{\Big\langle}\expval{\tilde{g}_k.\tilde{p}^{\bullet}(nt),\tilde{\varphi}^k}_n^2 \pmb{\Big\rangle_{\rho}}^{\frac12}\pmb{\langle}\hat{\tilde{p}}_k(nt)^2 \pmb{\rangle_{\rho}}^{\frac12} + \pmb{\Big\langle} \expval{\tilde{g}_k.\tilde{r}^{\bullet}(nt),\phi^k}_{n-1}^2 \pmb{\Big\rangle_{\rho}}^{\frac12} \pmb{\langle} \hat{\tilde{r}}_k(nt)^2 \pmb{\rangle_{\rho}}^{\frac12} 
	\\	&  \leq \frac{\sqrt{C}}{n} \sum_{k \in I(\alpha)} \pmb{\Big\langle}\expval{\tilde{g}_k.\tilde{p}^{\bullet}(nt),\tilde{\varphi}^k}_n^2 \pmb{\Big\rangle_{\rho}}^{\frac12} + \pmb{\Big\langle} \expval{\tilde{g}_k.\tilde{r}^{\bullet}(nt),\phi^k}_{n-1}^2 \pmb{\Big\rangle_{\rho}}^{\frac12}  \leq \frac{\sqrt{C}}{n}\frac{2\sqrt{c}}{n^{\frac12-\frac{3\eta}{2}}} |I(\alpha)|  \\ & \leq \frac{C_0}{n^{\frac12-\frac{3\eta}{2}}} \to 0,
\end{aligned} 
\end{equation}
where in the first inequality, we used Cauchy Schwartz \eqref{CSineq}. In the second inequality, we exploited the bounds $\pmb{\langle} \hat{\tilde{p}}_k(nt)^2 \pmb{\rangle_{\rho}} \leq 4C ,\pmb{\langle} \hat{\tilde{r}}_k(nt)^2 \pmb{\rangle_{\rho}} \leq 4C$, as we already did in \eqref{gkbound2} and \eqref{rkbound2}. In the third inequality, we benefited from the bounds in \eqref{gkbound}. Hence, if one takes $\eta<\frac13$, one can deduce $|\tilde{\mathcal{U}}_n^g(t)| \to 0$, almost surely. \\
Lastly, recall that we expressed $\mathcal{U}_n^g(t)=\bar{\mathcal{U}}_n^g(t)+ \tilde{\mathcal{U}}_n^g(t)$, we sum up the result of this step as follows: 
\begin{equation}
	\begin{aligned}
			|\mathcal{U}_n^g(t) - \mathcal{U}_n^g(0)|= |\bar{\mathcal{U}}_n^g(t)+ \tilde{\mathcal{U}}_n^g(t)-(\bar{\mathcal{U}}_n^g(t)+ \tilde{\mathcal{U}}_n^g(t))|= 
			|\tilde{\mathcal{U}}_n^g(t)-\tilde{\mathcal{U}}_n^g(0)|  \leq |\tilde{\mathcal{U}}_n^g(t))| +|\tilde{\mathcal{U}}_n^g(0))| \to 0,
	\end{aligned}
\end{equation}
almost surely, 
thanks to $\bar{\mathcal{U}}_n^g(t)=\bar{\mathcal{U}}_n^g(0)$  from \eqref{constanintimeU}, and   \eqref{tildelimit}.
This finishes the proof of this step \eqref{highmodelimit}.	\\

\textit{Step3. Summing up.} \\
In this step, we finish the proof of Lemma \ref{Thermallimitlemma},(the limit \eqref{Thermallimitmain}), by combining the results from previous steps \eqref{lowmodelimit}, \eqref{highmodelimit}. In fact, the later 	expressions let us conclude that $|\mathcal{U}_n^g(t)+\mathcal{L}_n^g- \mathcal{U}_n^g(0)-\mathcal{L}_n^g(0)| \to 0$,  almost surely. By comparing $T_n^g(t)$, and $\mathcal{U}_n^g(t)+\mathcal{L}_n^g$, one can see it is sufficient to control the following term in order to obtain \eqref{Thermallimitmain} \footnote{Recall that all the expressions corresponding to the $n$th component of the $r$ operator is zero, and we bring them in the same sum just to lighten the notation.}.
\begin{equation} \label{errorterm}
\mathcal{E}_n^g(t) := \frac{1}{n}\sum_{x=1}^n g(\frac{x}{n})\Big(\frac{\pmb{\langle} \tilde{p}_x^o(nt) \tilde{p}_x^{\bullet}(nt) \pmb{\rangle_{\rho}}}{m_x}+ \pmb{\langle}  \tilde{r}_x^o(nt) \tilde{r}_x^{\bullet}(nt) \pmb{\rangle_{\rho}} \Big).
\end{equation}
	However, this term can be treated as usual, using the Cauchy Schwartz inequality \eqref{CSineq}:
	
\begin{equation} \label{errortermlimit}
	\begin{aligned}
		&|\mathcal{E}_n^g(t)| \leq \Big|\frac{1}{n} \sum_{x=1}^n \big|\frac{g(\frac{x}{n})}{m_x}\big||\pmb{\langle} \tilde{p}_x^o(nt) \tilde{p}_x^{\bullet}(nt) \pmb{\rangle_{\rho}}|\Big| + \Big|\frac{1}{n}\sum_{x=1}^{n-1} |g(\frac{x}{n})||\pmb{\langle} \tilde{r}_x^o(nt) \tilde{r}_x^{\bullet}(nt) \pmb{\rangle_{\rho}}|\Big|
\\ & \leq  \Big|\frac{1}{n} \sum_{x=1}^n \big|\frac{g(\frac{x}{n})}{m_x}\big||\pmb{\langle} (\tilde{p}_x^o(nt))^2 \pmb{\rangle_{\rho}}^{\frac12} \pmb{\langle}(\tilde{p}_x^{\bullet}(nt))^2 \pmb{\rangle_{\rho}}^{\frac12}|\Big| + \Big|\frac{1}{n}\sum_{x=1}^{n-1} |g(\frac{x}{n})||\pmb{\langle} (\tilde{r}_x^o(nt))^2 \pmb{\rangle_{\rho}}^{\frac12} \pmb{\langle} (\tilde{r}_x^{\bullet}(nt))^2 \pmb{\rangle_{\rho}}^{\frac12} |\Big|	 \\
& \leq  \Big|\frac{1}{n}\sum_{x=1}^n \big|\frac{g(\frac{x}{n})}{m_x}\big| \pmb{\langle} (\tilde{p}_x^o(nt))^2  \pmb{\rangle_{\rho}}\Big|^{\frac12}  \Big|\frac{1}{n}\sum_{x=1}^n |\frac{g(\frac{x}{n})}{m_x}| \pmb{\langle} (\tilde{p}_x^{\bullet}(nt))^2  \pmb{\rangle_{\rho}}\Big|^{\frac12} 
\\ &+  \Big|\frac{1}{n}\sum_{x=1}^{n-1} |g(\frac{x}{n})| \pmb{\langle} (\tilde{r}_x^o(nt))^2  \pmb{\rangle_{\rho}}\Big|^{\frac12}  \Big|\frac{1}{n}\sum_{x=1}^{n-1} |g(\frac{x}{n})| \pmb{\langle} (\tilde{r}_x^{\bullet}(nt))^2  \pmb{\rangle_{\rho}}\Big|^{\frac12}  \\
& \\&  \leq \sqrt{ 2\tilde{c}_0C_1}\Bigg(\frac{1}{n}\sum_{x=1}^n |g(\frac{x}{n})| \Big( \frac{ \pmb{\langle} (\tilde{p}_x^o(nt))^2  \pmb{\rangle_{\rho}}}{m_x} + \pmb{\langle} (\tilde{r}_x^o(nt))^2 \pmb{\rangle_{\rho}} \Big) \Bigg)^{\frac12} \leq \Big(\frac{\tilde{c}_0C_1C'n^{(1-\alpha)}}{n}\Big)^{\frac12} \to 0, 
\end{aligned}
\end{equation}	
	The first inequality is obtained by \eqref{CSineq}. In the second inequality, the Cauchy-Schwartz has been used. Then by taking $\tilde{c}_0=\max \{c_0,c'_0 \}$, we take advantage of the following bounds: $\frac{1}{n}\sum_{x=1}^{n-1} \pmb{\langle} (\tilde{r}_x^{\bullet}(nt))^2 \pmb{\rangle_{\rho}} \leq c_0'$, and $\frac{1}{n}\sum_{x=1}^{n} \pmb{\langle} (\tilde{p}_x^{\bullet}(nt))^2 \pmb{\rangle_{\rho}} \leq c_0$ from \eqref{rkbound2} and \eqref{gkbound2}. Then, we used the inequality $\sqrt{a}+\sqrt{b} \leq \sqrt{2(a+b)}$. The last inequality can be obtained exactly similar to \eqref{lowmodesbound1} and \eqref{lowmodecint}, where we bounded $|\mathcal{L}_n^g(t)| \leq C'\frac{n^{(1-\alpha)}}{n}$.  \\

	Finally, we prove $|T_n^g(t) - T_n^g(0)| \to 0$, almost surely, by using \eqref{lowmodelimit}, \eqref{highmodelimit}, and \eqref{errortermlimit}. First, recall that we defined  $\tilde{p}^{o}_x(nt), \tilde{r}^{o}_x(nt)$ in \eqref{lowmodeportion},  and $\tilde{p}^{\bullet}_x(nt), \tilde{r}^{\bullet}_x(nt)$ in \eqref{highmodedef},  such that $\tilde{p}_x(nt)= \tilde{p}_x^o(nt)+\tilde{p}_x^{\bullet}(nt)$, and $\tilde{r}_x(nt)= \tilde{r}_x^o(nt)+\tilde{r}_x^{\bullet}(nt)$, for all $x \in \mathbb{I}_n$. By using the later, and comparing the definition of $T_n^g(t)$ \eqref{Tn}, $\mathcal{L}_n^g(t)$ \eqref{lowmodelimit}, $\mathcal{U}_n^g(t)$ \eqref{highmodeportion}, and $\mathcal{E}_n^g(t)$  \eqref{errorterm}, it is evident that: 
	\begin{equation}
		T_n^g(t)= \mathcal{L}_n^g(t) + \mathcal{U}_n^g(t)+\mathcal{E}_n^g(t).
	\end{equation}
	Therefore, we can conclude: 
	\begin{equation}
		\begin{aligned}		
		|T_n^g(t)-T_n^g(0)| =&|\mathcal{L}_n^g(t) + \mathcal{U}_n^g(t)+\mathcal{E}_n^g(t)-(\mathcal{L}_n^g(0) + \mathcal{U}_n^g(0)+\mathcal{E}_n^g(0))| \leq |\mathcal{U}_n^g(t)-\mathcal{U}_n^g(0)|+ \\ & |\mathcal{L}_n^g(0)|+ |\mathcal{L}_n^g(t)| +|\mathcal{E}_n^g(0)| +|\mathcal{E}_n^g(t)| \to 0,
		\end{aligned}	
	\end{equation}
	almost surely, where $|\mathcal{U}_n^g(t)-\mathcal{U}_n^g(0)| \to 0$ almost surely, by \eqref{highmodelimit}. Moreover, $|\mathcal{L}_n^g(0)|+ |\mathcal{L}_n^g(t)| \to 0$, since \eqref{lowmodelimit} is valid for any $t$, and $|\mathcal{E}_n^g(0)| +|\mathcal{E}_n^g(t)| \to 0$ as a consequence of \eqref{errortermlimit}, which holds for all $t \in [0,T]$. This finishes the proof of \eqref{Thermallimitmain} and Lemma \ref{Thermallimitlemma}. 
	\end{proof} 
	\begin{remark} \label{rmkthermaleqavg3}
		For a clean chain in thermal equilibrium, we have: $\expval{\tilde{e}_{[ny]}(nt)}_{\rho}-\expval{\tilde{e}_{[ny]}(0)}_{\rho}$ vanishes as $n \to \infty$. One can observe this by using the fact that $A_r^\beta$ and $A_r^0$ can be diagonalized in the same basis for this chain, and hence this difference can be computed explicitly. Therefore, a simple calculation shows that in this case, we can obtain \eqref{Thermallimitmain}. Notice that the rest of the proof for a clean chain in thermal equilibrium is exactly similar to the disordered case; therefore, we can obtain the \eqref{maintheorem} for this chain, with corresponding $\mathrm{f}_{\beta}$ from \eqref{thermaleqcte}.
	\end{remark}
		\subsection{Decomposition}
		
			In this section, we finish  proof of \eqref{econv} for any test function $g \in C^1([0,1])$.  We will extend this result to continuous test functions in the next section. 
			\begin{proof}[Proof of \eqref{econv} with $C^1$ test function]
			 First, we can solve the  macroscopic equation explicitly. In fact, thanks to the regularity assumption: $\bar{r}(y), \bar{p}(y) \in C^1([0,1])$, the wave equation (two first equation in \eqref{pde1}, with boundary conditions \eqref{bc1}, \eqref{bc2}) for $\mathrm{r}$ and $\mathrm{p}$ can be solved explicitly, by expanding in the Fourier basis and, we have a smooth in time,  strong solutions, such that $\mathrm{p}(y,t) , \mathrm{r}(y,t) \in C^1([0,1])$. Moreover, one can observe that the solution to the equation \eqref{pde1}, \eqref{bc1}, and \eqref{bc2} for $\mathrm{e}$ is given by: 
		\begin{equation} \label{esolution}
			\mathrm{e}(y,t)=\frac{\mathrm{p}^2(y,t)}{2\bar{m}}+\frac{\mathrm{r}^2(y,t)}{2} + \mathrm{f}^{\mu}_{\beta}(y).
		\end{equation}		 
		One can justify this solution by simply taking the derivative and plug-in the solution to the wave equation. Notice that this formal argument is legitimate, thanks to the regularity of $\mathrm{r}$ and $\mathrm{p}$.\\
		Fix $g \in C^1([0,1])$, then the LHS of \eqref{econv} is given by: 
		\begin{equation} \label{macrodecomp}
			\begin{aligned}		
			\int_0^1 g(y)\mathrm{e}(y,t)dy=\underbrace{\int_0^1 g(y)\Big(\frac{\mathrm{p}^2(y,t)}{2\bar{m}}+\frac{\mathrm{r}^2(y,t)}{2}\Big)dy}_{K_g: \text{Macroscopic Mechanical energy}} + \underbrace{\int_0^1 g(y)\mathrm{f}^{\mu}_{\beta}(y)dy}_{T_g: \text{Macroscopic Thermal energy}},
			\end{aligned}		
		\end{equation}
		where we defined $K_g(t)$ and $T_g(t) \equiv T_g$, to be the Macroscopic Mechanical energy, and macroscopic Thermal energy, respectively.   
		At the macroscopic level, we already observed the following decomposition \footnote{Notice that we come back to the notation where we denote the thermal average by $\expval{.}_{\rho}$ instead of $\pmb{\expval{.}_{\rho}}$, since there is no confusion here.}: 
		\begin{equation} \label{microdecomp}
			\begin{aligned}				
				\frac{1}{n} \sum_{x=1}^n g(\frac{x}{n}) \expval{e_x(nt)}_{\rho}	= & \underbrace{\frac{1}{n} \sum_{x=1}^n g(\frac{x}{n})\Big(\frac{\expval{(p_x(nt))^2}_{\rho}}{m_x}+\expval{(r_x(nt))^2}_{\rho}\Big)}_{K_n^g(t): \text{Microscopic Mechanical Energy}}  \\
				&+ \underbrace{\frac{1}{n}\sum_{x=1}^n g(\frac{x}{n})\Bigg(\frac{\expval{(p_x(nt)-\expval{p_x(nt)}_{\rho})^2}_{\rho}}{m_x} + \expval{(r_x(nt)-\expval{r_x(nt)}_{\rho})^2}_{\rho}\Bigg)}_{T_n^g(t): \text{Miscroscopic Thermal Energy}},
		\end{aligned}		
		\end{equation}		  
		
	where we defined $K_n^g(t)$ to be the microscopic Mechanical energy. Recalling the definition $\tilde{p}_x(nt)= p_x(nt)-\expval{p_x(nt)}_{\rho}$, and $\tilde{r}_x(nt)-\expval{r_x(nt)}_{\rho}$, the second term $T_n^g(t)$ is the microscopic Thermal energy, which is defined in \eqref{Tn}.	\\ 
	Comparing these two expressions, \eqref{macrodecomp} and \eqref{microdecomp}, the rest of the proof becomes clear. In step one, we proved that the microscopic Mechanical energy converges to the macroscopic counterpart: $K_n^g(t) \to K_g(t)$ in \eqref{kinetic1}. In the previous section, we proved that the microscopic Thermal energy is frozen in time i.e. $T_n^g(t)-T_n^g(0) \to 0$, in \eqref{Thermallimitmain}. Finally, in Section \ref{SLLNsection}, we proved that $T_n^g(0) \to T_g$, in \eqref{e0conv1}. All these limit are almost surely w.r.t $\mathbb{P}$. Combining these three argument finishes the proof: 
	\begin{equation} \label{mainlimitregular}
		\begin{aligned}		
		&\frac{1}{n} \sum_{x=1}^n g(\frac{x}{n}) \expval{e_x(nt)}_{\rho}= K_n^g(t)+T_n^g(t)= K_n^g(t)+T_n^g(0) +(T_n^g(t)-T_n^g(0)) \to K_g(t)+T_g +0 = \\
		&\int_0^1 g(y) \Big(\Big(\frac{\mathrm{p}^2(y,t)}{2\bar{m}}+\frac{\mathrm{r}^2(y,t)}{2}\Big) + \mathrm{f}^{\mu}_{\beta}(y)\Big)dy = \int_0^1 g(y)\mathrm{e}(y,t)dy,
		\end{aligned}	
	\end{equation}
	almost surely, where these three limits have been deduced from \eqref{kinetic1}, \eqref{e0conv1} and \eqref{Thermallimitmain} respectively. This finishes the proof of \eqref{econv} for $g \in C^1([0,1])$.
	 \end{proof}
	 Notice that the only limit among those, where we needed the stronger assumption $g \in C^1([0,1])$, rather than $g \in C^0([0,1])$, was the last one: 	$T_n^g(t)-T_n^g(0) \to 0$. We will circumvent this obstacle in the next section, using the energy estimate: $\frac{1}{n}\sum_{x=1}^n \expval{e_x(nt)}_{\rho} \leq C$. 
	 
	 \subsection{From $C^1$ to $C^0$}
	 
 	This section extends the previous result and omits the additional regularity assumption on $f$ and finish the proof of Theorem \ref{maintheorem}. As we already mentioned, the essential tool for this purpose is the following energy estimate: There exists a constant $C>0$ independent of $n$, such that for all $t \in [0,T]$: 
 	\begin{equation} \label{energyestimate}
 		\frac{1}{n} \sum_{x=1}^n \expval{e_x(nt)}_{\rho} \leq C.
	\end{equation} 	 
The later is rather straightforward due to our previous calculations. First, notice that \\ $\sum_{x=1}^n e_x(nt) = H(nt) = \sum_{x=1}^n e_x(0)=: \sum_{x=1}^n e_x$ is conserved by the evolution, hence we have: $\frac{1}{n} \sum_{x=1}^n \expval{e_x(nt)}_{\rho}= \frac{1}{n} \sum_{x=1}^n \expval{e_x}_{\rho}$. On the other hand we have: 
\begin{equation} \label{energyestimate2}
\begin{aligned}
\frac{1}{n}\sum_{x=1}^n \expval{e_x}_{\rho}=\frac{1}{n}\sum_{x=1}^n \Big(\frac{\expval{p_x}_{\rho}^2}{2m_x} + \frac{\expval{r_x}_{\rho}^2}{2}\Big) + \frac{1}{n}\sum_{x=1}^n\Big(\frac{\expval{\tilde{p}_x^2}_{\rho}}{2m_x} + \frac{\expval{\tilde{r}_x^2}_{\rho}}{2}\Big),
\end{aligned}
\end{equation} 		
	where $\tilde{p}_x= p_x -\expval{p_x}_{\rho}$, $\tilde{r}_x=r_x-\expval{r_x}_{\rho}$. 
	 The first sum in \eqref{energyestimate2}, is bounded by a constant $C_1$, since we have  $\expval{p_x}_{\rho} =\frac{m_x}{\bar{m}}\bar{p}_x -\mathscr{E}_x^n$, $\expval{r_x}_{\rho} = \bar{r}_x$ from \eqref{momentumelongationaverage2}, where $\bar{p}_x=\bar{p}(\frac{x}{n})$, $\bar{r}_x=\bar{r}(\frac{x}{n})$, and $\mathscr{E}_x^n$ is given in \eqref{ERROR!}.  Notice that we used the fact that $|\mathscr{E}_n^x|$ is bounded by a deterministic constant, which is evident from its definition and properties of $m_x$, $\beta$ and $\bar{p}$.  \\ 
	
	 The second sum in \eqref{energyestimate2}, is bounded by $C_2$ independent of $n$, which is given in \eqref{thermalboundinitial} in Remark \ref{themalboundinitialremark}.  Therefore $\frac{1}{n} \sum_{x=1}^n \expval{e_x}_{\rho} $ is bounded by another deterministic constant $C$, uniform in $n$. \\
	
	 \begin{proof}[Proof of \eqref{econv} for $f \in C^0$] 
	 After all, we can state the proof of \eqref{econv} for a test function $f \in C^0([0,1])$. Fix $f \in C^0([0,1])$ and recall the definition of the regularizing family $\zeta_{\epsilon}$, for any $\epsilon>0$ as follows: Let  $\zeta \in C^{\infty}_c (\mathbb{R})$, with $(\zeta)=[-1,1]$, $\int_{\mathbb{R}} \zeta(y')dy'=1$, $\zeta(y) \geq 0 $, and $\zeta(y)=\zeta(-y)$. Let $\zeta_{\epsilon}:= \frac{1}{\epsilon}\zeta(\frac{y}{\epsilon})$, for $0<\epsilon< 1$. Notice that we have $\zeta_{\epsilon} \in C^{\infty}_c(\mathbb{R})$, $supp(\zeta_{\epsilon})=[-\epsilon,\epsilon]$, $\int_{\mathbb{R}} \zeta_{\epsilon}(y')dy'=1$, and $\zeta_{\epsilon}(y)\geq 0$ for any $0<\epsilon<1$.\\
	  For any $\delta>0$, let $f_{\delta} :[0,1] \to \mathbb{R}$ be defined as $f_{\delta}(y):=(f*\zeta_{\delta})(y)= \int f(y')\zeta(y-y')dy'$. By properties of convolution, we have $f_{\delta} \in C^1([0,1]), \forall \delta>0$.\\ 
	  Fix $t \in [0,T]$, and  $\epsilon>0$, we are going to introduce  a proper $\delta$:
	  Recall the solution to the macroscopic equation $\mathrm{e}(y,t)=\frac{\mathrm{p}^2(y,t)}{2\bar{m}}+\frac{\mathrm{r}^2(y,t)}{2} + \mathrm{f}^{\mu}_{\beta}(y)$, as we already observed $\frac{\mathrm{p}^2(y,t)}{2\bar{m}}+\frac{\mathrm{r}^2(y,t)}{2} $ is continuous, thanks to the regularity assumption on $\bar{r}$ and $\bar{p}$. Moreover, we proved the continuity of $ \mathrm{f}^{\mu}_{\beta}(y)$ in Proposition \ref{fcont}, hence $\mathrm{e}(y,t)$ is positive and continuous on $[0,1]$ (positivity of $\mathrm{f}^{\mu}_{\beta}(y)$ is evident from the construction), and $\int_0^1 \mathrm{e}(y,t)$ is bounded by a constant $C_0>0$. Let $\tilde{C}= \max \{C_0 ,C \}$, where $C$ is the uniform bound on $\frac{1}{n} \sum_{x=1}^n \expval{e_x}_{\rho}$. Since $f$ is continuous  on $[0,1]$, it will be uniformly continuous, so there exists $\delta(\epsilon)>0$, such that if $y,y' \in [0,1]$, $|y-y'|\leq 2\delta$, then $|f(y)-f(y')|\leq \frac{\epsilon}{4\tilde{C}}$. This choice of $\delta$, besides the properties   of $\zeta_{\delta}$, in particular, the fact that $supp(\zeta_{\delta})=[-\delta,\delta]$, and $\int \zeta_{\delta}(y) dy =1$, lead to the following estimate: $\sup_{y}|f_\delta(y) -f(y)| \leq \frac{\epsilon}{4\tilde{C}} $, this can be observed from the definition of $f_{\delta}$:
	  
		$$|(f-f_{\delta})(y)|= \Big|f(y)-\int f(y')\zeta_{\delta}(y-y')dy'\Big|= \Big|\int f(y) \zeta(y-y')dy' - \int f(y')\zeta_{\delta}(y-y')dy' \Big| \leq $$
		$$ \int |f(y)-f(y')| \zeta_{\delta}(y-y')dy' = \int_{y-\delta}^{y+\delta} |f(y)-f(y')| \zeta_{\delta}(y-y')dy' \leq \sup_{|y-y'| \leq 2\delta}(|f(y)-f(y')|) \leq \frac{\epsilon}{4\tilde{C}}.  $$ 
	 Notice that this bound is valid for any $y$, so we can take the $\sup$ over $y$.  By using the later, again, thanks to the choice of $\delta$, definition of $\zeta_{\delta}$, and $\tilde{C}$ we obtain 
		\begin{equation} \label{estimateregural2}
		\begin{aligned}	
	&\Big|\frac{1}{n}\sum_{x=1}^n (f-f_{\delta})(\frac{x}{n}) \expval{e_x(nt)}_{\rho}\Big| \leq \sup_{y} |(f-f_{\delta})(y)| \frac{1}{n} \sum_{x=1}^n \expval{e_x(nt)}_{\rho} \leq \frac{\epsilon}{4\tilde{C}} C \leq \frac{\epsilon}{4},
		\\ &\Big|\int_0^1 (f-f_{\delta})(y)\Big| \mathrm{e}(y,t)dy  \leq \sup_{y} |(f-f_{\delta})(y)| \int _0^1 \mathrm{e}(y,t)dy \leq \frac{\epsilon}{4\tilde{C}} C_0 \leq \frac{\epsilon}{4},
		\end{aligned}		
		\end{equation}

	uniformly in $n$. Now let us bound: 
	 \begin{equation}
	 \begin{aligned}
		&\Big|\frac{1}{n} \sum_{x=1}^n f(\frac{x}{n}) \expval{e_x(nt)}_{\rho} -\int_0^1 f_{\delta}(y) \mathrm{e}(y,t)dy	 \Big| = \Big|(\frac{1}{n} \sum_{x=1}^n (f(\frac{x}{n}) \expval{e_x(nt)}_{\rho} + \expval{e_x(nt)}_{\rho}(f-f_{\delta})(\frac{x}{n}) ) - \\ &(\int_0^1 f_{\delta}(y)\mathrm{e}(y,t) + \int_0^1 (f-f_{\delta})(y) \mathrm{e}(y,t)dy)\Big| \leq  \\
		& \Big|\frac{1}{n} \sum_{x=1}^n f_{\delta}(\frac{x}{n}) \expval{e_x(nt)}_{\rho} - \int_0^1 f_{\delta}(y) \mathrm{e}(y,t)dy\Big| + \Big|\frac{1}{n}\sum_{x=1}^n (f-f_{\delta})(\frac{x}{n}) \expval{e_x(nt)}_{\rho}\Big| + \\ &\Big|\int_0^1 (f-f_{\delta})(y) \mathrm{e}(y,t)dy\Big|  \leq \frac{\epsilon}{2}+\frac{\epsilon}{4} + \frac{\epsilon}{4}.
	 \end{aligned}
	 \end{equation}
	 
The last two term were bounded by \eqref{estimateregural2}. In order to treat the first term, note that as we observed $f_{\delta} \in C^1([0,1])$, so we can use the result of the previous section i.e. \eqref{mainlimitregular} for $f_{\delta}$:
	  \begin{equation}
	   \lim_{n \to \infty} \frac{1}{n} \sum_{x=1}^n f_{\delta}(\frac{x}{n}) \expval{e_x(nt)}_{\rho} \to \int_0^1 f_{\delta}(y) \mathrm{e}(y,t)dy,
	  \end{equation}
	   almost surely. Hence, there exists $N$, such that for $n>N$, we have $|\frac{1}{n} \sum_{x=1}^n f_{\delta}(\frac{x}{n}) \expval{e_x(nt)}_{\rho} - \int_0^1 f_{\delta}(y) \mathrm{e}(y,t)dy|  \leq \frac{\epsilon}{2}$, almost surely, and this finishes the proof of Theorem \ref{maintheorem}.
\end{proof}	  

		\newpage
		\appendix
		
		\section{Properties of the Limiting Function } \label{app1}    

In this section, we study the properties of $\mathrm{f}^{\mu}_{\beta}$, which is defined in \eqref{fdef}. We prove a couple of lemmas to facilitate the proof of Theorem \ref{SLLN}. In particular, we  prove the existence of the limit \eqref{fdef}, this means that $\mathrm{f}^{\mu}_{\beta}$ is well-defined. Moreover, we show that this function is continuous. Furthermore, we treat the equilibrium case and demonstrate that the function $\mathrm{f}^{\mu}(\beta_{eq})$ is well defined, i.e. $\mathrm{f}^{\mu}_{\beta_{eq}}(y)$ does not depend on $y$. Then we demonstrate the relation \eqref{stefano}. Finally, we prove a lemma which has been needed in \ref{SLLN}. All the proofs in this section share the same spirit: We represent $\expval{\tilde{e}_x}_{\rho}$ in terms of the Taylor series, then cut the series similar to \eqref{taylorcut}, and control the expressions depending on the first part of the Taylor series, using the fact that the number of these terms is finite,  $\beta$ is continuous and bounded, and the distribution of the masses is compactly supported. Finally, we use the fact that the terms depending on the remainder of the series is small.   

 Notice that since  $n$ is not fixed here, we denote the ensemble average with $\expval{}_{\rho^n}$, in order to emphasize the dependence on $n$. 
Before proceeding,  recall the definition of  $A_p^{\beta}$,  and  $A_r^{\beta}$ from \eqref{thermalmatrices}, since here $n$ is not fixed and we study matrices with different sizes, we change our notations only in this section and denote these matrices by $A_n^p$, and $A_n^r$, respectively i.e. $$A_n^p:=M_{\beta}^{-\frac12}(-\nabla_-\beta^0 \nabla_+)M_{\beta}^{-\frac12}, \quad A_n^r:=(\beta^o)^{\frac12}(-\nabla_+ M_{\beta}^{-1} \nabla_-)(\beta^o)^{\frac12},$$  where $M_{\beta}=M\beta^{-1}$, with $M=\diag(m_1,\dots,m_n)$, $\beta=\diag(\beta(\frac{1}{n}),\dots,\beta(\frac{n}{n}))$, and  $\beta^o=\diag(\beta(\frac{1}{n}),\dots,\beta(\frac{n-1}{n}))$. \\
	   Since all these proofs share the same spirit, we set  a handful of notation here:
	 
	 Recall the average expression $\expval{\tilde{e}_x}_{\rho^n}$ from \eqref{averageintermoff}:
	 \begin{equation} \label{app22}
	 \expval{\tilde{e}_x}_{\rho} = \frac{1}{\beta_x}\big(\langle x,\mathfrak{f}(A_n^r) x \rangle_{n-1} + \langle x, \mathfrak{f}(A_n^p) x \rangle_n +\epsilon_n^x\big).
	 \end{equation}
	Thanks to \eqref{Taylorexpansionmatrix}, the part corresponding to $p$ can be written as follows: 
	\begin{equation} \label{app23}
		\langle x , \mathfrak{f}(A_n^p) x\rangle_n= \sum_{k=0}^{\infty} a_k \langle x, (A_n^p-\alpha I_n)^k x \rangle_n. 
	\end{equation}
	As we argued in Section \ref{SLLNsection}, there exists a constant $c_0>0$, such that for every realization of the masses, and $\forall n$,  $||A_p^n||_2 \leq c_0$. Therefore, using the properties of Taylor series as in \eqref{taylorremainder}, we observe that $\forall \epsilon>0$, there exists $K_*(\epsilon) \in \mathbb{N}$, such that $\forall n $,  $\forall x \in \mathbb{I}_n$, and any realization of the masses we have: 
	\begin{equation} \label{app24}
	\Big|\sum_{k>K_*(\epsilon)} a_k\langle x, (A_n^p-\alpha I_n)^k x \rangle_n \Big| \leq \epsilon.
\end{equation}		
	Notice that we can choose $K_*(\epsilon)$, such that the same bound \eqref{app24} holds when we substitute $A_n^p$ with $A_n^r$. Hence, given $\epsilon>0$ one can define $\mathfrak{f}_{\prec}^{\epsilon}(.)$ and $\mathfrak{f}_{\succ}^{\epsilon}(.)$ as follows:	
	\begin{equation} \label{fepsilon}
		\mathfrak{f}_{\prec}^{\epsilon}(A_n^p):= \sum_{k=0}^{K_*(\epsilon)} a_k(A_n^p-\alpha I_{n})^k,  \qquad \mathfrak{f}_{\succ}^{\epsilon}(A_n^p):= \sum_{k>K_*(\epsilon)}^{\infty} a_k(A_n^p-\alpha I_{n})^k.
	\end{equation}
In particular, we can rewrite \eqref{app24} in the following way: For any $\epsilon>0$ there exits $K_*(\epsilon)$ such that for any $n$ and $x \in \mathbb{I}_n$ we have:
\begin{equation} \label{taylorbound}
\begin{split}
	&|\langle x, \mathfrak{f}^{\epsilon}_{\succ}(A_n^p) x \rangle_n| \leq \epsilon, \quad |\mathbb{E}(\langle x, \mathfrak{f}^{\epsilon}_{\succ}(A_n^p) x \rangle_n)| \leq \epsilon, \\
	&|\langle x, \mathfrak{f}^{\epsilon}_{\succ}(A_n^r) x \rangle_{n-1}| \leq \epsilon, \quad |\mathbb{E}(\langle x, \mathfrak{f}^{\epsilon}_{\succ}(A_n^r) x \rangle_{n-1})| \leq \epsilon.
\end{split}
\end{equation}
	Fix $(k,n) \in \mathbb{N}^2$, denote $A_n^p-\alpha I_{n}$ by $\tilde{A}_n^p$,  take $x \in \mathbb{I}_n$, and consider the following term: $\langle x,(\tilde{A}_n^p)^k x\rangle$. Here we represent this term in a more appropriate manner, introducing following notations. First, recall the random walk representation of 	$\langle x,(\tilde{A}_n^p)^k x\rangle$: 
\begin{equation} \label{rwrepapp}
		\begin{split}
				\langle x, (\tilde{A}_n^p)^k x \rangle=\sum_{x_1,\dots x_{k-1}=1}^n \langle x,\tilde{A}_p^{\beta} x_1 \rangle \langle x_1 ,\tilde{A}_p^{\beta} x_2 \rangle \dots \langle x_{k-1} ,\tilde{A}_p^{\beta} x \rangle.
		\end{split}
	\end{equation}
	Denote the set of indices with non-zero contribution, in RHS of \eqref{rwrepapp} by $\mathcal{I}_{n,k}^{x,p}$: 
\begin{equation} \label{Idef}
\mathcal{I}_{n,k}^{x,p}:= \{ (x_1,\dots x_{k-1}) \in \mathbb{I}_n^{k-1}  |  \langle x,(A^p_n-\alpha I_n) x_1 \rangle \langle x_1 ,(A^p_n-\alpha I_n x_2 \rangle \dots \langle x_{l-1} ,(A^p_n-\alpha I_n) x \rangle \neq 0 \},
\end{equation} 
	and denote the element of $\mathcal{I}_{n,k}^{x,p}$ by $\underline{x}:=(x_1,\dots,x_{k-1})$. Notice that there is a bijection between $\mathcal{I}_{n,k}^{x,p}$ and the set of paths in $[0,k] \times [0,n] \cap \mathbb{Z}^2$ from the point $(x,0)$ to $(x,k)$, consisting of the following vectors: $(-1,1),(0,1),(1,1) \in \mathbb{Z}^2$. In fact, this bijection is given as follows: $\forall \underline{x} \in \mathcal{I}_{n,k}^{x,p}$, assign to $\underline{x}$ the path which is given by the following points: $(0,x)$, $(1,x_1)$,$\dots$, $(j,x_j)$, $\dots$,$(k-1,x_{k-1}),(k,x)$.  
	
	Let $\tilde{\mathcal{I}}_{n,k}^{x,p}$ be the set where every element of $\mathcal{I}_{n,k}^{x,p}$ shifted by the vector $(x,\dots,x) \in (\mathbb{I}_n)^{k-1}$, precisely define: 
	\begin{equation} \label{Idef2}
		\begin{split}		
		\tilde{\mathcal{I}}_{n,k}^{x,p} := \{(x_1-x,\dots,x_{k-1}-x) | \underline{x} \in \mathcal{I}_{n,k}^{x,p} \}.
		\end{split}	
		\end{equation}	 
	We denote each element of $\tilde{\mathcal{I}}_{n,k}^{x,p}$ by $\underline{\eta}=(\eta_1,\dots, \eta_{k-1})$. Notice that $\underline{\eta} $ corresponds to a path from $(0,0)$ to $(0,k)$ in $[0,k]\times[-x,n-x] \cap \mathbb{Z}^2$ consisting of the aforementioned vectors.  Finally, define  $\tilde{\mathcal{I}}_k$ as follows:
	\begin{equation}
\begin{split}				
					\tilde{\mathcal{I}}_k:= \{ \underline{\eta} \in \mathbb{Z}^{k-1}| \forall i \in \mathbb{I}_{k}, \: |\eta_i-\eta_{i-1}| \leq 1,\:  \text{with } \eta_0=\eta_k=0\}.	
\end{split}	
	\end{equation}
	Observe that each $\underline{\eta} \in \mathcal{I}_k$,  corresponds to a path from $(0,0)$ to $(0,k)$ in $\mathbb{Z}^2$ consisting of  $(-1,1),(0,1),(1,1) \in \mathbb{Z}^2$. \\
	Having in mind the geometric interpretation of $\tilde{\mathcal{I}}_{n,k}^{x,p}$ and $\tilde{\mathcal{I}}_k$, one can observe\footnote{This bound is obvious, since $|\tilde{\mathcal{I}}_k|$ is the solution to the problem of the number of path from $(0,0)$ to $(k,0)$ consisting of the vectors $(1,1),(1,-1),(1,0)$, this is equivalent to the number of solution of $s_1+\dots+s_k=0$, for $s_i \in \{-1,0,1\}$, which is obviously bounded by $3^k$. This bound is not sharp and a better asymptotic will be $\frac{c3^k}{\sqrt{k}}$ for $c$ around $\frac12$, but $3^{n_*}$ is sufficient for our purposes.}: 
	\begin{equation} \label{setcardinal}
		\begin{split}		
		&\tilde{\mathcal{I}}_{n,k}^{x,p} \subset \tilde{\mathcal{I}}_k, \quad|\tilde{\mathcal{I}}_{n,k}^{x,p}| \subset |\tilde{\mathcal{I}}_k| \leq 3^k,
	\\
		&\tilde{\mathcal{I}}_{n,k}^{x,p} = \tilde{\mathcal{I}}_k \quad \text{iff} \quad [\frac{k}{2}]\leq x \leq n-[\frac{k}{2}].
		\end{split} 	
 	\end{equation}
	By using \eqref{Idef2}, we can rewrite the sum in \eqref{rwrepapp}, as a sum over the set $\tilde{\mathcal{I}}_{n,k}^{x,p}$. Here we introduce a set of notations in order to rewrite each term in  \eqref{rwrepapp} in a more suitable way. \\
	Fix $k \in \mathbb{N}$, and take $\underline{\eta} \in \tilde{\mathcal{I}_k}$, then consider the set of indices $\underline{j}=(j_{-[\frac{k}{2}]},\dots,j_0,\dots, j_{[\frac{k}{2}]+1}) \in \mathbb{Z}^{2[\frac{k}2]+2}$, where $j_i\neq j_{i'}$ for $i \neq {i'}$. Correspondingly, let $\underline{m}^k$ denotes a vector of $2[\frac{k}{2}]+2$ masses indexed by $\underline{j}$, i.e.  $\underline{m}^k=(m_{j_{-[\frac{k}{2}]}},\dots,m_{j_{[\frac{k}{2}]+1}})$, notice that we extended the set of i.i.d random variables $\{ m_x \}_{x=1}^n$ to the set of i.i.d random variables $\{m_x\}_{x \in \mathbb{Z}}$\footnote{This extension is not necessary, it is done to make our notation coherent; however we do not use of this extension.}. Moreover, let $\underline{b}^k \in [\beta_{min},\beta_{max}]^{2[\frac{k}{2}]+2}$ denotes the following vector: $\underline{b}^k=(b_{-[\frac{k}{2}]},\dots,b_{[\frac{k}{2}]+1})$. We define $\mathcal{F}_{k,\underline{\eta}}(\underline{m}^k,\underline{b}^k)$ as follows:
	
	\begin{equation} \label{FFdef}
		\begin{split} 	
 	&\mathcal{F}_{k,\underline{\eta}}(\underline{m}^k,\underline{b}^k)= \theta_{\eta_0\eta_1} \dots \theta_{\eta_i \eta_{i+1}} \dots \theta_{\eta_{k-1}\eta_k}, \\
 	&  \theta_{\eta_i\eta_{i+1}} = 
 	\begin{cases}
		b_{\eta_i}(\frac{b_{\eta_i}}{m_{j_{\eta_i}}}+\frac{b_{(\eta_i+1)}}{m_{j_{(\eta_i+1)}}}) -\alpha \quad \text{if} \quad \eta_i=\eta_{i+1},	\\
	-\frac{b_{\hat{\eta_i}}}{m_{j_{\hat{\eta}_i}}} \sqrt{b_{\hat{\eta}_i}b_{(\hat{\eta}_i+1)}} \quad \text{if} \quad \eta_i \neq \eta_{i+1},
	\end{cases} 
		\end{split}	
	\end{equation} 	  
	where we denoted $\hat{\eta_i}=\min \{ \eta_i,\eta_{i+1} \}$, and $\eta_0=\eta_k=0$. The following properties of $\mathcal{F}_{k,\underline{\eta}}$ are straightforward, since the distribution of the masses is compactly supported and $\mathcal{F}_{k,\underline{\eta}}$ is continuous on a compact set:
	\begin{itemize}
	\item $\mathcal{F}_{k,\underline{\eta}}(\underline{m}^k,\underline{b}^k)$ is uniformly continuous in the second component, uniformly in the distribution of the masses. More precisely, $\forall \epsilon>0$, there exists $\delta>0$ such that if $|\underline{b}^k_1-\underline{b}^k_2|< \delta$,\footnote{Here $|.|$ denotes the Euclidean distance in $\mathbb{R}^{2[\frac{k}{2}]+2}$.} then for any realization of the masses,    $|\mathcal{F}_{k,\underline{\eta}}(\underline{m}^k,\underline{b}^k_1)-\mathcal{F}_{k,\underline{\eta}}(\underline{m}^k,\underline{b}^k_2)| \leq \epsilon$.
	\item  Consider two different set of masses $\underline{m}^k_1$ and $\underline{m}^k_2$, since masses are i.i.d, we have: $\mathbb{E}(\mathcal{F}_{k,\underline{\eta}}(\underline{m}^k_1,\underline{b}^k))=\mathbb{E}(\mathcal{F}_{k,\underline{\eta}}(\underline{m}^k_2,\underline{b}^k))$. Notice that we used the assumption $j_i \neq j_{i'}$ for $i \neq i'$. 	 
	\item From the above properties, it is clear that taking two set of masses $\underline{m}^k_1$ and $\underline{m}^k_2$ we have: $\forall \epsilon>0$, there exists $\delta>0$ such that if $|\underline{b}^k_1-\underline{b}^k_2|< \delta$, then $|\mathbb{E}(\mathcal{F}_{k,\underline{\eta}}(\underline{m}^k,_1\underline{b}^k_1))-\mathbb{E}(\mathcal{F}_{k,\underline{\eta}}(\underline{m}^k_2,\underline{b}^k_2))|< \epsilon$.
\end{itemize}	 
	
	 Recall that we fixed $k$ and $\underline{\eta} \in \tilde{\mathcal{I}}_k$. Let us take $n \in \mathbb{N}$ and $x \in \mathbb{I}_n$, then we define the vectors $\underline{m}^k(x,n)$, $\underline{b}^k(x,n)$ as follows: 
	 \begin{equation} \label{mbdef}
	 \begin{split}
	 &\underline{m}^k(x,n)_{j_i}= m_{x+i}, \quad -[\frac{k}{2}] \leq i \leq [\frac{k}{2}]+1, \quad  \\
	 & \underline{b}^k(x,n)_i= \begin{cases} 
	 \beta(\frac{x+i}{n}) \quad \text{if  } 0\leq x+i \leq n, \\
			\beta(1) \quad \text{if  } n<x+i,\\
			\beta(0) \quad  \text{if  } 0<x+i,
	 \end{cases} \quad	-[\frac{k}{2}] \leq i \leq [\frac{k}{2}]+1.
	\end{split}
\end{equation}	  
	
	Finally, by combining the above notations and definitions, in particular \eqref{Idef2}, \eqref{FFdef}, and \eqref{mbdef}, we end-up with the following identity for $[\frac{k}{2}]+1 \leq x \leq n- [\frac{k}{2}]+1$: 
	\begin{equation}\label{rwrepF}
			\langle x, (\tilde{A}^p_n)^k x \rangle_n= \sum_{\underline{\eta} \in \tilde{\mathcal{I}}_{n,k}^{x,p}} \mathcal{F}_{k,\underline{\eta}}\big(\underline{m}^k(x,n),\underline{b}^k(x,n)\big)=\sum_{\underline{\eta} \in \tilde{\mathcal{I}}_k} \mathcal{F}_{k,\underline{\eta}}\big(\underline{m}^k(x,n),\underline{b}^k(x,n)\big),	
	\end{equation}

where we have the second equality thanks to \eqref{setcardinal} and the choice of $x$. 	Notice that to check this identity one should compare the definition of $\mathcal{F}_{k,{\eta}}$ with the definition of the matrix $\tilde{A}_n^p$ in \eqref{aprep}. Moreover, this identity holds for $x<[\frac{k}{2}]+1$ and $x>n-[\frac{k}{2}]+1$ if one slightly modifies the definition of $\mathcal{F}_{k,\underline{\eta}}$. However, the current form of this identity is sufficient for our purposes. It is worth mentioning that $\langle x, (\tilde{A}^r_n)^k x \rangle_{n-1}$ can be written in the similar fashion, where one should define $\tilde{\mathcal{F}}_{k,\underline{\eta}}$ similar to $\mathcal{F}_{k,\underline{\eta}}$. One can check that $\tilde{\mathcal{F}}_{k,\underline{\eta}}$ has  the three aforementioned properties. Since this task is rather straightforward, we only treat the terms corresponding to $p$ and the terms corresponding to $r$ can be treated similarly.\\
	Thanks to \eqref{fepsilon}  and \eqref{rwrepF}, we can establish the existence of the following limit:  $ \lim_{n \to \infty} \mathbb{E}(\expval{\tilde{e}_{[ny]}}_{\rho^n})$, for every $y \in [0,1]$. We prove this fact by showing that the sequence $\mathbb{E}(\expval{\tilde{e}_{[ny]}}_{\rho^n})$ is a Cauchy sequence.
\begin{lemma} \label{cauchyseqofexpectation}
		Recall the assumption on the distribution of the masses, where $\mu(x)$ is smooth  and supported on $[m_{min},m_{max}]$, $0< m_{min}< m_{max} < \infty$. We have $\forall y \in [0,1]$, and $\forall \epsilon>0$, there exists $N_0$, such that $\forall n,l>N_0$, we have $|\mathbb{E}(\expval{\tilde{e}_{[ny]}}_{\rho^n})-\mathbb{E}(\expval{\tilde{e}_{[ly]}}_{\rho^l})| \leq \epsilon$. 
\end{lemma}
	\begin{proof}
		Take $\epsilon>0$, and recall \eqref{app22} then we have: 
		\begin{equation}\label{app71}
\begin{split}			
			&\big|\mathbb{E}\big(\expval{\tilde{e}_{[ny]}}\big)-\mathbb{E}\big(\expval{\tilde{e}_{[ly]}}\big)\big|= \Big|\mathbb{E}\Big(\frac{1}{\beta_{[ny]}}\big(\langle [ny],\mathfrak{f}(A_n^r) [ny] \rangle_{n-1} + \langle [ny], \mathfrak{f}(A_n^p) [ny] \rangle_n + \epsilon^{[ny]}_n\big)\Big)-
			\\ & \mathbb{E}\Big(\frac{1}{\beta_{[ly]}}\big(\langle [ly],\mathfrak{f}(A_n^r) [ly] \rangle_{l-1} + \langle [ly], \mathfrak{f}(A^p_l) [ly] \rangle_l + \epsilon^{[ly]}_l\big)\Big)\Big|.
\end{split}		
		\end{equation} 
Since $\beta$ is continuous, and $0<\beta_{min} \leq \beta(y)\leq \beta_{max}$, we have $\beta_{[ny]}=\beta(\frac{[ny]}{n})$ and $\beta(\frac{[ly]}{l})$ are sufficiently close. In addition, $\mathfrak{f}(A^r_n)$ ,$\mathfrak{f}(A^p_n)$ are uniformly bounded in $n$. Hence, it is enough to show 
$$ \big|\mathbb{E}\big(\langle [ny],\mathfrak{f}(A_n^r) [ny] \rangle_{n-1} + \langle [ny], \mathfrak{f}(A_n^p) [ny] \rangle_n + \epsilon^{[ny]}_n\big)-
			 \mathbb{E}\big(\langle [ly],\mathfrak{f}(A_l^r) [ly] \rangle_{l-1} + \langle [ly], \mathfrak{f}(A^p_l) [ly] \rangle_l + \epsilon^{[ly]}_l\big)\big|<\epsilon,$$ 
			 for $N_0$ large enough. But the terms $|\epsilon^{[ny]}_n|$ and $|\epsilon^{[ly]}|$ are bounded by $\frac{\mathcal{C}}{n}$ and $\frac{\mathcal{C}}{l}$, respectively. Therefore, for $N_0$ large enough, they will be small, and it is enough to show:
			 $$\big|\mathbb{E} \big( \langle [ny],\mathfrak{f}(A_n^r) [ny] \rangle_{n-1} + \langle [ny], \mathfrak{f}(A_n^p) [ny] \rangle_n \big) -
			 \mathbb{E} \big( \langle [ly],\mathfrak{f}(A_l^r) [ly] \rangle_{l-1} + \langle [ly], \mathfrak{f}(A^p_l) [ly] \rangle_l \big)\big|<\epsilon,$$ for proper $N_0$. Actually, we prove that there exists $N_0$, such that for $n,l>N_0$, 
\begin{equation} \label{app8}
\big|\mathbb{E}\big(\langle [ny],\mathfrak{f}(A_n^p) [ny] \rangle_{n}\big) -
			 \mathbb{E}\big(\langle [ly],\mathfrak{f}(A_n^p) [ly] \rangle_{l}\big) |<\epsilon.
\end{equation}			
			 The term $|\mathbb{E}( \langle [ny], \mathfrak{f}(A_n^r) [ny] \rangle_{n-1} - \mathbb{E}( \langle [ly], \mathfrak{f}(A^r_l) [ly] \rangle_{l-1}|$ can be treated exactly the same way.	
In order to demonstrate \eqref{app8}, recall the definition of $\mathfrak{f}_{\prec}^{\frac{\epsilon}{4}}(A_n^p)$ and $\mathfrak{f}_{\succ}^{\frac{\epsilon}{4}}(A_n^r)$  from the expression \eqref{fepsilon}, and recall $K_*(\frac{\epsilon}{4})$, which is given in this definition. Taking advantage of \eqref{taylorbound}, we get $$\Big|\mathbb{E}\Big(\langle [ny] , \mathfrak{f}_{\succ}^{\frac{\epsilon}{4}}(A_n^p) [ny] \rangle_n\Big)-\mathbb{E}\Big(\langle [ly] , \mathfrak{f}_{\succ}^{\frac{\epsilon}{4}}(A_l^p) [ly] \rangle_l\Big)\Big| \leq \frac{\epsilon}{2}.$$ 
By using the fact that $\mathfrak{f}(A_n^p)=\mathfrak{f}_{\prec}^{\frac{\epsilon}{4}}(A_n^p)+\mathfrak{f}_{\succ}^{\frac{\epsilon}{4}}(A_n^p)$, and $\mathfrak{f}(A_l^p)=\mathfrak{f}_{\prec}^{\frac{\epsilon}{4}}(A_l^p)+\mathfrak{f}_{\succ}^{\frac{\epsilon}{4}}(A_l^p)$, it is sufficient to prove that for $n,l>N_0$:
\begin{equation} \label{app10}
	\Big|\mathbb{E}\Big(\langle [ny] , \mathfrak{f}_{\prec}^{\frac{\epsilon}{4}}(A_n^p) [ny] \rangle\Big)-\mathbb{E}\Big(\langle [ly] , \mathfrak{f}_{\prec}^{\frac{\epsilon}{4}}(A_l^p) [ly] \rangle\Big)\Big| \leq \frac{\epsilon}{2}.
\end{equation}
Since $K_*(\frac{\epsilon}{4})$ is independent of $n$ and $l$, it is enough to prove that $\forall \tilde{\epsilon}>0$, there exist $N_0$ such that $\forall k \in \{0,\dots,K_*(\frac{\epsilon}{4})\}$ and $ \forall n,l>N_0$:
\begin{equation} \label{app11}
	\big|\mathbb{E}\big(\langle [ny] , (A_n^p-\alpha I_{n})^k [ny] \rangle\big)-\mathbb{E}\big(\langle [ly] , (A_l^p-\alpha I_{l})^k [ly] \rangle\big)\big|  <\tilde{\epsilon}.
\end{equation}
	Then, taking $\tilde{\epsilon}=\frac{\epsilon}{2cK_*(\frac{\epsilon}{4})}$, where $c$ is the bound on $|a_0|,\dots,|a_{K_*(\frac{\epsilon}{4})}|$, completes the proof.
We can obtain \eqref{app11} by using \eqref{rwrepF} and properties of $\mathcal{F}_{k,\underline{\eta}}$ as follows: Let us assume $y \in (0,1)$\footnote{The case $y=0$ or $y=1$, corresponds to the paths which are constructed by vectors of the form $(1,0),(1,-1)$, for $y=0$, and $(1,0),(1,1)$ for $y=1$. In either case, our argument is similar, where we can modify the set $\tilde{\mathcal{I}}_k$ and function $\mathcal{F}$ accordingly.}, fix $k \in \{1, \dots K_*(\frac{\epsilon}{4}) \}$, and take $N'$ such that for $n,l>N'$ we have: $K_*(\frac{\epsilon}{4})<[ny]<n-K_*{\frac{\epsilon}{4}}$ and $K_*(\frac{\epsilon}{4})<[ly]<l-K_*(\frac{\epsilon}{4})$. By this choice we can use \eqref{rwrepF} and observe:
	\begin{equation} \label{app12}
	\begin{split}
	&	\big|\mathbb{E}\big(\langle [ny] , (A_n^p-\alpha I_{n})^k [ny] \rangle_n\big)-\mathbb{E}\big(\langle [ly] , (A_l^p-\alpha I_{l})^k [ly] \rangle_l\big)\big| = \\
	& \bigg|\sum_{\underline{\eta} \in \tilde{\mathcal{I}}_k} \bigg( \mathbb{E} \Big( \mathcal{F}_{k,\underline{\eta}}\big(\underline{m}^k([ny],n),\underline{b}^k([ny],n)\big) \Big ) - \mathbb{E}\Big( \mathcal{F}_{k,\underline{\eta}}\big(\underline{m}^k([ly],l),\underline{b}^k([ly],l)\big) \Big ) \bigg ) \bigg|.
 	\end{split}
	\end{equation}
		Thanks to the third property of $\mathcal{F}_{k,\underline{\eta}}$, $\forall$ $\epsilon'>0$, there exists $\delta_{\underline{\eta}}(\epsilon')$ such that for $|\underline{b}^k([ly],l)-\underline{b}^k([ny],n)| \leq \delta_{\underline{\eta}}(\epsilon')$, we have $$\Big|\mathbb{E} \Big( \mathcal{F}_{k,\underline{\eta}}\big(\underline{m}^k([ny],n),\underline{b}^k([ny],n)\big) \Big ) - \mathbb{E}\Big( \mathcal{F}_{k,\underline{\eta}}\big(\underline{m}^k([ly],l),\underline{b}^k([ly],l)\big) \Big )\Big| \leq \epsilon'.$$
On the other hand, since $\beta(.)$ is continuous, there exist $N^k_{\underline{\eta}}$ such that for $n,l>N^k_{\underline{\eta}}$ and for all $-[\frac{k}{2}]-1<i<[\frac{k}{2}]+1$, we have  $$|\beta(\frac{[ny]+i}{n})-\beta(\frac{[ly]+i}{l})| \leq \frac{\delta_{\underline{\eta}}(\frac{\tilde{\epsilon}}{3^{K_*(\frac{\epsilon}{4})}})}{\sqrt{k+3}}.$$ Hence, thanks to the definition of $\underline{b}^k$ \eqref{mbdef}, for $n,l > N_{\underline{\eta}}^k$ we get $|\underline{b}^k([ly],l)-\underline{b}^k([ny],n)| \leq \delta_{\underline{\eta}}(\frac{\tilde{\epsilon}}{3^{K_*(\frac{\epsilon}{4})}})$.  Consequently, if we take $N_k= \max_{ \{ \underline{\eta} \in \tilde{\mathcal{I}}_k \} } \{ N_{\underline{\eta}}^k , N' \}$,  $\forall n,l > N_k$, we have $\forall \underline{\eta} \in \tilde{\mathcal{I}}_k $:

 $$\Big|\mathbb{E} \Big( \mathcal{F}_{k,\underline{\eta}}(\underline{m}^k([ny],n),\underline{b}^k([ny],n)) \Big ) - \mathbb{E}\Big( \mathcal{F}_{k,\underline{\eta}}(\underline{m}^k([ly],l),\underline{b}^k([ly],l)) \Big )\Big| \leq \frac{\tilde{\epsilon}}{3^{K_*(\frac{\epsilon}{4})}}. $$
 Combining the later with the estimate $|\tilde{\mathcal{I}}_k| \leq 3^k \leq 3^{K_*(\frac{\epsilon}{4})}$,  we get \eqref{app11}. Finally, taking $N_0 = \max_{\{ k \in \mathbb{I}_{K_*(\frac{\epsilon}{4})}\}} \{N_k, N'\}$ finishes the proof. \\ 
 Notice that in order to deal with the term $|\mathbb{E}( \langle [ny], \mathfrak{f}(A_n^r) [ny] \rangle_{n-1} - \mathbb{E}( \langle [ly], \mathfrak{f}(A^r_l) [ly] \rangle_{l-1}|$  one should properly modify the definition of $\mathcal{F}_{\underline{\eta},k} $ and $\tilde{\mathcal{I}}^{x,p}_{n,k}$. In particular, in this case $\mathcal{F}_{\underline{\eta},k}$ is given by: 
 \begin{equation} \label{app16}
		\begin{split} 	
 	&\tilde{\mathcal{F}}_{n,\underline{\eta}}(\underline{m}^k,\underline{b}^k)= \theta_{\eta_0\eta_1} \dots \theta_{\eta_i \eta_{i+1}} \dots \theta_{\eta_{k-1}\eta_k}, \\
 	&  \theta_{\eta_i,\eta_{i+1}} = 
 	\begin{cases}
		\underline{b}^k_{\eta_i}(\frac{\underline{b}^k_{\eta_i}}{m_{\eta_i+k}}+\frac{\underline{b}_{\eta_i+1}}{(m_{\eta_i+1)}} -\alpha \quad \text{if} \quad \eta_i=\eta_{i+1},	\\
	-\frac{\underline{b}^k_{hat{\eta_i}+1}}{m_{\hat{\eta}_i}} \sqrt{\underline{b}^k_{\hat{\eta}_i}\underline{b}^k_{\hat{\eta}_i+1}} \quad \text{if} \quad \eta_i \neq \eta_{i+1}.
	\end{cases} 
		\end{split}	
	\end{equation} 
		Since this function satisfies the same properties, the rest of the proof is exactly similar to the previous case.

 \end{proof}

As an obvious consequence of Lemma \ref{cauchyseqofexpectation} we have:
\begin{corollary} \label{fdeflimit}
	$\forall y \in [0,1]$,  the limit $\lim_{n \to \infty} \mathbb{E}(\expval{\tilde{e}_{[ny]}}_{\rho^n})$ exists, and the function $\mathrm{f}^{\mu}_{\beta}$ is well-defined. 
\end{corollary}
Moreover, following the proof of Lemma \ref{cauchyseqofexpectation}, we can deduce the following corollary as well:

\begin{corollary} \label{feqdef}
In thermal equilibrium i.e. when for $\beta_{eq} \in (0,\infty)$, $\beta(y)=\beta_{eq}$ is constant in $y$, we have: 
\begin{equation} \label{feqwd}
\forall y,y' \in (0,1), \quad \quad \mathrm{f}^{\mu}_{\beta}(y)= \mathrm{f}^{\mu}_{\beta}(y'). 
\end{equation} 
In particular, the function $\mathrm{f}^{\mu}(.)$ in \eqref{thermaleqfunc} is well defined. 
\end{corollary}
\begin{proof}
In order to proof \eqref{feqwd} it is enough to show that $\lim_{n \to \infty} \mathbb{E}(\expval{\tilde{e}_{[ny]}}_{\rho^{n,\beta_{eq}}})-\mathbb{E}(\expval{\tilde{e}_{[ny']}}_{\rho^{n,\beta_{eq}}})=0$. (We omit the subscript of $\rho$ since it is clear that we are in thermal equilibrium). Take $\epsilon>0$, first recall the expression of  $\expval{\tilde{e}_{[ny]}}_{\rho}$ \eqref{app22}, then rewrite $\mathfrak{f}(.)=\mathfrak{f}_{\prec}^{\frac{\epsilon}{4}}(.)+\mathfrak{f}_{\succ}^{\frac{\epsilon}{4}}(.)$ as it has been defined in \eqref{fepsilon} and observe:
\begin{equation} \label{app120}
\begin{split}
&\big|\mathbb{E}\big(\expval{\tilde{e}_{[ny]}}_{\rho^{n,\beta_{eq}}}\big)-\mathbb{E}\big(\expval{\tilde{e}_{[ny']}}_{\rho^{n,\beta_{eq}}}\big)\big| \leq\\
&\frac{1}{\beta_{eq}}\Big|\mathbb{E} \Big ( \langle [ny] , \mathfrak{f}_{\prec}^{\frac{\epsilon}{4}}(A_n^p) [ny] \rangle )+ \langle [ny] , \mathfrak{f}_{\prec}^{\frac{\epsilon}{4}}(A_n^r) [ny] \rangle \Big)-\mathbb{E} \Big( \langle [ny] , \mathfrak{f}_{\prec}^{\frac{\epsilon}{4}}(A_n^p) [ny] \rangle+ \langle [ny'] , \mathfrak{f}_{\prec}^{\frac{\epsilon}{4}}(A_n^r) [ny'] \rangle \Big) \Big|+ \\
&\epsilon
\end{split}
\end{equation}
where we bounded the terms involving $\mathfrak{f}_{\succ}^{\frac{\epsilon}{4}}$ by $\frac{\epsilon}{2}$, thanks to \eqref{taylorbound}.  Moreover,  $|\epsilon_{n}^{[ny]}|+|\epsilon_n^{[ny']}|$  is bounded by $\frac{C}{n}$; therefore,  it has been bounded by $\frac{\epsilon}{2}$, by taking $n>N_1$, for proper $N_1$.  Lastly, recall $K_*(\frac{\epsilon}{4})$ from \eqref{fepsilon}, and choose $N_2$ such that for $n>N_2$, $[\frac{K_*(\frac{\epsilon}{4})}{2}]+1<[ny]<n-[\frac{K_*(\frac{\epsilon}{4})}{2}]-1$, and $[\frac{K_*(\frac{\epsilon}{4})}{2}]+1<[ny']<n-[\frac{K_*(\frac{\epsilon}{4})}{2}]-1$. Thanks to this choice, and by using \eqref{rwrepF}, for any $k \in \{ 1,\dots , K_*(\frac{\epsilon}{4}) \}$ we have: 
\begin{equation} \label{app121}
	\begin{split}
	&	|\mathbb{E}(\langle [ny] , (A_n^p-\alpha I_{n})^k [ny] \rangle_n)-\mathbb{E}(\langle [ny'] , (A_n^p-\alpha I_{n})^k [ny'] \rangle_n)| = \\
	& \bigg|\sum_{\underline{\eta} \in \tilde{\mathcal{I}}_k} \bigg( \mathbb{E} \Big( \mathcal{F}_{k,\underline{\eta}}(\underline{m}^k([ny],n),\underline{b}^k([ny],n)) \Big ) - \mathbb{E}\Big( \mathcal{F}_{k,\underline{\eta}}(\underline{m}^k([ny'],n),\underline{b}^k([ny'],n)) \Big ) \bigg ) \bigg|=0,
 	\end{split}
	\end{equation}
where, first, we used  the fact that in thermal equilibrium we have $\underline{b}^k([ny],n)=\underline{b}^k([ny'],n)$, then we took advantage of the second property of $\mathcal{F}_{k,\underline{\eta}}$. Therefore, by using the definition of $\mathfrak{f}_{\prec}^{\frac{\epsilon}{4}}$, the term in second line of \eqref{app120} is zero, for $n>N_2$ (the part corresponding to  $A_n^r$ is completely analogous).  Hence taking $n> \max \{ N_1,N_2 \}$ gives us the desirable result.
\end{proof}

\begin{proposition} \label{fcont}
The function $\mathrm{f}^{\mu}_{\beta}(y): (0,1) \to \mathbb{R}$ is continuous.  
\end{proposition}
\begin{proof}
Fix $\epsilon>0$, and observe
\begin{equation} \label{app161}
|\mathrm{f}_{\beta}^{\mu}(y)- \mathrm{f}_{\beta}^{\mu}(y')| \leq |\mathrm{f}_{\beta}^{\mu}(y)- \mathrm{f}_{\beta,n}^{\mu}(y)| +|\mathrm{f}_{\beta,n}^{\mu}(y)- \mathrm{f}_{\beta,n}^{\mu}(y')|+ |\mathrm{f}_{\beta,n}^{\mu}(y')- \mathrm{f}_{\beta}^{\mu}(y')| ,
\end{equation}
where $\mathrm{f}^{\mu}_{\beta,n}(y):=\mathbb{E}(\langle \tilde{e}_{[ny]} \rangle_{\rho^n})$. Since $\lim_{n \to \infty} \mathrm{f}_{\beta,n}^{\mu}(y)=\mathrm{f}_{\beta}^{\mu}(y)$, if we take $n>N_1(\epsilon)$, then $|\mathrm{f}_{\beta}^{\mu}(y)- \mathrm{f}_{\beta,n}^{\mu}(y)| + |\mathrm{f}_{\beta,n}^{\mu}(y')- \mathrm{f}_{\beta}^{\mu}(y')| \leq \frac{\epsilon}{2}$. Moreover, we claim that there exist   $N_2(\epsilon)$, such that for $n>N_2(\epsilon)$, there exists $\delta$ such that for  $|y-y'| < \delta$, $|\mathrm{f}_{\beta,n}^{\mu}(y)- \mathrm{f}_{\beta,n}^{\mu}(y')| < \frac{\epsilon}{2}$. Proving this claim completes the proof, since we can take $n > \max \{N_1(\epsilon), N_2(\epsilon) \}$ and observe that for $|y-y'|< \delta$, we have $|\mathrm{f}_{\beta}^{\mu}(y)-\mathrm{f}_{\beta}^{\mu}(y')|<\epsilon$.

 However, the proof of this statement follows the same lines of Lemma \ref{cauchyseqofexpectation} and Corollary \eqref{feqdef}. Similar to \eqref{app120}, we divide  $\mathfrak{f}(.)=\mathfrak{f}_{\prec}^{\frac{\epsilon}{4}}(.)+\mathfrak{f}_{\succ}^{\frac{\epsilon}{4}}(.)$, and take $N_3(\epsilon)$ such that for $n>N_2(\epsilon)$ we have $[\frac{K_*(\frac{\epsilon}{4})}{2}]+1<[ny]<n-[\frac{K_*(\frac{\epsilon}{4})}{2}]-1$, and $[\frac{K_*(\frac{\epsilon}{4})}{2}]+1<[ny']<n-[\frac{K_*(\frac{\epsilon}{4})}{2}]-1$. Therefore, we have: 
  \begin{equation} \label{app1201}
\begin{split}
&\big|\mathbb{E}(\expval{\tilde{e}_{[ny]}}_{\rho^{n}})-\mathbb{E}(\expval{\tilde{e}_{[ny']}}_{\rho^{n}})\big| \leq\\
&\frac{1}{\beta(\frac{[ny]}{n})}\Big|\mathbb{E} \Big ( \langle [ny] , \mathfrak{f}_{\prec}^{\frac{\epsilon}{4}}(A_n^p) [ny] \rangle )+ \langle [ny] , \mathfrak{f}_{\prec}^{\frac{\epsilon}{4}}(A_n^r) [ny] \rangle \Big)-\mathbb{E} \Big( \langle [ny] , \mathfrak{f}_{\prec}^{\frac{\epsilon}{4}}(A_n^p) [ny] \rangle+ \langle [ny'] , \mathfrak{f}_{\prec}^{\frac{\epsilon}{4}}(A_n^r) [ny'] \rangle \Big) \Big| \\
&+\frac{\epsilon}{2}+|\epsilon_n^{[ny]}| + |\epsilon_n^{[ny']}|+C\Big(\beta^{-1}(\frac{[ny']}{n})-\beta^{-1}(\frac{[ny']}{n})\Big),
\end{split}
\end{equation}
 where we can find $N_4$  and $\delta_0$ such that  the last line will be bounded by $\frac{3 \epsilon}{4}$, for $n>N_4$ and $|y-y'|<\delta_0$. Let us take $N_2(\epsilon)= \max \{N_3(\epsilon),N_4 \}$. Recall the definition of $\mathfrak{f}_{\prec}^{\frac{\epsilon}{4}}(.)$ \eqref{fepsilon} as a Taylor sum up to $K_*(\frac{\epsilon}{4})$ terms. By using the choice of $N_2(\epsilon)$ rewrite each term of this sum as a sum over the paths $\underline{\eta} \in \tilde{\mathcal{I}}_k$ as in \eqref{rwrepF}. The rest of the  proof boils down to demonstrating the fact that for $n>N_2(\epsilon)$,  for all $1\leq k \leq K_*(\frac{\epsilon}{4})$, and  for all $\underline{\eta} \in \tilde{\mathcal{I}}_k$, we have: $\forall \hat{\epsilon}>0$  there exist $\delta_{k,\underline{\eta}}(\hat{\epsilon})>0$, such that if $|y-y'| < \delta_{k,\underline{\eta}}(\hat{\epsilon})$ then 
\begin{equation}\label{app171}
\Big|\mathbb{E} \Big( \mathcal{F}_{k,\underline{\eta}}(\underline{m}^k([ny],n),\underline{b}^k([ny],n)) \Big)-\mathbb{E} \Big ( \mathcal{F}_{k,\underline{\eta}}(\underline{m}^k([ny'],n),\underline{b}^k([ny'],n)) \Big)\Big| < \hat{\epsilon}.
\end{equation}
 However, recalling the definition of $\underline{b}^k([ny],n)$ from \eqref{mbdef}, since $\beta(.)$ is continuous, one can observe that if $|y-y'|< \tilde{\delta}$, $|\underline{b}^k([ny],n)-\underline{b}^k([ny'],n)|$ is sufficiently small for a proper choice of $\tilde{\delta}$. Therefore, by using the third property of $\mathcal{F}_{k,\eta}$, we obtain  the desired $\delta_{k,\underline{\eta}}(\hat{\epsilon})$. Finally, taking $\hat{\epsilon}=\frac{\epsilon}{3^{K_*(\frac{\epsilon}{4})}K_*(\frac{\epsilon}{4})4c}$, and $\delta= \min_{0<k \leq  K_*(\frac{\epsilon}{4})} \{\delta_0, \min_{\underline{\eta} \in \tilde{\mathcal{I}}_k} \{ \delta_{k,\underline{\eta}} \} \}$ finishes the proof. Notice that here $c$ is the uniform bound on Taylor coefficients $|a_0|,\dots |a_{K_*({\frac{\epsilon}{4}}})|$. Moreover, the choice of $\hat{\epsilon}$ is justified by the bound $|\tilde{\mathcal{I}}_k| \leq 3^k$. Furthermore, as usual the part corresponding to $r$ can be treated exactly in the same way. 
 \end{proof}

The next proposition proves \eqref{stefano}, and illustrates the fact that $\mathrm{f}_{\beta}^{\mu}(y)$ is in fact a function of inverse temperature at point $y$ i.e. $\beta(y)$. Since the proof of this proposition is similar to the previous lemma and proposition, we just sketch the proof and only highlight the differences:   
	
	\begin{proposition} \label{spro}
				Let $\beta \in C^0([0,1])$ satisfying the assumptions stated in the definition \eqref{initalstate}. Recall the definition of $\mathrm{f}^{\mu}_{\beta}:[0,1] \to \mathbb{R}$ from \eqref{fdef}, and $\mathrm{f}^{\mu}:(0,\infty) \to \mathbb{R}$ from \eqref{thermaleqfunc}, then we have $\forall y \in (0,1)$: 
				\begin{equation} \label{stefano2}
					\mathrm{f}^{\mu}_{\beta}(y)=\mathrm{f}^{\mu}(\beta(y)).
				\end{equation}
		\end{proposition}
\begin{proof}
	Fix $y \in (0,1)$ and recall that we denote the average in Gibbs state in thermal equilibrium at inverse temperature $\beta_{eq}$, with $\expval{.}_{\rho^{n,\beta_{eq}}}$. Since in thermal equilibrium, we have translation invariance in the limit thanks to 
\eqref{feqwd} in Corollary \ref{feqdef}, it is enough to prove that $\forall \epsilon>0$, there exist $N(\epsilon)$, such that for $n>N(\epsilon)$: 
\begin{equation} \label{b121}
	\big|\mathbb{E}\big(\expval{\tilde{e}_{[ny]}}_{\rho^n}\big)-\mathbb{E}\big(\expval{\tilde{e}_{[ny]}}_{\rho^{n,\beta(y)}}\big)\big| \leq 2\epsilon.
\end{equation} 
Let us denote the matrices corresponding to thermal averages at temperature profile $\beta(.)$ \eqref{thermalmatrices} by  $A_n^{p,\beta(.)}$ and $A_n^{r,\beta(.)}$, only for the sake of this proposition. Similarly, denote the same matrices in thermal equilibrium at temperature $\beta(y)$ by $A_n^{p,\beta(y)}$, $A_n^{r,\beta(y)}$. The proof goes as follows: we rewrite \eqref{b121} in terms of $\mathfrak{f}(A_n^{p,\beta(.)})$, $\mathfrak{f}(A_n^{r,\beta(.)})$, $\mathfrak{f}(A_n^{p,\beta(y)})$, and $\mathfrak{f}(A_n^{r,\beta(y)})$ thanks to \eqref{app22}, up to a vanishing error. Then we decompose $\mathfrak{f}(.)=\mathfrak{f}_{\prec}^{\frac{\epsilon}{4}}(.)+\mathfrak{f}_{\succ}^{\frac{\epsilon}{4}}(.)$ and we bound all the terms corresponding to $\mathfrak{f}_{\succ}^{\frac{\epsilon}{4}}(.)$ by $\frac{\epsilon}{2}$, thanks to \eqref{taylorbound}. Notice that the definition of $\mathfrak{f}$, $\mathfrak{f}_{\succ}^{\epsilon}$ and $\mathfrak{f}_{\prec}^{\epsilon}$ depends on the matrices through the constants $\alpha$ and $K_*(\epsilon)$. Here, we take the definition which is given by matrices $A_n^{p,\beta(.)}$ and $A_n^{r,\beta(.)}$. Then it is straightforward to check that  $\mathfrak{f}(A_n^{r,\beta(y)})$ and $\mathfrak{f}(A_n^{p,\beta(y)})$, are well defined and they satisfy the same bounds as in \eqref{taylorbound}, since $\beta_{\min}\leq\beta(y) \leq \beta_{max}$, and the same uniform bound $c_0$ in Lemma \ref{decaylemma} holds for $||A_n^{p,\beta(y)}||$ and $||A_n^{r,\beta(y)}||$. The terms involving $\mathfrak{f}^{\frac{\epsilon}{4}}_{\prec}$ can be treated similar to Proposition \ref{fcont}, we sketch the terms related to $p$, the other ones is similar. First, we choose $N_0(\epsilon)$ such that for $n> N_0(\epsilon)$, $K_*(\frac{\epsilon}{4}) \leq [ny] \leq n-K_*(\frac{\epsilon}{4})$, then we expand each term of the sum appearing in $\mathfrak{f}^{\frac{\epsilon}{4}}_{\prec}(.)$ by using the random walk representation in \eqref{rwrepF}, similar to \eqref{app171}, it is enough to show that for any $\hat{\epsilon}>0$, $1 \leq k \leq  K_*(\frac{\epsilon}{4})$  and $\underline{\eta} \in \tilde{\mathcal{I}}_k$ there exists $N$ such that for $n>N$, we have: 
\begin{equation} \label{app172}
 \Big|\mathbb{E} \Big( \mathcal{F}_{k, \underline{\eta}}(\underline{m}^k([ny],n),\underline{b}^k([ny],n)) \Big )-\mathbb{E} \Big (\mathcal{F}_{k, \underline{\eta}}(\underline{m}^k([ny],n),\tilde{\underline{b}}^k([ny],n)) \Big)\Big|  \leq \hat{\epsilon},
\end{equation}  
where $\tilde{\underline{b}}^k([ny],n)$ is defined analogous to $\underline{b}^k([ny],n)$ \eqref{mbdef} for a constant profile of temperature at inverse temperature $\beta(y)$ i.e. $\tilde{\underline{b}^k}([ny],n)_i=\beta(y)$ for $-[\frac{k}{2}] \leq i \leq [\frac{k}{2}]+1$. However, existence of $N$ such that for $n>N$ \eqref{app172} holds is evident from the third property of $\mathcal{F}_{k,\underline{\eta}}$ and the fact that $\beta(.)$ is continuous. 
\end{proof}
	Recall the definition of $Y^n_x=g(\frac{x}{n})(\expval{\tilde{e}_x}_{\rho^n}-\mathbb{E}(\expval{\tilde{e}_x}_{\rho^n}))$, and the sequence $n_k$ in the proof of \ref{SLLN}, where $n_k \to \infty$ with $\frac{n_{k+1}}{n_k} \to 1$. The following lemma was an essential part of the proof:
	\begin{lemma} \label{app2}
		Fix $\epsilon>0$, then for every realization of the masses, there exists $N_*$ such that, for every $n_k>N_*$ and every $n$ with $n_k \leq n < n_{k+1}$, we have: $\forall x \in \mathbb{I}_{n_k}$, with $n_*<x<n_k-n_*$, $|Y^n_x -Y^{n_k}_x| \leq \epsilon$. Here, $n_*$	 is a constant only depending on $\epsilon$.
		 \end{lemma}

\begin{proof}
	Fix $\epsilon>0 $, and recall the definition of $Y_x^n$, we have 
	\begin{equation} \label{edit3}
		\begin{split}	
	&\big|Y_n^x-Y_{n_k}^x\big|=\Big|g(\frac{x}{n})\expval{\tilde{e}_x}_{\rho^n}-g(\frac{x}{n})\mathbb{E}\big(\expval{\tilde{e}_x}_{\rho^n}\big)-g(\frac{x}{n_k})\expval{\tilde{e}_x}_{\rho^{n_k}}+g(\frac{x}{n_k})\mathbb{E}\big(\expval{\tilde{e}_x}_{\rho^{n_k}}\big)\Big|\leq \\ & \Big|g(\frac{x}{n})\expval{\tilde{e}_x}_{\rho^n}-g(\frac{x}{n_k})\expval{\tilde{e}_x}_{\rho^{n_k}}\Big|+\Big|g(\frac{x}{n})\mathbb{E}\big(\expval{\tilde{e}_x}_{\rho^n}\big)-g(\frac{x}{n_k})\mathbb{E}\big(\expval{\tilde{e}_x}_{\rho^{n_k}}\big)\Big|. 
		\end{split}	
	\end{equation}
	The second term in \eqref{edit3} can be treated by using Lemma \ref{cauchyseqofexpectation}, and continuity of $g$ (note that we used the choice of $n$ and $n_k$, where $\frac{n}{n_k} \to 1$, as well). Hence, there exits $N_1$, such that for $n_k>N_1$, we have: 
	$$\Big|g(\frac{x}{n})\mathbb{E}\big(\expval{\tilde{e}_x}_{\rho^n}\big)-g(\frac{x}{n_k})\mathbb{E}\big(\expval{\tilde{e}_x}_{\rho^{n_k}}\big)\Big| \leq c_1\Big|g(\frac{x}{n})-g(\frac{x}{n_k})\Big|+c_2\Big|\mathbb{E}\big(\expval{\tilde{e}_x}_{\rho^n}\big)-\mathbb{E}\big(\expval{\tilde{e}_x}_{\rho^{n_k}}\big)\Big|\leq \frac{\epsilon}{2},$$ 
	where $c_1$ is the bound on $|g|$, and $c_2$ is the bound on $|\mathbb{E}(\expval{\tilde{e}_x}_{\rho^{n_k}})|$.

	Similarly, for the first term in \eqref{edit3}  $|g(\frac{x}{n})\expval{\tilde{e}_x}_{\rho^n}-g(\frac{x}{n_k})\expval{\tilde{e}_x}_{\rho^{n_k}}|$, it is enough to deal with  $|\expval{\tilde{e}_x}_{\rho^n}-\expval{\tilde{e}_x}_{\rho^{n_k}}|$. 
	Thanks to the expression of $\expval{\tilde{e}_x}_{\rho^n}$ in \eqref{app22}, for proper $N_2$, we have for $n>N_2$: 
	\begin{equation} \label{app3}
		\begin{split}
			&c_1\big|\expval{\tilde{e}_x}_{\rho^n}-	\expval{\tilde{e}_x}_{\rho^{n_k}}\big| \leq c_1C_0\Big(\big|\langle x, \mathfrak{f}(A^p_n) x \rangle_{n}-\langle x, \mathfrak{f}(A^p_{n_k}) x \rangle_{n_k}\big|+\big|\langle x, \mathfrak{f}(A^r_n) x \rangle_{n-1}-\langle x, \mathfrak{f}(A^r_{n_k}) x \rangle_{n_k-1}\big| \\
			&+\frac{\epsilon}{6}\Big),
		\end{split}	
	\end{equation}
	where we chose $N_2$ such that $|\epsilon^x_n|+|\epsilon^x_{n_k}| \leq \frac{\epsilon}{6}$. From now on, let us show $\frac{\epsilon}{c_1C_0}$ by $\epsilon$. Now it is enough to find $N_3$, such that for $n_k>N_3$, $$\big|\langle x, \mathfrak{f}(A^p_n) x \rangle_{n}-\langle x, \mathfrak{f}(A^p_{n_k}) x \rangle_{n_k}\big| \leq \frac{\epsilon}{6}.$$   
Let us decompose $\mathfrak{f}(.)=\mathfrak{f}^{\frac{\epsilon}{12}}_{\prec}(.)+\mathfrak{f}^{\frac{\epsilon}{12}}_{\succ}(.)$ as in \eqref{fepsilon}, recall $K_*(\frac{\epsilon}{12})$ from \eqref{fepsilon} and let $n_*=K_*(\frac{\epsilon}{12})$, notice that $n_*$ only depends on $\epsilon$. 
Therefore, thanks to \eqref{taylorbound} we have:

\begin{equation} \label{app31}
\big|\langle x, \mathfrak{f}(A^p_n) x \rangle_{n}-\langle x, \mathfrak{f}(A^p_{n_k}) x \rangle_{n_k}\big| \leq  \big|\langle x, \mathfrak{f}_{\prec}^{\frac{\epsilon}{12}}(A^p_n) x \rangle_{n}-\langle x, \mathfrak{f}_{\prec}^{\frac{\epsilon}{12}}(A^p_{n_k}) x \rangle_{n_k}\big| + \frac{\epsilon}{12}.
\end{equation}
 Now, it is sufficient to show that 
\begin{equation} \label{tempapp}
\big|\langle x, \mathfrak{f}_{\prec}^{\frac{\epsilon}{12}}(A^p_n) x \rangle_{n}-\langle x, \mathfrak{f}_{\prec}^{\frac{\epsilon}{12}}(A^p_{n_k}) x \rangle_{n_k}\big| \leq \frac{\epsilon}{12},
\end{equation} 
for $n_k$ sufficiently large. In order to control this term, recall the definition of $\underline{b}^k$ from \eqref{mbdef} and notice that for any $\delta>0$, there exists $N_4(\delta)$ such that for $n_k>N_4(\delta)$, $|\underline{b}^k(x,n)-\underline{b}^k(x,n_k)| < \delta$, since $\beta$ is continuous, $n_k\leq n < n_{k+1}$, and $\frac{n_{k+1}}{n_k} \to 1$. By using the first property of $\mathcal{F}_{k,\underline{\eta}}$, we choose $\delta>0$ such that $|\mathcal{F}_{k,\underline{\eta}}(\underline{m}^k(x,n),\underline{b}^k(x,n))-\mathcal{F}_{k,\underline{\eta}}(\underline{m}^k(x,n_k),\underline{b}^k(x,n_k))| \leq \frac{\epsilon}{12c_1n_*3^{n_*}}$, where $c_1$ is the uniform bound on coefficients $a_0,\dots ,a_{n_*}$ in \eqref{fepsilon}.\\ Taking $n>\max\{N_1,N_2,N_3,N_4(\delta), n_* \}$, and combining this last estimate with the  expression of $\mathfrak{f}_{\prec}^{\frac{\epsilon}{12}}$ in \eqref{fepsilon} as well as the random walk representation relation \eqref{rwrepF}, gives us \eqref{tempapp} for any $n_*<x<n-n_*$. The term corresponding to $r$ can be bounded by $\frac{\epsilon}{6}$ similarly.    
\end{proof}

\section{Alternative Coordinates} \label{appendix0}
Since it is more fashionable to treat quantum harmonic chains in terms of $p$ and $q$ coordinates, in this section, we introduce our model in terms of these coordinates and rewrite the main transformations in terms of $q$ coordinates, in order to illustrate the link between our setup and the conventional one. 
  Define on the Hilbert space $L^2(\mathbb{R}^n)$, the following Hamiltonian operator: 
	\begin{equation} \label{Ham1}
	H_n=\frac{1}{2}\sum_{x=1}^n (\frac{p_x^2}{m_x} + (q_{x+1}-q_{x})^2),
 \end{equation}
	where for each $x \in \mathbb{I}_n$, $q_x$ denotes the position operator of the particle $x$, i.e. denoting the space variable by $\pmb{\zeta} \in \mathbb{R}^n$, for any $\pmb{\ket{\psi}} \in L^2(\mathbb{R}^n) $,  $q_x \pmb{\ket{\psi(\zeta_1,\dots,\zeta_x,\dots,\zeta_n)}}=\pmb{\zeta_x} \pmb{\ket{\psi(\zeta_1,\dots,\zeta_x,\dots,\zeta_n)}}$, and $p_x$ denotes the corresponding momentum operator $p_x=-i \partial \setminus \partial_{\pmb{\zeta}_x}$. Moreover, we assume the free boundary condition: $q_0=q_1$ and $q_n=q_{n+1}$. Here $m_x$ are positive i.i.d random variables. \\
	The Hamiltonian \eqref{Ham1} can be diagonalized with a linear transformation and written as sum of free bosons:
	\begin{equation} \label{Hamtemp}
		H_n = \hat{p}_0^2+\sum_{k=1}^{n-1}\omega_{k}(\hat{b}_k^* \hat{b}_k+\frac12),
	\end{equation}
	where $b_k$ and $b_k^*$ are bosonic annihilation and creation operators with commutation relations $[\hat{b}_k^*,\hat{b}_{k'}]=\delta_{k,k'}, [\hat{b}_k^*,\hat{b}_{k'}^*]=[\hat{b}_k,\hat{b}_{k'}]=0 $.  Here $\hat{p}_o=\sum_{x=1}^n p_x$ is the total momentum operator. Since  $[\hat{p}_0,H_n]=0$, and the Hamiltonian $H_n$ is translation invariant, after a straightforward analysis, the expression \eqref{Hamtemp} indicates that $H_n$ has a purely continuous spectrum (see Remark 3.3 of \cite{BSS2},  and Remark 3.6 of \cite{BSS3} for more details). However, one can observe that the Heisenberg dynamics generated by \eqref{Hamtemp} on $L^2(\mathbb{R}^n)$ is similar to \eqref{bogoliinv2}. Up to a $\hat{p}_0=(\sum_{x=1}^n m_x)^{-1} p_x$ operator which is constant in time.\\
	  Another technical problem arising in this description, concerns the density operator
	  $\rho^n_{\bar{p},\bar{r},\beta}=\exp(-\tilde{H}_{\beta}^n)$ \eqref{initalstate}.  The
	   pseudo-Hamiltonian $\tilde{H}_{\beta}^n$ can be written as 
	   \begin{equation} \label{tempinitial}
	   	\tilde{H}_{\beta}^n= \frac{\tilde{\mathfrak{p}}_0^2}{2}+ 
	   	\sum_{k=1}^{n-1} \gamma_k(\tilde{\mathfrak{b}}_k^*\tilde{\mathfrak{b}}_k +\frac12 
	   	),
	   \end{equation}
		where $\tilde{\mathfrak{b}}_k^*$ and $\tilde{\mathfrak{b}}_k$ are another set of 
		bosonic operators \eqref{bosonoperatorthermal}, \eqref{tempboson}, and 
		\begin{equation} \label{pothermal}
	\tilde{\mathfrak{p}}_0=(\sum_{x=1}^n\frac{m_x}{\beta_x})^{-\frac12}
	\big(\sum_{x=1}^n p_x -\Pi_0\big)= (\sum_{x=1}^n\frac{m_x}{\beta_x})^{-\frac{1}{2}}			(\hat{p}_o-\Pi_0),
	\end{equation}
	    Thanks to the definition of $\tilde{\mathfrak{p}}_0$ one can observe that similar 
	    to $H_n$, $\tilde{H}_{\beta}^n$ has a continuous spectrum as an operator on 
	    $L^2(\mathbb{R}^n)$ in this coordinates. Moreover, $\rho^n_{\bar{p},\bar{r},\beta}$
	    is not trace-class \textit{in this description}, anymore. 
	    Although, $\rho^n_{\bar{p},\bar{r},\beta}$ seems to be
	    a natural choice for our locally Gibbs state corresponding to $\bar{r}, \bar{p},
	    \beta$, we need to slightly modify it, in order to circumvent the above mentioned
	    technicalities.  
	    Recall that total momentum $\hat{p}_o= \sum_{x=1}^n p_x$
is conserved by the dynamics. Ideally, one would be
 tempted to fix the total momentum apriori to
the value prescribed by the macroscopic profile i.e.,
\begin{equation} \label{initialtotalmomentum2}
 \Pi_0 = \sum_{x=1}^n \bar{p}(\frac{x}{n})\frac{m_x}{\bar{m}}.
\end{equation}
However, this is not convenient for technical reasons, instead we modify our initial locally Gibbs state as follows: Let us denote $L^{2}(\mathbb{R}^n)$ by $\cH_n$, only in this section. Inspiring from diagonalization \eqref{hambosontemp}, we decompose $\cH_n$ as follows: Let 
$\tilde{\pmb{\zeta}}_0:= (\sum_{x=1}^n \frac{m_x}{\beta_x})^{-\frac12}
\sum_{x=1}^n \frac{m_x}{\beta_x} \pmb{\zeta}_x$, and 
$V_{\tilde{\pmb{\zeta}}_0}^{\perp} \subset \bR^n$, be the orthogonal complement of $\Span(\tilde{\pmb{\zeta}}_0)$. Denote the Lebesgue measure on  $V_{\hat{\pmb{\zeta}}_0}^{\perp} \subset \bR^n$, by 
$d\nu_{n-1}$, then we have: 
$\cH_n= L^2(\bR,d\tilde{\pmb{\zeta}}_0) \otimes L^2 (V_{\tilde{\pmb{\zeta}}_0}^{\perp}, d\nu_{n-1}) \equiv \cH_o \otimes \cH_{n-1}^-$. Again appealing to \eqref{hambosontemp}, 
observe that 
$\tilde{H}_{\beta}^{n}=\frac{1}{2}\tilde{\mathfrak{p}}_0^2+ H_{\beta}^{n,-}$, where 

\begin{equation} \label{pothermal00}
\tilde{\mathfrak{p}}_0=(\sum_{x=1}^n\frac{m_x}{\beta_x})^{-\frac12}
\big(\sum_{x=1}^n p_x -\Pi_0\big)= (\sum_{x=1}^n\frac{m_x}{\beta_x})^{-\frac{1}{2}}(\hat{p}_o-\Pi_0),
\end{equation}

only acts on $\cH_o$ and  
\begin{equation} \label{hamiltonianthermal2}
H_{\beta}^{n,-}:= \tilde{H}_{\beta}^{n}-\frac{1}{2}\tilde{\mathfrak{p}}_0^2,
\end{equation} 
only acts on $\cH_{n-1}^-$. \\
Take $\pmb{\ket{\phi}} \in L^2(\bR,d\tilde{\pmb{\zeta}}_0)$ such that 
$\pmb{\braket{\phi}{\phi}}=1$, 
$\pmb{\bra{\phi}} \tilde{\mathfrak{p}}_0 \pmb{\ket{\phi}}=0$,
$\pmb{\bra{\phi}} \tilde{\mathfrak{p}}_0^2 \pmb{\ket{\phi}}=1$, this means the total momentum
has the following average and uncertainty:
$\pmb{\bra{\phi}} \hat{p}_o \pmb{\ket{\phi}} = \Pi_0 $ and
 
$\pmb{\bra{\phi}} \hat{p}_o^2 \pmb{\ket{\phi}}- \Pi_0^2 = (\sum_{x=1}^n \frac{m_x}{\beta_x})$,
(See Remark \ref{initialstateremark}), then we define the locally Gibbs state with "fixed total momentum" as:
\begin{equation} \label{initalstatefixp}
\rho_{\bar{p},\bar{r},\beta}^n= \frac{1}{Z_n} \Big( \pmb{\dyad{\phi}} \otimes \exp(-H_{\beta}^{n,-})\Big),
\end{equation} 
where $\pmb{\dyad{\phi}}$ denotes the projection operator into the 
subspace spanned by the pure state  $\pmb{\ket{\phi}}$.
Observe that $H_{\beta}^{n,-}=\tilde{H}_{\beta}^{n}- \frac{\tilde{\mathfrak{p}}_0^2}{2}$ is essentially self-adjoint on $\mathcal{S}(\mathbb{R}^{n-1})$ (see e.g. \cite{simon}, \cite{NSS}, \cite{RB2}), and denote its closure with the same symbol. Notice that $H_{\beta}^{n,-}$ can be mapped into $H_{\beta}^n$ \eqref{hamiltoniantemp} by a unitary transformation.
Consequently,  one can check that $H_{\beta}^{n,-}$  has a discrete spectrum with non-negative eigenvalues (with a process similar to  Section \ref{pb}). We can express
$H_{\beta}^{n,-}$ in terms of the sum of free bosonic operators, and obtained the spectrum explicitly similar to what we did in Section \ref{pb}.  Hence, using spectral theory, one can observe that $\rho^n_{\bar{p},\bar{r},\beta}$ is well defined  and trace-class. 
 \\
 	Therefore, for every operator $a$, if $a\rho$ is a trace class operator, we can define the "average of the observable $a$ in the state $\rho$", i.e. $\expval{a}_{\rho}$ as:
\begin{equation} \label{mixedaverage00}
	\expval{a}_{\rho}=\Tr(\rho a).
\end{equation} 
As before, one can observe that $\expval{p_x}_{\rho}$, $\expval{r_x}_{\rho}$, and $\expval{e_x}_{\rho}$ are well defined, this suggests that \eqref{initalstatefixp} is an appropriate modification of \eqref{initalstate} in this coordinate. \\
In order to avoid the aforementioned difficulties, we describe our model in terms of elongation operators. One can argue that elongation operators are "physically" relevant, since in the classical counterpart of our system the elongation is the "real" physical variable, rather than the position of the particles. Let us highlight the relation between these models by a couple of remarks:

\begin{remark} \label{remarkdef}
Notice that same result of Theorem \ref{maintheorem} holds for this system, as well. In fact, initially the previous description can be mapped into this new description via a unitary 
transformation (See Remark 3.3 in \cite{BSS2}).\\
 The proof in this new coordinate is basically identical, except for some considerations concerning center of mass, which makes the proof even simpler. For example $\mathscr{E}_x^n$ in \eqref{momentumelongationaverage1} does not appear anymore thanks to the proper choice of $\pmb{\ket{\phi}}$.
  \\
In the previous description \eqref{Ham0}, by definition we have $\sum_{x=1}^n p_x=0$. This is because  we begin the description of the system by quantizing the classical description corresponding to the observer in center of the mass. Consequently, we should take $\int_0^1 \bar{p}(y) dy=0$. In this new coordinates, since we describe the center of mass separately by $\pmb{\ket{\phi}}$, we can take any $\bar{p} \in C^1([0,1])$ with non zero average. 
\end{remark}

\begin{remark} \label{initialstateremark}
Physically  the initial state \eqref{initalstatefixp}, means that initially we prepare our system in the lab such that our system's center of the mass is known and is given 
by the wave function 
$\pmb{\ket{\phi}}$, such that the above mentioned averages (momentum and kinetic energy 
contribution) is prescribed by the macroscopic profile of momentum and temperature, and  other degrees of freedom
are subjected to the thermal and mechanical fluctuations. Mathematically, this state is more 
convenient. 
We should emphasize  
that the assumption, $\pmb{\bra{\phi}} \hat{p}_o^2 \pmb{\ket{\phi}}- \Pi_o^2 = (\sum_{x=1}^n 
\frac{m_x}{\beta_x})$ is not crucial. Our result holds as long as we replace $(\sum_{x=1}^n 
\frac{m_x}{\beta_x})$ with any constant  of order $n$. One can replace the pure state 
$\pmb{\dyad{\phi}}$ with any mixed state $\rho_o$  acting on $\mathcal{H}_o$, with 
$\Tr(\rho_o)=1$, such that 
$$\Tr(\rho_o \hat{p}_o)=\Pi_0, \quad \quad \Tr(\rho_o \hat{p}_o^2) -\Pi_o^2 \sim \mathcal{O}(n), $$
and obtain the similar result in Theorem \ref{maintheorem}.\\
Notice that the aforementioned constants are random and they have been  
defined for a realization
of the masses. However, in the thermodynamic limit $\frac{\Pi_0}{n} \to \int_0^1 \bar{p}(y)dy,
$ and $\frac{1}{n} \sum_{x=1}^n \frac{m_x}{\beta_x} \to \int_{0}^1 \frac{\bar{m}}{\beta(y)}dy,
$ almost surely, thanks to law of large numbers, which further justifies our choice. \\
Finally, one can construct such $\pmb{\ket{\phi}}$ easily, via an inverse Fourier transform (in $\pmb{\tilde{\zeta}_0}$ variable) of a Gaussian function.
 
\end{remark}

	
	
	We show our main transformations in terms of $q$ coordinates: Rewrite the Hamiltonian in $q$ coordinate as:
	$$H_n=\frac12(\langle p, M^{-1} p \rangle_n + \langle \nabla_+ q , \nabla_+ q \rangle_{n-1}))= \frac12\langle p, M^{-1} p \rangle_n + \langle q,-\Delta q \rangle_n).$$
The proper transformation of $q$ in order to diagonalize the Hamiltonian is 
	$$\hat{q}_k= \langle \varphi^k, M^{\frac12}q \rangle_n = \sum_{x=1}^n \sqrt{m_x}\varphi^k_x q_x,$$
where $\hat{p}_k$ is defined as before. We can find the following relation between $\hat{q}_k$ and $\hat{r}_k$:
\begin{equation}
\begin{split}
\hat{r}_k = \omega_k\hat{q}_k.	
\end{split}
\end{equation}
 Canonical commutation relation (CCR) in terms of $q$ reads:
	\begin{align} \label{ccrtimeq}
	[\hat{q}_k,\hat{p}_{k'}]=i\delta_{k,k'}, \qquad [\hat{q}_k,\hat{q}_{k'}]=[\hat{p}_k,\hat{p}_{k'}]=0.  
	\end{align}

	The inverse is given by 
$$q=M^{-\frac12}O\hat{q}, \qquad q_x= \frac{1}{\sqrt{m_x}}\sum_{k=0}^{n-1} \varphi_x^k \hat{q}_k, $$
and the Hamiltonian reads
\begin{equation} 
		\begin{split}
			H_n = \frac{\hat{p}_0}{2}\frac12\sum_{k=1}^{n-1}( \hat{p}_k^2+\omega_k^2 \hat{q}_k^2),		
		\end{split}
	\end{equation} 

The bosonic operators have the following form in terms of $\hat{q}$ coordinates:
 $$\hat{b}_k=\frac{1}{\sqrt{2}}(\sqrt{\omega_k}\hat{q}_k+i\frac{1}{\sqrt{\omega_k}}\hat{p}_k), \quad \hat{b}_k^*= \frac{1}{\sqrt{2}}					(\sqrt{\omega_k}\hat{q}_k-i\frac{1}	{\sqrt{\omega_k}}\hat{p}_k).$$
 
 In order to expose the aforementioned link further, we introduce the coordinate $\tilde{\mathfrak{q}}_k$ similar to $\hat{q}_k$:  

	Define $\bar{q}_x$ as follows: First, construct  $\bar{q}_x$, for $x \in \mathbb{I}_n$,  from $\bar{r}_x$, by defining $\bar{q}_1=\bar{q}_0 = c$ ($c$ is an arbitrary constant, corresponding to the macroscopic position of the first particle) and let $\bar{q}_x= \sum_{y=1}^{x-1}  \bar{r}_y + \bar{q}_1$. 
Then, we have $ \tilde{q}_x= q_x-\bar{q}_x$, which gives us: 
$$\tilde{H}_{\beta}^n=\frac12 (\langle \tilde{p},M_{\beta}^{-1} \tilde{p} \rangle_n + \langle \nabla_+ \tilde{q}, 		\beta^o \nabla_+ \tilde{q} \rangle_{n-1} ) = \frac12 (\langle \tilde{p},M_{\beta}^{-1} \tilde{p} \rangle_n + \langle  \tilde{q},-\nabla_- \beta^o \nabla_+ \tilde{q} \rangle_n ).  $$
Therefore, $\tilde{\mathfrak{q}}_k$ is defined as: 
$$ \tilde{\mathfrak{q}}= O_{\beta}^{\dagger} M_{\beta}^{\frac12}\tilde{q}, \qquad \tilde{\mathfrak{q}}_k=\langle \psi^k,M_{\beta}^{\frac12} \tilde{q} \rangle_n = \sum_{x=1}^{n}  \sqrt{\frac{m_x}{\beta_x}}	\psi^k_x \tilde{q}_x. $$
Moreover, it is illuminating to know the relation between $\tilde{\mathfrak{r}}_k$ and $\tilde{\mathfrak{q}}_k$: 
$$ 
	\tilde{\mathfrak{r}}_k=\gamma_k \tilde{\mathfrak{q}}_k.	$$
 The inverse relation for $\tilde{\mathfrak{q}}_k$  is given by:
$$		\tilde{q}=M_{\beta}^{-\frac12} O_{\beta}\tilde{\mathfrak{q}}, \qquad \tilde{q}_x= \sqrt{\frac{\beta_x}{m_x}}\sum_{k=0}^{n-1} \psi^k_x\tilde{\mathfrak{q}}_k. $$
 Finally, the pseudo-Hamiltonian is diagonalized as follows: 
\begin{equation} \label{hambosontemp1}
		\begin{split}		
		\tilde{H}_{\beta}^n=\frac{\tilde{\mathfrak{p}}}{2}+\frac12\sum_{k=1}^{n-1} (\tilde{\mathfrak{p}}_k^2+\gamma_k^2\tilde{\mathfrak{q}}_k^2), 
		\end{split}
	\end{equation}

Commutation relation is given by $$[\tilde{\mathfrak{q}}_k,\tilde{\mathfrak{q}}_{k'}]=[\tilde{\mathfrak{p}}_k,\tilde{\mathfrak{p}}_{k'}]=0, \qquad   [\tilde{\mathfrak{q}}_k,\tilde{\mathfrak{p}}_{k'}] =i\delta_{k,k'}, \quad \forall k \in \mathbb{I}_{n-1}^0,$$
and the bosonic operators are given by: 
\begin{align} \label{tempboson}
&\tilde{\mathfrak{b}}_k=\frac{1}{\sqrt{2}}(\sqrt{\gamma_k} \tilde{\mathfrak{q}}_k +i \frac{1}{\sqrt{\gamma_k}}\tilde{\mathfrak{p}}_k), \\
	  	&\tilde{\mathfrak{b}}_k^*=\frac{1}{\sqrt{2}}(\sqrt{\gamma_k} \tilde{\mathfrak{q}}_k -i \frac{1}{\sqrt{\gamma_k}}\tilde{\mathfrak{p}}_k).
\end{align}

\section{Average of bosonic operators} \label{avgbos}
Recall the average of bosonic operators \eqref{bosonicthermalaverage} from Lemma \ref{avgboslem}. One can compute these averages and proof the lemma as follows (Notice 
that since $E_0$ is a constant, we can simply omit it from our computation and it does not 
change the desired averages) :
\begin{proof}
	First we compute $Z_n:=\Tr(\exp(-H_{\beta}^n))$, using spectral theory, we expand the trace in the basis of eigenvalues of $H_{\beta}^n$ i.e. $\{ \pmb{\ket{\bar{\theta}}} | \bar{\theta} \in \mathbb{N}^{n-1} \} $.  Then, the following computation gives us the result: (we abbreviate $\sum_{k=1}^{n-1} \theta_k \gamma_k$ by $\bar{\theta}.\bar{\gamma}$, where $\bar{\gamma}$ stands for the vector $\bar{\gamma}:=(\gamma_1,\dots,\gamma_{n-1})$) \footnote{We do not explain the proof in details since this is a classical result one can find  \cite{RB2}.}
		\begin{equation}
			\begin{split}
				Z_n&= \Tr(\exp(-H_{\beta}^n))= \sum_{\bar{\theta} \in \mathbb{N}^{n-1}} \pmb{\bra{\bar{\theta}}} \exp(-H_{\beta}^n) \pmb{\ket{\bar{\theta}}}
				= \sum_{\bar{\theta} \in \mathbb{N}^{n-1}} \pmb{\bra{\bar{\theta}}} \exp(-\bar{\theta}.\bar{\gamma}) \pmb{\ket{\bar{\theta}}} \pmb{\braket{\bar{\theta}}} \\
				&=\sum_{\bar{\theta} \in \mathbb{N}^{n-1}} \exp(-\bar{\theta}.\bar{\gamma})=\prod_{k=1}^{n-1} (\sum_{\theta_k=0}^{\infty} e^{-\theta_k \gamma_k})= \prod_{k=1}^{n-1} \frac{1}{1-e^{-\gamma_k}}.
			\end{split}
		\end{equation}
		
		 Now from \eqref{ccrtemp}, \eqref{energyterm}, observe that  $$\tilde{\mathfrak{b}}_k \pmb{\ket{(\theta_1,\dots,\theta_k, \dots,\theta_{n-1})}}=\sqrt{\theta_k}\pmb{\ket{(\theta_1,\dots,\theta_k-1, \dots,\theta_{n-1})}},$$
		 $$ \tilde{\mathfrak{b}}_k^* \pmb{\ket{(\theta_1,\dots,\theta_k, \dots,\theta_{n-1})}}=\sqrt{\theta_k+1}\pmb{\ket{(\theta_1,\dots,\theta_k+1, \dots,\theta_{n-1})}}.$$ 
		 Therefore, since $\pmb{\ket{\bar{\theta}}}$ are orthonormal, we have $$\forall \pmb{\ket{\bar{\theta}}},\quad \forall k, \quad  \pmb{\bra{\bar{\theta}}} \tilde{\mathfrak{b}}_k \pmb{\ket{\bar{\theta}}}=\pmb{\bra{\bar{\theta}}} \tilde{\mathfrak{b}}_k^* \pmb{\ket{\bar{\theta}}}=\pmb{\bra{\bar{\theta}}} \tilde{\mathfrak{b}}_k \tilde{\mathfrak{b}}_{k'} \pmb{\ket{\bar{\theta}}}=\pmb{\bra{\bar{\theta}}} \tilde{\mathfrak{b}}_k^* \tilde{\mathfrak{b}}_{k'}^*  \pmb{\ket{\bar{\theta}}}=0.$$ 
Moreover, if $k\neq k'$, we have $\quad \pmb{\bra{\bar{\theta}}} \tilde{\mathfrak{b}}_k \tilde{\mathfrak{b}}_{k'}^* \pmb{\ket{\bar{\theta}}}=\pmb{\bra{\bar{\theta}}} \tilde{\mathfrak{b}}_{k'}^* \tilde{\mathfrak{b}}_k \pmb{\ket{\bar{\theta}}}=0$. Since $\pmb{\ket{\bar{\theta}}}$ are orthonormal, if we expand $\Tr(\exp(-H_{\beta}^n))$ in the basis of $H_{\beta}^n$, we  deduce the first set of equalities in \eqref{bosonicthermalaverage}. Furthermore, for $k \neq k'$, we deduce$\expval{\tilde{\mathfrak{b}}_k \tilde{\mathfrak{b}}_{k'}^*}_{\rho}=\expval{\tilde{\mathfrak{b}}_{k'}^* \tilde{\mathfrak{b}}_k}_{\rho}=0$ as well.\\ 
On the other hand, if $k=k'$, the same relations \eqref{4terms}, \eqref{ccrtemp}imply: 
		
		$$\tilde{\mathfrak{b}}_k^*  \tilde{\mathfrak{b}}_k\pmb{\ket{(\theta_1,\dots,\theta_k, \dots,\theta_{n-1})}}=\theta_k \pmb{\ket{(\theta_1,\dots,\theta_k, \dots,\theta_{n-1})}}, $$ thus, we have $\pmb{\bra{\bar{\theta}}} \tilde{\mathfrak{b}}_k^* \tilde{\mathfrak{b}}_k\pmb{\ket{\bar{\theta}}} =\theta_k$. Consequently, we can compute 
	$\expval{\tilde{\mathfrak{b}}_k^* \tilde{\mathfrak{b}}_{k}}_{\rho}$:
	\begin{equation} 
	\begin{split}		
		\expval{\tilde{\mathfrak{b}}_k^* \tilde{\mathfrak{b}}_k}_{\rho} &= \Tr(\rho\tilde{\mathfrak{b}}_k^* \tilde{\mathfrak{b}}_{k'})=\frac{1}{Z_n}\Tr(e^{-H_{\beta}^n}\tilde{\mathfrak{b}}_k^* \tilde{\mathfrak{b}}_{k'})= \frac{1}{Z_n}\sum_{\bar{\theta} \in \mathbb{N}^{n-1}} \pmb{\bra{\bar{\theta}}} \exp(-H_{\beta}^n)\tilde{\mathfrak{b}}_k^* \tilde{\mathfrak{b}}_{k'}  \pmb{\ket{\bar{\theta}}} \\
		 &= \frac{1}{Z_n} \sum_{\bar{\theta} \in \mathbb{N}^{n-1}} \theta_k \pmb{\bra{\bar{\theta}}} \exp(-H_{\beta}^n) \pmb{\ket{\bar{\theta}}}= \frac{1}{Z_n}\sum_{\bar{\theta} \in {\mathbb{N}^{n-1}}} \theta_k \exp(-\sum_{j=1}^{n-1} \theta_j \gamma_j)\\
		 &=\frac{1}{Z_n} \frac{-\partial Z_n}{\partial \gamma_k} = \frac{1}{e^{\gamma_k}-1}.
	\end{split}
		 \end{equation}
	Since we have the commutator relation $[\tilde{\mathfrak{b}}_k ,\tilde{\mathfrak{b}}_k^*]=1$, we obtain the last equality: $\expval{\tilde{\mathfrak{b}}_k \tilde{\mathfrak{b}}_{k}^*}_{\rho}=\frac{1}{e^{\gamma_k}-1}+1$.
	\end{proof}

\medskip
\section*{Acknowledgement}
I would like to thank Stefano Olla for proposing this problem, his support and help during this project. I am also deeply grateful to  Fran\c{c}ois Huveneers for numerous enlightening discussions, his help and support. Moreover, I should mention that this work was partially supported by ANR-15-CE40-0020-01 LSD of the French National Research Agency. Finally, I would like to thank B. Doyon for pointing out \cite{doyon}.
\bibliographystyle{amsplain}
\bibliography{bibilo}

\end{document}